\documentclass[structabstract]{aa}  
\usepackage{graphicx}
\usepackage{lscape}
\usepackage{longtable}
\usepackage[]{natbib}
\usepackage{txfonts}
\usepackage{color}
\usepackage{url}
\begin{document}
\title{Radio jet emission from GeV-emitting narrow-line Seyfert 1 galaxies\thanks{Data displayed in
    Figs.~\ref{fig:lc_lmh_03n08} and \ref{fig:lc_lmh_09n15} (Table~\ref{tab:lc_sample} is an example) are
    only available in electronic form at the CDS via anonymous ftp to cdsarc.u-strasbg.fr (130.79.128.5) or
    via http://cdsarc.u-strasbg.fr}} \titlerunning{Radio jet emission from GeV-emitting NLSy1s}
% \subtitle{I. Overviewing the $\kappa$-mechanism}

\author{E.~Angelakis\inst{1}
%\fnmsep
%\thanks{Just to show the usage
 %   of the elements in the author field}
  \and   
  L.~Fuhrmann\inst{1}
  \and
  N.~Marchili\inst{2}
  \and
  L.~Foschini\inst{3}
  \and
  I.~Myserlis\inst{1}
  \and
  V.~Karamanavis\inst{1}
  \and
  S.~Komossa\inst{1}
  \and
  D.~Blinov\inst{4,5}
  \and
  T.~P.~Krichbaum\inst{1}
  \and 
  A.~Sievers\inst{6}
  \and 
  H.~Ungerechts\inst{6}
  \and 
  J.~A.~Zensus\inst{1}
}

\institute{Max-Planck-Institut f\"ur Radioastronomie, Auf dem H\"ugel 69, 53121 Bonn, Germany\\
  \email{angelaki@mpifr-bonn.mpg.de}
  %\thanks{The university of heaven temporarily does not
  %accept e-mails}
  \and
  Dipartimento di Fisica ed Astronomia, Universit\`{a} di Padova, via Marzolo 8, 35131
  Padova, Italy
  \and
  INAF – Osservatorio Astronomico di Brera, via E. Bianchi 46, 23807 Merate (LC), Italy
  \and
  Department of Physics and Institute of Theoretical \& Computational Physics, University of Crete, GR-710 03, Heraklion, Greece
  \and
  Astronomical Institute, St. Petersburg State University,Universitetsky pr. 28, Petrodvoretz, 198504 St. Petersburg, Russia  \and
  Instituto de Radio Astronom\'{i}a Milim\'{e}trica, Avenida Divina Pastora 7, Local 20 E 18012, Granada, Spain
  }

\authorrunning{Angelakis et al.} 

\date{Received:; accepted: }

% \abstract{}{}{}{}{} 
% 5 {} token are mandatory
 
  \abstract
  {With the current study we aim at understanding the properties of radio emission and the assumed
    jet from four radio-loud and $\gamma$-ray-loud narrow-line Seyfert 1 galaxies. These are Seyfert 1 galaxies with
    emission lines at the low end of the FWHM distribution.}
   {%The ultimate goal is the quantification of their variability properties at these
     %wavelengths and the investigation of whether this emission is indeed attributed to a
     %jet similar to those seen in typical blazars. At the same time we bring the
     %monitoring data to the disposal of the community for further investigations.  
     The ultimate goal is twofold: first we investigate whether a relativistic jet is operating at the source
      producing the radio output, and second, we quantify the jet characteristics to understand possible
     similarities with and differences from the jets found in typical blazars.}
   {We relied on the most systematic monitoring of radio-loud and $\gamma$-ray-detected narrow-line Seyfert 1
     galaxies in the cm and mm radio bands conducted with the Effelsberg 100~m and IRAM 30~m telescopes. It
     covers the longest time-baselines and the most radio frequencies to date. This dataset of
     multi-wavelength, long-term radio light-curves was analysed from several perspectives. We developed a
     novel algorithm to extract sensible variability parameters (mainly amplitudes and time scales) that were
     then used to compute variability brightness temperatures and the corresponding Doppler factors. The jet
     powers were computed from the light curves to estimate the energy output and compare it with that of typical
     blazars. The dynamics of radio spectra energy distributions were examined to understand the mechanism
     causing the variability.}
   {The length of the available light curves for three of the four sources in the sample allowed a firm
     understanding of the general behaviour of the sources. They all display intensive variability that
     appears to be occurring at a pace rather faster than what is commonly seen in blazars. The flaring events
     become progressively more prominent as the frequency increases and show intensive spectral evolution that
     is indicative of shock evolution. The variability brightness temperatures and the associated Doppler
     factors are moderate, implying a mildly relativistic jet. The computed jet powers show very
     energetic flows. The radio polarisation in one case clearly implies a quiescent jet
     underlying recursive flaring activity. Finally, in one case, the sudden disappearance of a $\gamma$-ray
       flare below some critical frequency in our band needs a more detailed investigation of the possible
       mechanism causing the evolution of broadband events.}
   {Despite the generally lower flux densities, the sources appear to show all typical characteristics seen in
     blazars that are powered by relativistic jets, such as intensive variability, spectral evolution across the different
     bands following evolutionary paths explained by travelling shocks, and Doppler factors indicating mildly
     relativistic speeds.}
  

   \maketitle
% ATTENTION notes just before submission: In th esecond version of th epaper we msut inlude:
% - compute the D fro equipartition ass TPK suggested 
% - compute jet power in more frequencies 
% - add references whenever you compare with blazard (stefanies comment)
% - plots the light curves that have the spectral indices as well 
% - put online aslo the spectral indices 
% - make suer that the values averag fluxes is really what ones gets from the online punblishe data and the
% indices too
%===================================================================
\section{Introduction}
\label{sec:introduction}
% TODO 140712:
%\red{(im: focus on radio properties so that (a) shorten introduction and (b) attract the reader's attent%ion)}
Seyfert galaxies were first identified as a distinct class of extragalactic systems by
\cite{1943ApJ....97...28S}, who studied the nuclear emission of six ``extragalactic nebulae'' with ``high
excitation nuclear emission lines superposed on a normal G-type spectrum''. The lines appeared to be
broadened, reaching widths of about 8500~km~s$^{-1}$. \citeauthor{1943ApJ....97...28S} also noted that the
maximum width of the Balmer emission lines increased with the luminosity of the nucleus as well as the ratio
between the light from the nucleus and the total light of the object, setting the scene for the class of
Seyfert galaxies that in summary show, bright, star-like nucleus, and broad emission lines that cover a wide
range of ionisation states. \cite{1974ApJ...192..581K} much later classified Seyfert galaxies into types 1 and
2 depending on whether the H$\beta$ lines (Balmer series) are broader than the forbidden ones, or of
approximately the same width. Typically, the width of forbidden lines in both classes are of the order of
``narrow emission lines'', that is, of about 300--800~km~s$^{-1}$, while the width of ``broad emission lines''
-- including the hydrogen Balmer lines -- is about 1000--6000~km~s$^{-1}$ \citep{Osterbrock:1984tl}.

\cite{1978ApJ...225..776D} found that MRK\,359 appeared, to be lying at the low end of the line-width
distribution and to have H$\beta$ and forbidden lines of comparable widths that were similar to those in
Seyfert 2 galaxies ($\approx 300$~km~s$^{-1}$); furthermore, MRK\,359 showed strong featureless continuum, and
strong high-ionisation lines (e.g. [\ion{Fe}{vii}] and [\ion{Fe}{x}]); these properties are common in Seyfert
1 galaxies, which led to the definition of yet another sub-class of AGN: the ``narrow-line Seyfert 1'' --
hereafter NLSy1 -- galaxies
\citep{1984ApL....24...43G,1985ApJ..297...166,1983ApJ...273..478O}. \cite{Koski:1978dp} and
\cite{1978ApJS...38..187P} noted that MRK\,42 showed a similar behaviour. \cite{1985ApJ..297...166} studied
eight such sources
% TODO 140712:
%\red{(im: 8 of them, by a quick view of the paper abstract.  careful though: only 2 of
%  them have strong FeII lines so maybe you need to edit the criteria below)} 
and concluded that they are characterised by (a) unusually narrow H$\beta$ lines, (b) strong \ion{Fe}{ii}
emission, (c) normal luminosities, and (d) H$\beta$ slightly weaker than in typical Seyfert 1
galaxies. Conventionally, today sources are categorised as NLSy1 if they show (a) a narrow width of the broad
Balmer emission line with a FWHM(H$\beta$) $\lesssim 2000$~km~s$^{-1}$, and (b) weak forbidden lines with
[\ion{O}{iii}]$\lambda$5007/H$\beta$ $<$ 3
\citep[][]{1985ApJ..297...166,1989ApJ...342..224G,2006ApJS..166..128Z}.

The first attempts to investigate the radio properties of these systems were undertaken by
\cite{1995AJ....109...81U}, among others, who studied seven NLSy1s with the Very Long Array (VLA) in A
configuration at 5~GHz. They found that the radio power was moderate ($10^{20-23}$~W~Hz$^{-1}$), the emission
is compact ($<300$~pc), and in the two of three cases where radio axes could be found and high optical
polarisation was detected, the radio axes were oriented perpendicularly to the electric vector position Angle
(EVPA); the third one showed EVPA nearly parallel to the radio axis.
% TODO 140712:
%\red{(im: correct: assuming NO intrinsic continuum polarisation. Also for the third galaxy
%  EVPA is nearly parallel to the radio axis)}. 
Later, \cite{Moran:2000cu} also studied 24 NLSy1s with the VLA in A configuration to confirm that most of the
sources were unresolved and that they showed relatively steep spectra. \cite{2003ApJ...588..746S} investigated
26 NLSy1 galaxies and found that only 9 were detected in the FIRST catalogue \citep{White1997ApJ}, and all
were radio-quiet. \cite{2006AJ....132..531K} studied 128 NLSy1s in a dedicated search for radio-loud (RL)
NLSy1s and concluded, among others, that (a) their morphology is similar to compact steep-spectrum sources
\citep[CSS;][discussed PKS\,2004$-$447 as an example of NLSy1 that was also CSS]{2006MNRAS.370..245G}; (b) the
radio-loudness $R$ -- defined as the ratio of the 6~cm flux to the optical flux at 4400~$\AA$
\citep{Kellermann1989AJ} -- is distributed smoothly up to the critical value of 10 and covers about four
orders of magnitude; (c) almost 7~\% of the NLSy1 galaxies are formally RL, but only 2.5~\% of them exceed a
radio index $R > 100$; (d) most RL NLSy1 are compact steep-spectrum sources, than accreting close to or above
the Eddington luminosity, $L_\mathrm{Edd}$; (e) the black-hole masses are generally at the upper
observed end for NLSy1, but are still smaller than what is typically seen in other RL AGNs.
 
Even before the detection of the RL subset of this class at high energies, it was clear that they comprise a
special source type for a number of reasons. \cite{2008ApJ...685..801Y} studied a sample of 23 NLSy1s with a
radio-loudness exceeding 100 to find that these RL AGNs may be powered by black holes of moderate masses
($10^{6}$--$10^7$~M$_{\sun}$) accreting at high rates and that they show a variety of radio properties
reminiscent of blazars. \cite{2006AJ....132..531K} pointed out that NLSy1s provide an excellent probe for
studying the physics scaling towards lower black-hole masses given their systematically low black-hole masses
in the range $10^{6}$--$10^{8}$~M$_{\odot}$. Despite the fundamental differences between RL NLSy1s and blazars
-- in terms of black-hole masses and accretion disc luminosities ($0.2$--$0.9$~$L_\mathrm{Edd}$) -- their jets
seem to be behaving similarly and share the same properties \citep [see also][]{2012nsgq.confE..10F}. That
alone has implications on the energy production and dissipation at different scales.

According to \cite{2006AJ....132..531K}, an important question arising from the behaviour of these systems is
whether or not the black-hole mass and radio-loudness are related and especially whether there is a limiting
black-hole mass above which objects are preferentially RL, and whether or not RL and radio-quiet galaxies show
the same spread in their black-hole masses. On this ground, \cite{2008RMxAC..32...86K} argued that the study
of NLSy1s needs to (a) include the radio and infrared properties of NLSy1 galaxies when correlation analyses
are performed, (b) determine the sufficient and necessary conditions for the onset of NLSy1 activity, and (c)
the investigation of whether the low black-hole mass is enough to explain the typically observed NLSy1
characteristics.

The first report for the detection of $\gamma$-ray emission from a source classified as an RL NLSy1 was given
by \cite{2009ApJ...699..976A} and \cite{2010ASPC..427..243F}, who discussed the detection of significant GeV
emission by the {\it Fermi}/LAT instrument from PMN\,J0948$+$0022. Already in its first year of operations,
{\it Fermi}/LAT detected a total of four NLSy1s \citep{2009ApJ...707L.142A}. As of today, there are seven RL
NLSy1s detected in the MeV--GeV energy bands five of which with high significance \cite[TS$>25$, see
also][]{Dammando:2013te, 2014arXiv1409.3716F}. A thorough review of the recent discoveries is given by
\cite{2012nsgq.confE..10F}.

%\footnote{An updated summary of the activities in this field can found at: \url{http://tinyurl.com/gnls1s}}.

%TODO 140712: 
%\red{+++ add reference to Luigi's last paper on th survey
%  saying it is a complete review.}
%\red{(+++ Main characteristics of
%  RL, Gamma-loud NLSY1s i.e. Blazar-like SED, +++ refs )}.

The multi-wavelength campaigns that followed these discoveries \citep[e.g.][]{2009ApJ...707..727A,
  2009ApJ...699..976A,2011AnA...528L..11G,2012A&A...548A.106F,2011nlsg.confE..26F} and the study of
subsequently discovered NLSy1s \citep{2009ApJ...707L.142A} showed a clear blazar-like behaviour, indicating
the existence of a relativistic jet viewed at small angles.

Despite the research activity that their high-energy detection has motivated, several questions remain
open. Especially the exact nature and properties of NLSy1s and the properties of the jet that seems to be
present. Particularly in the radio regime, where a jet is expected to be dominant, the studies conducted so
far have mostly been based on non-simultaneous datasets.

In the current work we quantify some of the properties of the assumed radio jet emission. The motivation for
this is to compare them with those seen in other classes of AGNs -- especially blazars. To do this, we studied
the radio behaviour of four RL NLSy1s detected in $\gamma$ rays by {\it Fermi}/LAT through regular multi-band,
single-dish radio monitoring with the Effelsberg 100~m and IRAM 30~m telescopes. We focus on the variability
properties -- especially brightness temperatures and variability Doppler factors, which will (in a later
publication) be combined with long-baseline radio interferometric measurements to estimate the viewing angles
and Lorentz factors (Fuhrmann et al., in prep.). The energetics of the observed radio outbursts and the
variability patterns of the broadband radio spectra are seen as indicators of the variability and emission
mechanism at play. Very early results have been presented in several other studies, including
\cite{2012MNRAS.426..317D}, \cite{2011nlsg.confE..26F}, and \cite{2012A&A...548A.106F}.

Here we present the longest-term, multi-frequency radio monitoring datasets of the known $\gamma$-ray-detected
and radio-loud NLSy1s to date. As an example, the 4.85~GHz light-curve lengths, range between longer than 1.5
years and longer than 5 years. The monitoring was conducted at eight different frequencies for the two
brightest sources and at six frequencies for the faintest.

The paper is structured as follows: in Sect.~\ref{sec:sample} we review some archival information about the
sources in our sample that is necessary for the following discussion. After describing the observing methods
in Sect.~\ref{sec:observations}, we continue in Sect.~\ref{sec:rlcs} with the phenomenological description of
the obtained light curves. The light curves discussed there comprise the basis for the flare decomposition
method presented in Sect.~\ref{sec:flaredecomp}. Subsequently, we continue in the frequency domain by studying
the radio spectra in Sect.~\ref{sec:spec_evolution}. In Sect.~\ref{sec:jetpow} we compute the radio powers for
the 14.60~GHz datasets, while in Sect.~\ref{sec:polarization} we discuss our polarisation
measurements. Summarising comments are discussed in Sect.~\ref{sec:polarization}, where we describe the
general connection of all our findings. The main findings of our study are presented in
Sect.~\ref{sec:conclusions}, which concludes the paper.

Throughout the paper we assume a $\Lambda CDM$ cosmology with $H_{0}=70$~km~s$^{-1}$~Mpc$^{-1}$ and
$\Omega_{\Lambda}=0.73$ \citep{2011ApJS..192...18K}.

\section{The sample: what is already known }
\label{sec:sample}
%-----------------------------------------------------------------------
% DONE: 140506
% locked
\begin{table*}[]
  \caption{The four monitored NLSy1s with their observed positions, redshift, and
    classification.}     
  \label{tab:sample}  
  \centering                    
  \begin{tabular}{llcccc} 
    \hline\hline                 
    Source ID &Survey name &RA           &DEC           &z  &Class  \\    
              &            &(hh:mm:ss.s) &(dd:mm:ss.s)  &   &       \\    
    \hline\\                        
%    J0324$+$3410 &B2\,0321+33B      &03:24:41.2  &$+$34:10:45.1 &0.0610$^\mathrm{a}$, 0.0629$^\mathrm{b}$  &Seyfert 1$^\mathrm{1}$, NLRG$^\mathrm{2}$, NLSy1$^\mathrm{3}$ \\ 
    J0324$+$3410 &B2\,0321+33B      &03:24:41.2  &$+$34:10:45.1 &0.0610$^\mathrm{a}$                      &NLSy1$^\mathrm{1}$ \\ 
                 &                  &            &              &0.0629$^\mathrm{b}$                      &                  \\ 
    J0849$+$5108 &SBS\,0846$+$513   &08:49:58.0  &$+$51:08:29.0 &0.584701$^\mathrm{c}$                    &NLSy1$^\mathrm{2}$ \\ 
    J0948$+$0022 &PMN\,J0948$+$0022 &09:48:57.3  &$+$00:22:25.6 &0.585102$^\mathrm{c}$                    &NLSy1$^\mathrm{3}$ \\ 
    J1505$+$0326 &PKS\,1502$+$036   &15:05:06.5  &$+$03:26:31.0 &0.407882$^\mathrm{c}$                    &NLSy1$^\mathrm{4}$ \\ 
  \\  \hline                                  
  \end{tabular}
  \tablebib{(a) \cite{1996MNRAS.281..425M}; (b) \cite{2007ApJ...658L..13Z}; (c)
%    \cite{2010MNRAS.405.2302H}; (1) \cite{1993AJ....105.2079R}; (2)
    \cite{1996MNRAS.281..425M}; (1) \cite{2007ApJ...658L..13Z}; (2)
    \cite{2005ApJ...630..122G}; (3) \cite{2003ApJ...584..147Z}; (4) \cite{2006ApJS..166..128Z}}
\end{table*}
% -----------------------------------------------------------------------
We here investigate a sample of four sources, which are  
\begin{enumerate}
\item classified as NLSy1s,
\item radio-loud,
\item detected at $\gamma$ rays by the {\it Fermi}/LAT, and
\item satisfy certain observational constrains such as declination limits and a sufficiently high flux density
  to enable quality measurements at the Effelsberg ($\gtrsim 0.1$~Jy) and IRAM telescopes ($\gtrsim 0.3$~Jy).
\end{enumerate} 
%The goal of this effort is the systematic study of their radio emission -- assuredly
%attributed to relativistic jets operating at the sources -- in both time and spectral
%domain. 
We briefly summarise what is already known of these sources. Table~\ref{tab:sample} summarises their
positions, redshifts, and classifications.
% TODO 140702:
% \red{(+++ add reference to a summarising table)}.

\subsection{J0324$+$3410 (1H\,0323$+$342)} 
The source J0324$+$3410 is of special interest mostly because of the morphology of its host galaxy and the
estimated black-hole mass. From two HST archival images (each of 200~s exposure), \cite{2007ApJ...658L..13Z}
concluded that it is hosted by a one-armed spiral galaxy.  The authors showed that its radio and X-ray
emission can be explained by jet synchrotron radiation, while the infrared and optical light is dominated by
thermal emission from a Seyfert nucleus. \cite{2008A&A...490..583A} conducted $B$- and $R$-band observations
with the Nordic Optical Telescope (NOT).
%to find that the host seems to have a
%rather peculiar morphology in agreement with the previous authors. Interestingly, 
They suggested that its host resembles the morphology seen in the inner parts of Arp\,10 found by
\cite{1996ApJ...460..686C}, implying that J0324$+$3410 may be a merger remnant. Following the method discussed
by \cite{2005ApJ...630..122G}, they estimated the central black-hole mass to be $10^7~\mathrm{M}_{\sun}$; this
value lies in the overlapping region between the black-hole mass distributions for NLSy1s and blazars and
is similar to the value published earlier by \cite{2007ApJ...658L..13Z}.

Although the only case of an RL, {\it Fermi}-detected NLSy1 for which the host is well resolved, J0324$+$3410,
challenges the assumptions that powerful relativistic jets form only in giant elliptical galaxies
\citep[e.g.][]{2002ApJ...564...86B,2007ApJ...658..815S}.
%Classical, nearby, broad-line Seyfert 1 galaxies mostly reside in disk galaxies.
%TODO 140712: 
%\red{(+++ refs from Damandos 0846 paper with marscher et al.

\cite{2009ApJ...707L.142A} presented a thorough study including spectral energy distribution (hereafter SED)
fits. The presence of a jet was already apparent in these studies; but in an episodic manner and not as a
fully developed constantly broadband-emitting jet. The computed jet power placed the source in the BL\,Lac
range (see jet power computations in Sect.~\ref{sec:jetpow}). It is worth noting that they computed accretion
rates that reach extreme values of up to 90~\% of the Eddington luminosity, yet another peculiar property of
the RL, $\gamma$-ray-loud NLSy1s.

% TODO 140702: 
% \red{here focus on anwerin the question: everybody says that NLSy1s are in spiral etc
% etc so it is challenginf this and that. Is that so? Investigate the hists of these
% sources. See damandos papr 2 on 0846 as well. Damadno et al in paper 2013 has acomment
% about this for 0323 and another galaxy }
% \red{\url{http://arxiv.org/abs/1312.3118}}

% which have the bulk lying within 106–107 M, (e.g., Wang \& Lu 2001; Grupe \& Mathur
% 2004; Zhou et al. 2006) and 107–109 M, (e.g., Woo et al. 2005; Falomo et al. 2002),
% respectively.  2012ApJ...744..177L give the radio structure, redhsift the casse etc
% While the empirical relations of Vester-gaard \& Peterson (2006) give BH masses of 3 107
% M using ,the continuum luminosity at 5100 A ̊ and 1.8 107 M, usingthe Hb
% luminosity. These BH mass estimates are consistentwithin their uncertainties, giving M ∼
% 107 M .

\subsection{J0849$+$5108 (SBS\,0846$+$513)}
J0849$+$5108 was identified as an NLSy1 galaxy by \cite{2005ChJAA...5...41Z}. They found clear evidence for
emission from a relativistic jet and a stellar component. Additionally, the emission lines show
characteristics that classify it as a typical NLSy1 (FWHM(H$\beta$)$\simeq 1710$~km~s$^{-1}$,
[\ion{O}{iii}]$\lambda_{5007} \simeq 0.32\ \mathrm{H}\beta$ and strong \ion{Fe}{ii} emission). The fact that
the source was previously classified as a BL~Lac object \citep{1979ApJ...230...68A} makes the case especially
interesting, possibly providing a handle on the link between BL~Lac objects and NLSy1s.

\cite{2008ApJ...685..801Y} included the source in a very thorough study of a sample of 23 RL NLSy1 galaxies
and reported a black-hole mass of about $10^{7.4}$~M$_{\sun}$. On the basis of the detected significant
optical polarisation, they claimed that it has a jet. Finally, they found that during the high-energy states
the optical continuum is featureless, while at low states strong emission lines become obvious; this suggests
that the source maybe a transition between a quasar and a BL Lac that displays its latter character only at
flaring states.

% TODO 140712: 
% \red{(Arp et al. 1979; Stickel et al. 1989)} 

\cite {2011nlsg.confE..24F} was the first to report its detection at $\gamma$~rays. Later,
\cite{2012MNRAS.426..317D} discussed a flaring episode around June--July 2011. The multi-frequency datasets
collected during the campaigns following the detection are partly discussed in \cite{Dammando:2013hu}. Using
the Very Long Baseline Array (VLBA) at 5, 8.4 and 15~GHz \cite{2012MNRAS.426..317D} resolved the otherwise
compact source to reveal a core-jet structure. The feature attributed to the jet shows a steep spectrum. The
core has been decomposed into two compact components.  They discussed the detection of an apparent speed of
about 8~c, indicating a relativistic jet. In summary, the power output (isotropic $\gamma$-ray luminosity) of
$10^{48}$~erg~s$^{-1}$ on a daily scale (similar to that of luminous flat-spectrum radio quasars, FSRQs), the
apparent superluminal motions and the radio variability accompanied by spectral evolution indicate blazar-like
relativistic jet.

\cite{2012MNRAS.426..317D} also considered the suggestion of \cite{2008MNRAS.387.1669G} and
\cite{2011MNRAS.414.2674G} that the transition between FSRQs and BL\,Lac can be interpreted in terms of
different accretion rates. They found that the position of the source in a $\gamma$-ray photon index against
luminosity plot ($\Gamma_\mathrm{\gamma}-L_\mathrm{\gamma}$) also places it in the typical blazar territory.
 
\subsection{J0948$+$0022 (PMN\,J0948$+$0022)}
J0948$+$0022 shows the usual characteristics of an NLSy1 optical spectrum \citep{2002AJ....124.3042W}
% Indeed Williams et all were the first to classify this as NLSy1s
with a FWHM(H$\beta$) of 1500$\pm$55~km~s$^{-1}$, \ion{O}{iii}/H$\beta<3$, and strong optical \ion{F}{ii}
emission \citep{2003ApJ...584..147Z}. Although NLSy1s are usually radio quiet
\citep[e.g.][]{1995AJ....109...81U,2006AJ....132..531K}, J0948$+$0022 was found to be the first {\it very}
radio-loud NLSy1 with a $R>10^{3}$ \citep{2003ApJ...584..147Z}. In addition, \cite{2003ApJ...584..147Z}
reported an inverted radio spectrum and high brightness temperatures, which strongly supports that there is a
relativistic, Doppler-boosted jet seen at small viewing angles. The source has also been the first NLSy1
detected at high-energy $\gamma$ rays by {\it Fermi}/LAT during its first months of operation. {\it Fermi}/LAT confirmed
the relativistic jet emission from this source and established NLSy1s as a new class of $\gamma$-ray emitting
AGNs \citep{2010ASPC..427..243F,2009ApJ...699..976A,2009ApJ...707..727A}.

At pc scales, the source appears as one-sided VLBI structure dominated by a compact ($<55$~$\mu$as), bright
central component \citep{2006PASJ...58..829D,2011AnA...528L..11G}.  Kpc-scale radio emission has also been
found by \cite{2012ApJ...760...41D} with a two-sided extension of the core and a northern extent of 52~kpc in
projected distance. From these studies a relativistically boosted jet with Doppler factors $D>1$
\citep[e.g. $D>2.7-5.5$][]{2006PASJ...58..829D} has been inferred, in agreement with Doppler boosting
inferred from previous variability studies \citep[e.g.][]{2011nlsg.confE..26F}.

This NLSy1 is variable across the whole electromagnetic spectrum. Intense multi-wavelength campaigns performed
over the past years revealed continuous activity with an exceptional outburst in 2010
\citep[][]{2011arXiv1110.5649F} that occurred at all spectral bands. Together with detailed SED modelling and
the detection of polarised emission, these findings confirm the powerful relativistic jet in the source seen
at small viewing angles.

\subsection{J1505$+$0326 (PKS\,1502$+$036)}
Based on its optical characteristics (FWHM(H$\beta)=1082\pm113$~km~s$^{-1}$, [\ion{O}{iii}]/H$\beta$$\sim
1.1$, and a strong optical \ion{Fe}{ii} bump), this source was classified as NLSy1
\citep[e.g.][]{2008ApJ...685..801Y}. It exhibits one of the highest radio-loudness parameters among
  NLSy1s ($R=1549$,
  see also Sect.~\ref{sec:RL}). Early VLBI observations revealed a compact source (marginally resolved at
  milli-arcsecond scales) with an inverted radio spectral index \citep{1998A&AS..129..219D}.

  \cite{2009ApJ...707L.142A} reported the first detection of $\gamma$-ray
  emission from the source by {\it Fermi}/LAT. From SED modelling the authors inferred jet powers similar to
  those in J0948$+$0022
  and those typically observed in powerful quasars. A detailed multi-wavelength campaign carried out between
  2008 and 2012 did not reveal any significant variability at $\gamma$
  rays, in contrast to prominent flux density and spectral variability seen in the radio regime
  \citep{2013MNRAS.433..952D}.  The 15~GHz VLBI imaging showed a one-sided core-jet structure on pc scales. No
  significant proper motion of jet components was detected. The radio variability and VLBI findings together
  with an inferred high apparent isotropic $\gamma$-ray
  luminosity strongly support a relativistic, Doppler-boosted jet in this case as well.

%{\bf RECIPE: \\
%- classification\\
%- variability in radio and in optical\\
%- optical host morphology\\
%- radio morphology \\
%- jet presence\\
%- gamma power\\
%- optical mag \\
%- opticak polarization \\
%-  RL\\
%- SED\\ 
%- MBH\\ }   

%\blue{\\NOTES:
%\begin{itemize}
%\item Hyper: We must study if their properties lie between FSRQs and BLLacs,  Abdo et al 2009 : p.737: We have shown that the variability at multiple
 % wavebands and the physical parameters resulting from modelling the SEDs are typical of a
 % source midway between FSRQs and BL Lacs. -- Highlighted jul 12, 2013
%%\item start with what is know todate how many are found in gamm arays and why we shoe only
 % those 4.
%\item Their clasification and hosts
%\item find the hosts for every source and the classification it has received
%\item archival data for the sources (everything in NED xray gamma ray Fermi etc)
%\item \url{http://www.brera.inaf.it/utenti/foschini/gNLS1/catalog.html}\\
%\item integral observations by OMAR\\
%\end{itemize}
%}

%===================================================================
\subsection{Radio-loudness}
\label{sec:RL}
Of  the four sources in our study, J0948$+$0022, J0849$+$5108, and J1505$+$0326 are very RL, with
$R_{1.4}= 194-793$ \citep{2006AJ....132..531K},1445 and 1549 \citep{2008ApJ...685..801Y}, respectively;
where $R$ is the radio index according to \cite{Kellermann1989AJ}. J0324$+$3410 is only mildly RL with
$R_5=38-71$ \citep{2007ApJ...658L..13Z}. We note that each value of $R$ is uncertain by a factor of a
few to 10, reflecting uncertainties in extinction correction, optical host contribution, and
variability in the radio or optical bands.
% Unless the source undergoes changes beyond factors of the order of 100, the $R$ index would not change
% noticeably and reliably (due to an inherit uncertainty in $R$ of the order of 10, $R$ values would change
% noticeably in variability by factors greater than 100).
The monitoring presented here, reveals variability in the radio bands of up
to several ten of percent (see Table~\ref{tab:summary}), but not high enough to
modify the previous source classification as RL or radio-quiet. The values of the radio indices
reported in the literature are therefore adequate for the scope of this paper and are shown in Table~\ref{tab:Rindices}.
%Because none of our sources varied by more than a factor of 100, obviously, there is
%no need whatsoever to re-compute $R$.
% -----------------------------------------------------------------------
\begin{table}[]
  \caption{Radio-loudness of the sources in our sample. The moderate
    variability cannot cause remarkable changes in the $R$ indices, 
    making the published values representative enough.}
  \label{tab:Rindices}  
  \centering                    
  \begin{tabular}{llp{4cm}} 
    \hline\hline                 
    Source  &Radio Index \tablefoottext{a} &Reference\\
    \hline
    \\
J0324$+$3410   &$R_{5}  =38-71$    &(1) based on HST optical flux\\
J0948$+$0022        &$R_{1.4}=194-793$  &(2) based on USNO and GSC catalogues (note the variable optical flux)\\
J0849$+$5108    &$R_{1.4}=1445$     &(3) based on SDSS optical flux\\
J1505$+$0326     &$R_{1.4}=1549$               &(3) \ldots \\\\
    \hline                                  
  \end{tabular}
  \tablefoot{
    \tablefoottext{a}{With $R_{5}=S_{5\mathrm{GHz}}/S_{\mathrm{B}}$,
      $R_{1.4}=S_{1.4\mathrm{GHz}}/S_{\mathrm{4400A}}$ Kellermann's radio index. Under the
      assumptions of Kellermann et al.: $R_{1.4} = 1.9 R_{5}$.}
  }
\tablebib{
(1)~\citet{2007ApJ...658L..13Z}; (2)~\citet{2006AJ....132..531K}; (3)~\citet{2008ApJ...685..801Y}.
}
\end{table}
% -----------------------------------------------------------------------
%)1 

%
%\blue{\\NOTES:
%\begin{itemize}
%\item see Zhou et al 2005 amazing description of importance of the RL also for the
%  discussion part. See their discussion 
%\item see Foschini 2011 2011RAA....11.1266F for formulas for calculating the jet powers etc 
%\end{itemize}
%}

%===================================================================

%===================================================================
\section{Observations and data reduction}
\label{sec:observations}
%The current paper is concerned with the broad-band radio behaviour of the four RL NLSy1s. 
The light curves and radio SEDs presented here have been observed in the framework of the F-GAMMA programme
\citep{fuhrmann2007AIPC,2010arXiv1006.5610A}. They cover the frequency range from 2.64 to 142.33~GHz. Below
43.05~GHz the measurements were conducted with eight different heterodyne receivers mounted on the secondary
focus of the 100~m Effelsberg telescope. The observations at 86.24 and 142.33~GHz were obtained with the 30~m
IRAM telescope. In Table~\ref{tab:receivers} their most important operational characteristics are summarised .

\subsection{Effelsberg observations}
The Effelsberg station covers the band between 2.64 and 43.05~GHz. The systems at 4.85, 10.45 and 32~GHz are
equipped with multiple feeds allowing differential measurements meant to remove atmospheric effects
(e.g. emission or absorption fluctuations).
% For completeness it must be noted that this method is mostly
%effective in cases of linear tropospheric variations owing mostly to the only partial overlap between the
%atmosphere columns ``seen'' by the feeds.
The rest are equipped with only single feeds. All receivers have circularly polarised feeds. The observations
are made in cross-scanning mode, that is monitoring the telescope response while driving over the source
position in two different orthogonal directions (in our case, azimuth and elevation). Necessary
post-measurement corrections include the following:
\begin{enumerate}
\item Pointing-offset correction: meant to correct for the power loss caused by possible divergence of the
  actual source position from that observed by the telescope. This effect is of the order of a few percent.
\item Atmospheric-opacity correction: meant to correct for the attenuation caused by the transmission through
  the terrestrial atmosphere. This effect can be significant especially at higher frequencies where the
  atmospheric opacity becomes significantly high.
\item Elevation-dependent gain correction: correcting for sensitivity losses caused by small-scale departures
  of the primary reflector's geometry from that of an ideal paraboloid (as expected for a homology-designed
  reflector). The magnitude of this effect is constrained to within a few percent.
\item Absolute calibration: converting the measured antenna temperatures to SI units by reference to standard
  candles, that is, calibrators. The standard candles used for our programme along with the flux densities
  assumed for them are shown in Table~\ref{tab:calibrators}. Specifically, in the case of NGC\,7027, a partial
  power loss caused by its extended structure relative to the beam size above 10.45\,GHz was corrected for.
% the next table I produce by running the fit of a constant function on differences
% between steps f corrections. Eg. 
% fit [][0:] f(x)'LIST.13.all.sens' u :(100*(($12-$10)/$12)) via c
% 
% Th average offsets for each frequency I got them by 
% fit [][:100]f(x) 'test' u : (abs(($11))) via c
% -----------------------------------------------------------------------
\begin{table}[]
  \caption{Flux densities of the standard calibrators used at Effelsberg.}
  \label{tab:calibrators}  
  \centering                    
  \begin{tabular}{cccccc} 
    \hline\hline                 
    Source:    &3C\,48 &3C\,161&3C\,286&3C\,295&NGC\,7027\tablefootmark{a}\\
    \hline\\
    $S_{2.64}$ &9.51   &11.35    &10.69   &12.46    &3.75 \\
    $S_{4.85}$ &5.48   &6.62     &7.48    &6.56     &5.48 \\
    $S_{8.35}$ &3.25   &3.88     &5.22    &3.47     &5.92  \\
    $S_{10.45}$&2.60   &3.06     &4.45    &2.62     &5.92  \\
    $S_{14.60}$&1.85   &2.12     &3.47    &1.69     &5.85 \\
    $S_{23.05}$&1.14   &1.25     &2.40    &0.89     &5.65 \\
    $S_{32.00}$&0.80   &0.83     &1.82    &0.55     &5.49 \\
    $S_{43.00}$&0.57   &0.57     &1.40    &0.35     &5.34 \\
    \hline                                  
  \end{tabular}
  \tablefoot{
    \tablefoottext{a}{ The flux density of NGC\,7027 is corrected for
      beam-extension at frequencies above 10.45\,GHz.}}
  \tablebib{The flux densities of the calibrators are taken from
    \cite{Ott1994}, \cite{Baars1977AnA}, \cite{2008ApJ...681.1296Z}, and Kraus priv. comm.}
\end{table}
% -----------------------------------------------------------------------
\end{enumerate}
The five-year mean magnitudes of these effects as they are observed by the F-GAMMA programme are summarised in
Table~\ref{tab:corrections}.

\subsection{IRAM observations}
Observations with the IRAM 30~m telescope were made within the F-GAMMA monitoring programme and the more
general flux monitoring conducted by IRAM \citep[Institut de Radioastronomie
Millim\'{e}trique,][]{1998ASPC..144..149U}. Data from both programmes are included in this paper. The
observations were conducted with the newly installed EMIR receiver \citep[][]{2012A&A...538A..89C} using the 3
and 2~mm bands (each with linear polarisation feeds) tuned to 86.24 and 142.33~GHz and the narrow-band
continuum backends (1~GHz bandwidth) attached. Observations of J0324$+$3410 and J0948$+$0022 were performed
with cross-scans in azimuth and elevation direction and wobbler-switching along azimuth with a frequency near
2~Hz. Each cross-scan was preceded by a calibration scan to obtain instantaneous opacity information
\citep[e.g.][]{1989A&AS...79..217M}.
%and to relate the backend counts to the antenna
%temperature scale $T_{\mathrm{A}}^{*}$ 
%corrected for atmospheric attenuation and the forward efficiency of the antenna
After data quality control, the sub-scans of each scanning direction were averaged and fitted with Gaussian
curves. In the next step, each amplitude was corrected for small remaining pointing offsets and systematic
gain-elevation effects. This operation has an effect of about 1~\% at 86.24~GHz and 4~\% at 142.33~GHz (mean
pointing offsets are 1.8 \arcsec for both receivers). The latter correction, and given the IRAM 30~m
elevation-dependent gain curves, amounts to $\lesssim 5$~\%. The conversion to the standard flux density scale
was made using frequent observations of primary (Mars, Uranus) and secondary (W3OH, K3-50A, NGC\,7027)
calibrators.
%-----------------------------------------------------------------------
\begin{table*}[]
  \caption{Receiver characteristics.}     
  \label{tab:receivers}  
  \centering                    
  \begin{tabular}{rrcrrccc} 
    \hline\hline                 
    \multicolumn{1}{c}{$\nu$}   &\multicolumn{1}{c}{$\lambda$} &$\Delta\nu$ &\multicolumn{1}{c}{$T_\mathrm{sys}$}
    &\multicolumn{1}{c}{$FWHM$} &\multicolumn{1}{c}{Feeds}    &Polarisation &Aperture Efficiency\\    
    \multicolumn{1}{c}{(GHz)}   &\multicolumn{1}{c}{(mm)} &(GHz)            &\multicolumn{1}{c}{(K)}        &\multicolumn{1}{c}{(\arcsec)}  &                       & &(\%) \\    
    \hline\\                        
% NOTE: this values are from the website. The Tsys is the average over channels
    2.64  &110 &0.1       &17        &260 &1 &LCP, RCP  &58 \\ 
    4.85  &60 &0.5       &27        &146 &2 &LCP, RCP  &53 \\ 
    8.35  &36 &1.2       &22        &82  &1 &LCP, RCP  &45 \\ 
    10.45 &28 &0.3       &52        &68  &4 &LCP, RCP  &47 \\ 
    14.60 &20 &2.0       &50        &50  &1 &LCP, RCP  &43 \\ 
    23.05 &13 &2.7       &77        &36  &1 &\ldots &30 \\ 
    32.00 &9 &4.0       &64        &25  &7 &LCP    &32 \\ 
    43.05 &7 &2.8       &120       &20  &1 &\ldots &19 \\ 
\\         
     86.24&3 &8.0 (1~GHZ used)       &$\sim$65\tablefootmark{a}  &29  &1 &HLP, VLP    &63 \\ 
    142.33&2 &4.0 (1~GHZ used)      &$\sim$65\tablefootmark{a}  &16  &1 &HLP, VLP    &57 \\ 
  \\  \hline                                  
  \end{tabular}
  \tablefoot{The entries in each column is as follows: 1: central
    frequency, 2: receiver bandwidth, 3: system temperature,
    4: sensitivity, 5: full width at half maximum ($FWHM$), 6: 
    number of available feeds, 7: available polarisation channels,
    8: telescope effective area at the corresponding frequency.\\
\tablefoottext{a}{The values quoted for the IRAM receivers are typical values for the receiver temperatures
  $T_\mathrm{rx}$ and not system temperatures, hence they do not include atmospheric contributions, 
  background emission, etc.}
}
\end{table*}
% -----------------------------------------------------------------------
% -----------------------------------------------------------------------
\begin{table}[]
% \blue{140506-locked}
  \caption{
    The average fractional effect of each post-measurement correction applied to
    the data for each observing frequency. The numbers represent the 5-year average
    behaviour as it is observed by the F-GAMMA programme.}
  \label{tab:corrections}  
  \centering                    
  \begin{tabular}{rccc} 
    \hline\hline                 
    \multicolumn{1}{c}{Frequency}  &Pointing   &Opacity &Gain \\
    &Correction &Correction &Correction\\
    \multicolumn{1}{c}{(GHz)}     &(\%)       &(\%) &(\%)\\
    \hline
    \\
    \multicolumn{4}{c}{Effelsberg} \\
    2.64  &0.5 &2.3 &0.0 \\
    4.85  &0.4 &2.5 &1.2 \\
    8.35  &0.5 &2.5 &0.9 \\
    10.45 &1.2 &3.1 &1.2 \\
    14.60 &1.3 &2.9 &1.3 \\
    23.05 &1.6 &8.1 &2.1 \\
    32.00 &3.1 &7.9 &2.9 \\
    43.05 &5.1 &20. &2.1 \\
    \\
    \multicolumn{4}{c}{IRAM} \\
    86.24   &1.0 &\ldots &5.0 \\
    142.33  &4.0 &\ldots &5.0 \\
    \hline                                  
  \end{tabular}
\end{table}
% -----------------------------------------------------------------------

\subsection{Error estimates}
In every operational step the errors were propagated formally assuming Gaussianity. The exact details are
discussed in separate papers (Nestoras et al. in prep., for the IRAM observations; Angelakis et al. in prep.,
for the Effelsberg observations). An indicative empirical measure of the uncertainty in a measurement can be
given by the fractional fluctuations seen in sources known to be intrinsically stable, that is to say, the
calibrators. The datasets available for these sources cover the entire period of observations and include --
cumulatively -- all possible sources of fluctuations. This variability -- seen in sources intrinsically stable
-- must also be present in the light curves of the targets. In Table~\ref{tab:caksstat} we quote the mean flux
density at each frequency for each calibrator used at Effelsberg and NGC\,7027 used at IRAM.  Note that the
flux density for each such source was computed using the mean calibration factor of each session. There we
also show the modulation index defined as $m=100\cdot\frac{\sigma}{\left<S\right>}$ with $\sigma$ being the
standard deviation and $\left<S\right>$ the mean flux density.  $m$ remains at levels of a few percent even at
frequencies above 23.05~GHz where the troposphere becomes very disturbing.
 % --------------------------------------------------------------------
%
% The data show here are updated until October 2013 or something included. Th eIRAM I
% dunno. Update in january 2014. ONLY datapoint with S/N >=3 are sed! ALL OK
%
% UPADTE after the mepoch of 2 May 2014 and S/N 3 is only thre
\begin{table*}[]
  \caption{\label{tab:caksstat}The mean flux density and modulation index of the calibrators at each frequency.}
  \centering
  \begin{tabular}{lllrrrrrrrrrr}
    \hline\hline
    Source   &Observable       &Units& 2.64  &4.85  &8.35  &10.45 &14.60  &23.05 &32.00    &43.05    &86.24    &142.33  \\
    \hline\\
3C\,286  &N                               &              &102                            &108          & 108            & 111           & 106        & 101           & 89             & 46           & \ldots & \ldots     \\
                  &$\left<S\right>$  &(Jy)         &10.710                   & 7.477    & 5.2189     & 4.453    & 3.476  & 2.408     & 1.840    & 1.429  &\ldots      & \ldots   \\
                  &$m$                          &(\%)       &0.9                      & 0.5   & 0.7      & 1.1   & 1.9  & 2.9    & 3.5     & 4.4  &\ldots      & \ldots   \\\\
    3C\,48  &N                               &                        & 88                 & 91                & 92               & 92                & 93               & 83                 & 72             & 28           &\ldots & \ldots  \\
                    &$\left<S\right>$  &(Jy)                  & 9.579        & 5.509       & 3.267      & 2.613       & 1.872      & 1.162        & 0.817    & 0.599  &\ldots  & \ldots   \\
                    &$m$                        &(\%)                  & 0.7        & 0.6       & 0.9      & 1.2       & 1.7      & 2.9        & 4.6    & 4.5  &\ldots & \ldots \\\\
3C\,161 &N                               &                  & 30                   & 35                 & 29                 & 30                 & 30               & 22                 & 19                  & 4             &\ldots  & \ldots \\
                 &$\left<S\right>$  &(Jy)           & 11.361       & 6.597        & 3.793         & 2.961        & 2.053      & 1.221        & 0.822         & 0.597  &\ldots & \ldots  \\
                 &$m$                        &(\%)           & 1.0          & 0.8        & 1.4         & 2.2        & 3.3      & 5.4        & 7.3         & 2.3  &\ldots & \ldots \\\\
 3C\,295   &N                               &            & 38                & 38                  & 34                & 35              & 35                & 25                 & 24                     &\ldots & \ldots & \ldots \\
                    &$\left<S\right>$  &(Jy)       & 12.499       & 6.546       & 3.434      & 2.588      & 1.690       & 0.880        & 0.557  &\ldots & \ldots & \ldots \\
                    &$m$                        &(\%)      & 0.9        & 0.8       & 1.3      & 1.5      & 2.6       & 6.4        & 6.7        &\ldots & \ldots & \ldots \\\\
     NGC7027  &N                               &     & 64                  & 73              & 79                   & 75                & 71              & 62                 & 52              & 25                 & 61                   & 54         \\     
                       &$\left<S\right>$      &(Jy)               & 3.693       & 5.449       & 5.903       & 5.912        & 5.738       & 5.455        & 5.243      & 5.146       & 4.920         & 4.611  \\
                       &$m$                               &(\%)        & 1.2       & 0.5       & 1.1      & 1.2        & 2.1       & 3.8        & 5.3       & 5.3       & 2.7         & 4.6  \\\\
  \hline
  \end{tabular}
%  \tablefoot{The top panel shows likely members of Pismis~11. The second
%    panel contains likely members of Alicante~5. The bottom panel
%    displays stars outside the clusters.\\
%    \tablefoottext{a}{Frequency in GHz}\\
%    \tablefoottext{b}{Mean flux density in Jy}\\
%    \tablefoottext{c}{Standard Deviation of around the mean flux density in Jy: measure of the variability amplitude.}
%  }
\end{table*}
%--------------------------------------------------------------------

\subsection{Cross-telescope calibration}
\label{sec:cross_cal} 
\begin{figure}[] 
\centering
\begin{tabular}{c}
\includegraphics[trim=40pt 10pt 10pt 20pt  ,clip ,width=0.5\textwidth]{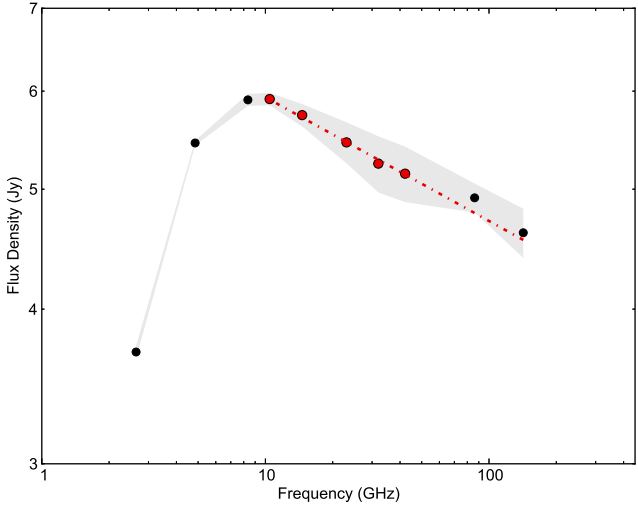} 
\end{tabular}
\caption{Observed radio SED of NGC\,7027 \citep{2008ApJ...681.1296Z} over the same period of time as the
  time baseline covered here. All the Effelsberg frequencies from 2.64 up to the highest IRAM frequency of
  142.33~GHz are shown. The filled circles (red and black) denote the measurements. The red symbols mark the
  Effelsberg measurements that where used in the fit, the red dotted-dashed line is the fitted spectrum. The
  grey area denotes the 1$\sigma$ region around each data point. The agreement of the extrapolated values with
  those measured by IRAM is better than 3~\%. }
\label{fig:ngcspec}
\end{figure}

Because below we discuss the dynamics of radio SEDs observed partly with the Effelsberg and partly with the IRAM 30~m
telescope, it is essential to address the cross-telescope calibration accuracy that might
introduce artefacts in the observed spectral shapes and their apparent temporal behaviour. An empirical yet
reliable measure of its goodness can be provided by NGC\,7027, which exhibits a flux density high
enough to be detected by both instruments with high signal-to-noise ratios (S/N) and repeatedly.  NGC\,7027
has a very well defined and analytically described time-dependent convex spectrum \citep{2008ApJ...681.1296Z}
that at frequencies above roughly $\sim 10$~GHz can be approximated by a power law of the form
$S\propto \nu^{-0.1}$. In Fig.~\ref{fig:ngcspec} we show the flux densities averaged over the entire period
discussed here and for all frequencies. Each circle denotes an average flux density. The red symbols (with
$\nu \ge 10.45$~GHz) are the Effelsberg measurements that were used for the model fit, and the red line
is the result of the fit. The grey area is confined by 1$\sigma$ of each average flux density.  Extrapolating the
fitted spectrum (red dotted-dashed line) towards the higher IRAM band and comparing these values with the
measured IRAM 30~m flux densities yields differences better than 3~\%,
%(that deviation normalised to the mean flux
% ea 140507: the next was done by Jainnis and is all correct check email
specifically, 2.7~\% at 86.22~GHz and 2.8~\% at 142.33~GHz), indicating a high-quality cross-telescope
calibration.
%% -----------------------------------------------------------------------
% -----------------------------------------------------------------------

\subsection{Analysis methods}
\label{subsec:analysis} 
The current section introduces notions and methods that are used below, even if they are repeated briefly in
the corresponding sections.
%is meant to set the scene for the studies that follow. Specifically, it is meant to

\paragraph {\bf Internal shocks causing variability:} One of the aims here is to study the mechanism that may
be causing the observed variability. Throughout the text it is assumed that this could well be caused by
internal shocks that propagate in the jet, imprinting a specific signature in the radio light curves. In this
model \citep{Marscher1985ApJ,turler2000AnA...361..850T}, the synchrotron self-absorbed component is undergoing
distinct evolutionary stages, each characterised by a different energy-loss mechanism. The followed path then
imprints a distinct phenomenology on the radio SEDs, making this model easily quantifiable and testable. It
has been argued that the implementation of this scenario in a system with simply a steep-spectrum quiescent
jet is enough to reproduce the observed plurality of phenomenologies \citep{2012arXiv1202.4242A}.

\paragraph {\bf The variability brightness temperature:} As discussed extensively in
Sect.~\ref{sec:flaredecomp}, the variability brightness temperature is a measure of the energetics of the
associated event. It is computed on the basis of the light travel-time argument and depends on the
magnitude of the flux density variation, $\delta S$ and the time span needed for that variation, $\delta
t$.
The flare decomposition method (Sect.~\ref{sec:flaredecomp}) aims at  separately estimating exactly
those parameters for each event. The variability brightness temperature at the source rest frame
{in K}, is given by
\begin{equation}
T_\mathrm{var} =  1.64\cdot10^{10}\cdot\frac{\delta S\cdot d^{2}_\mathrm{L}\cdot \lambda^2}{{\delta t}^2\cdot (1+z)^4}\nonumber,
\end{equation}
where
   \[
      \begin{array}{lp{0.8\linewidth}}
         \delta S      &is the increase in flux density (at the observer's frame) in units of Jy,\\
         \delta t       &is the time span needed for that increase (at the observer's frame) in units of days,\\
         d_\mathrm{L}     &is the source luminosity distance in units of Mpc,            \\
         \lambda&is the observing wavelength  in units of cm, and                             \\
         z        &is the source redshift,                    
      \end{array}
   \]
all measured at the observer's frame.

Any excess from an independently computed intrinsic limit is then interpreted in terms of
the Doppler-boosting factor $D$ as   
\begin{equation}
\label{eq:D}
D = (1+z)\sqrt[3+\alpha]{\frac{T_\mathrm{var}}{T_\mathrm{ref}}},
\end{equation}
where
   \[
      \begin{array}{lp{0.8\linewidth}}
        \alpha  &is the source spectral index\footnotemark[1]  with $S \propto \nu^{\alpha}$,   \\
        T_\mathrm{ref}  &is the limiting value of the brightness temperature. 
      \end{array}
   \]
   The value for the spectral index in the previous equation is chosen to be the mean values in the corresponding
   sub-band as given in Table~\ref{tab:specs}. In Sect.~\ref{sec:flaredecomp} we explain that the limiting value is based on the equipartition argument as
   proposed by \cite{Readhead1994ApJ}.  \footnotetext[1]{Here we assumed that the emission
    comes from a single blob. For a continuous jet the index $3+\alpha$ should be replaced
     by $2+\alpha$.  }

%===================================================================

\section{Radio light curves}
\label{sec:rlcs}
% -----------------------------------------------------------------------
% the plots here have ONLY 3 sigma data points UPDATED-OK: FINAL January 24
% done with 3 sigma points only on Sep  9 2014
\begin{figure*}[] 
\centering
\begin{tabular}{cc}
\includegraphics[width=0.3\textwidth,angle=-90]{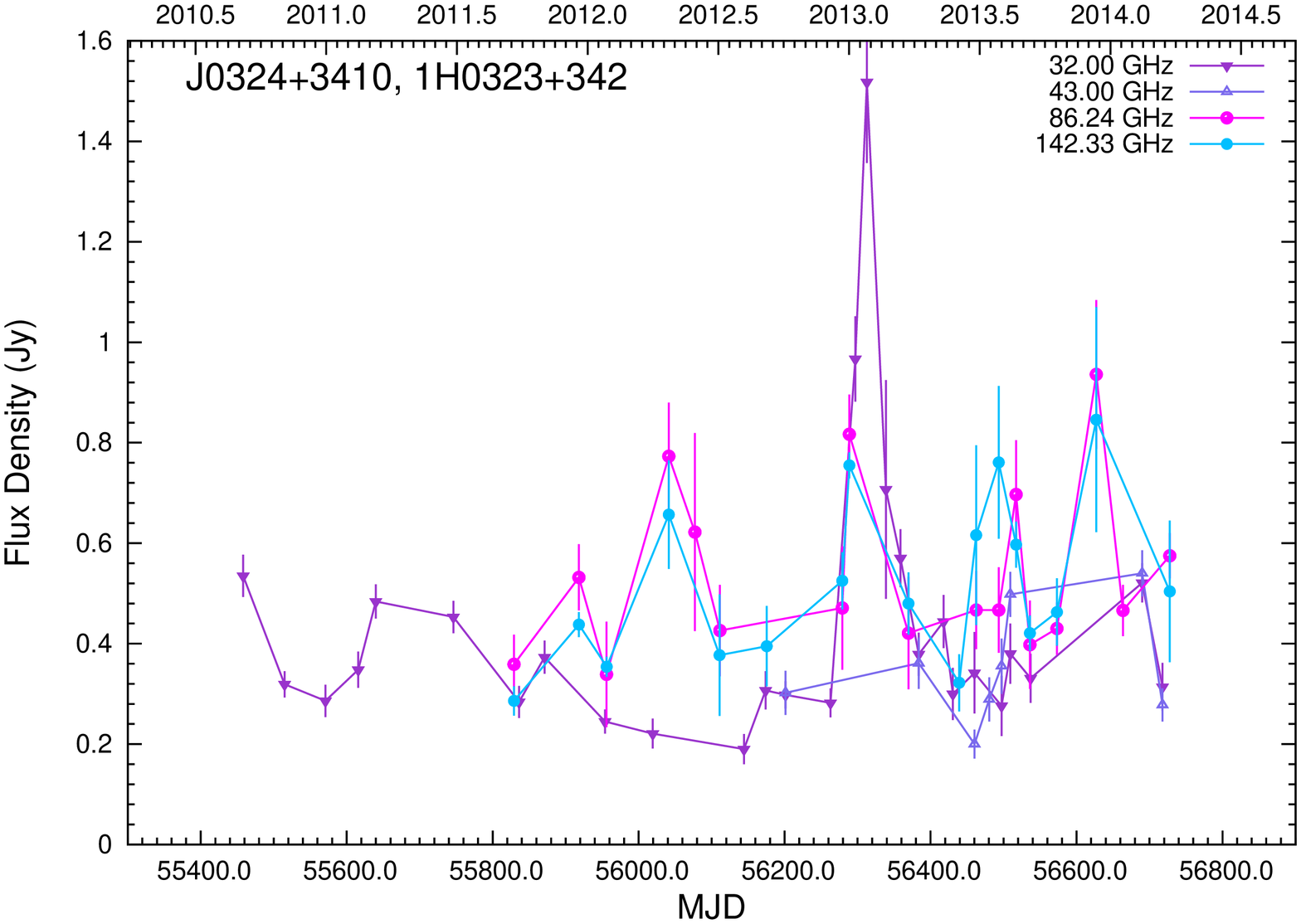} &\includegraphics[width=0.3\textwidth,angle=-90]{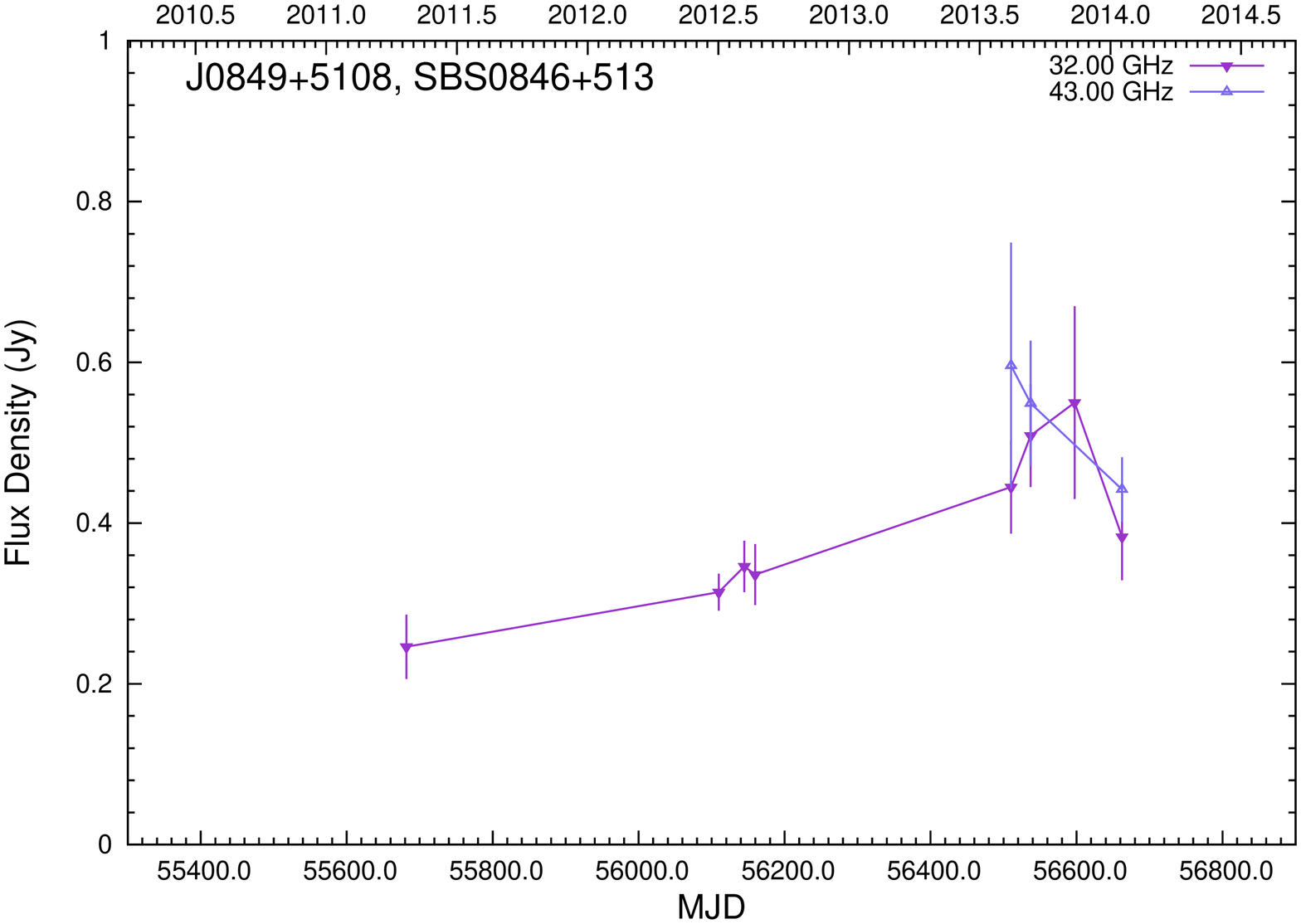}\\ 
\includegraphics[width=0.3\textwidth,angle=-90]{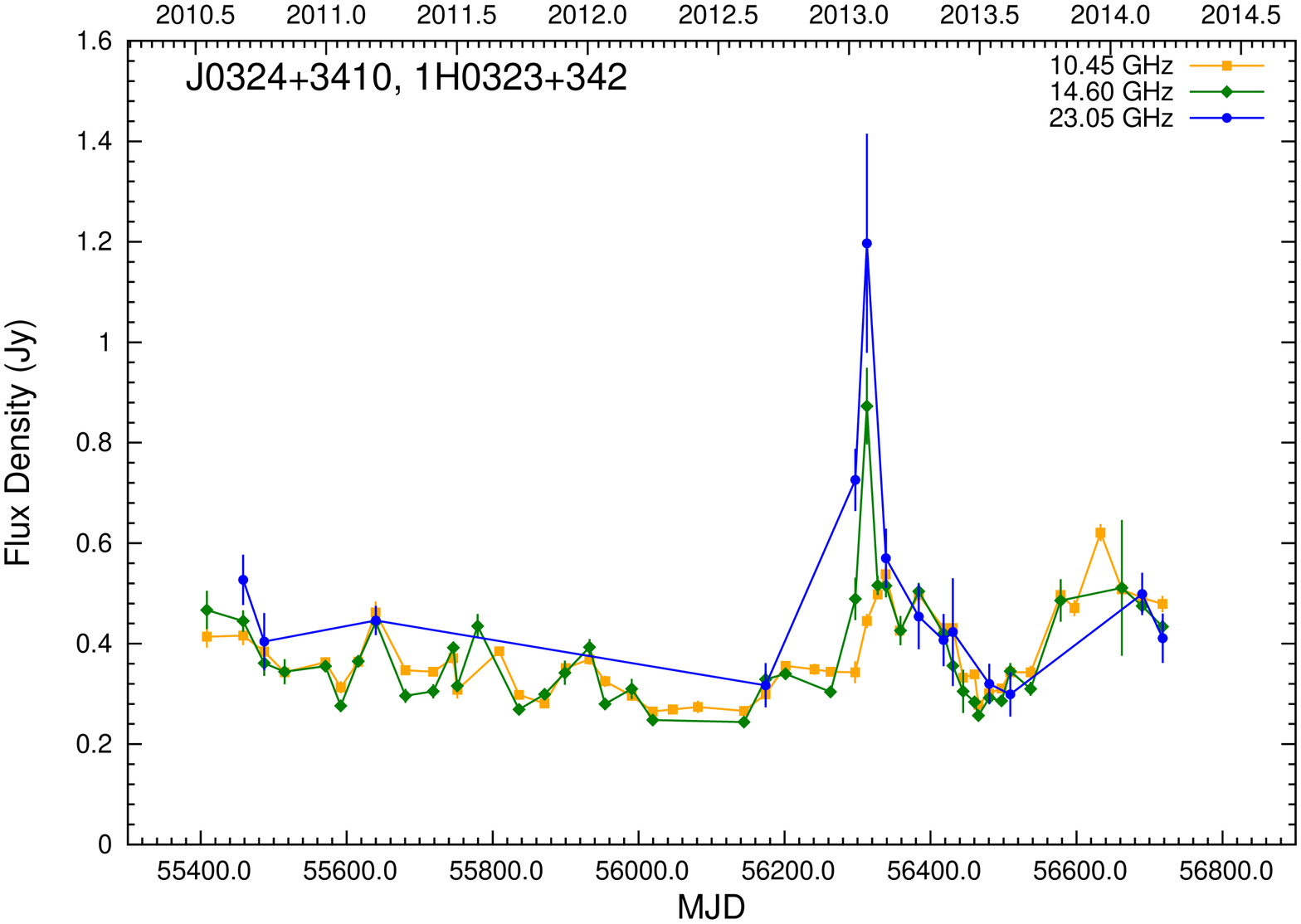} &\includegraphics[width=0.3\textwidth,angle=-90]{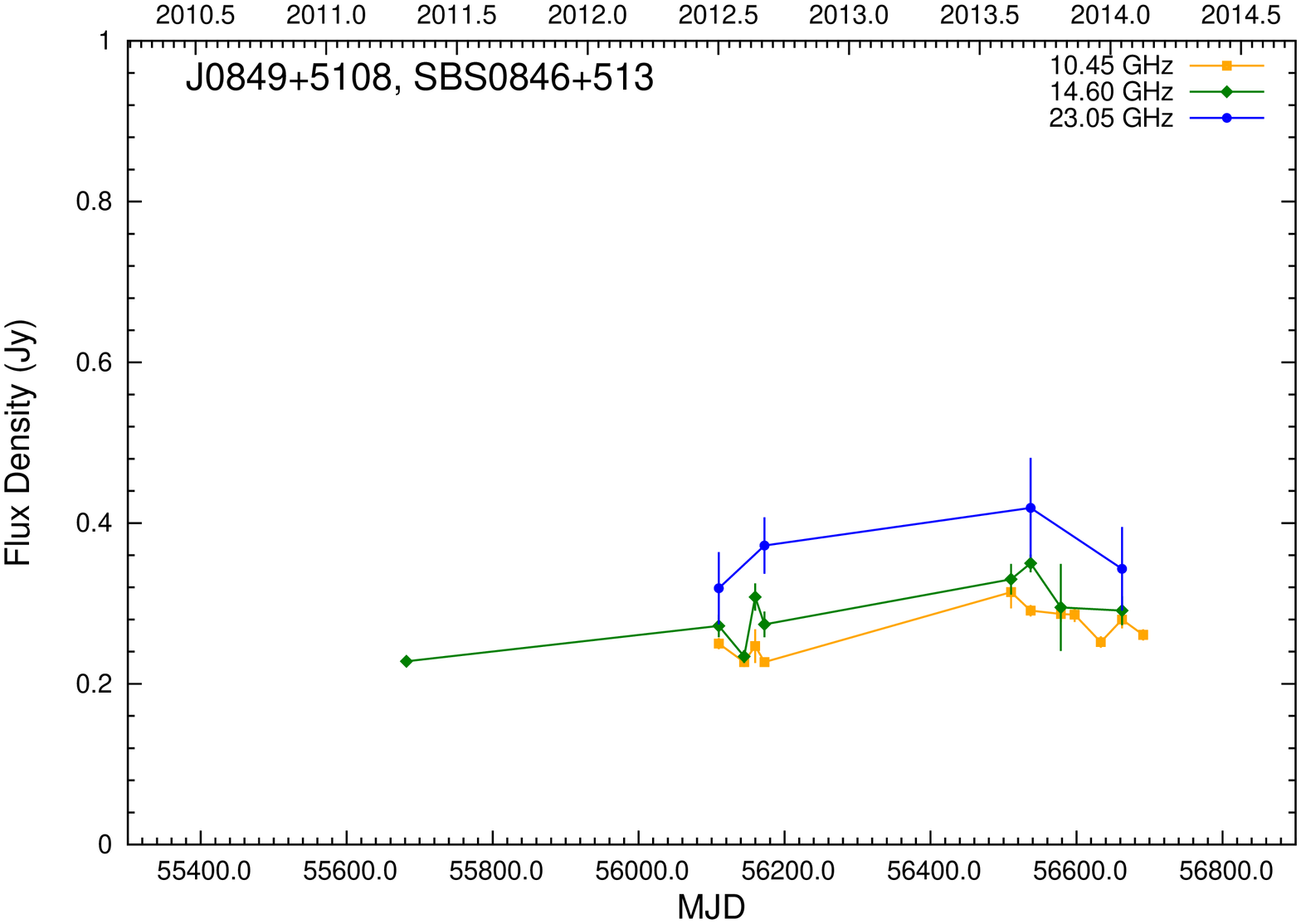} \\
\includegraphics[width=0.3\textwidth,angle=-90]{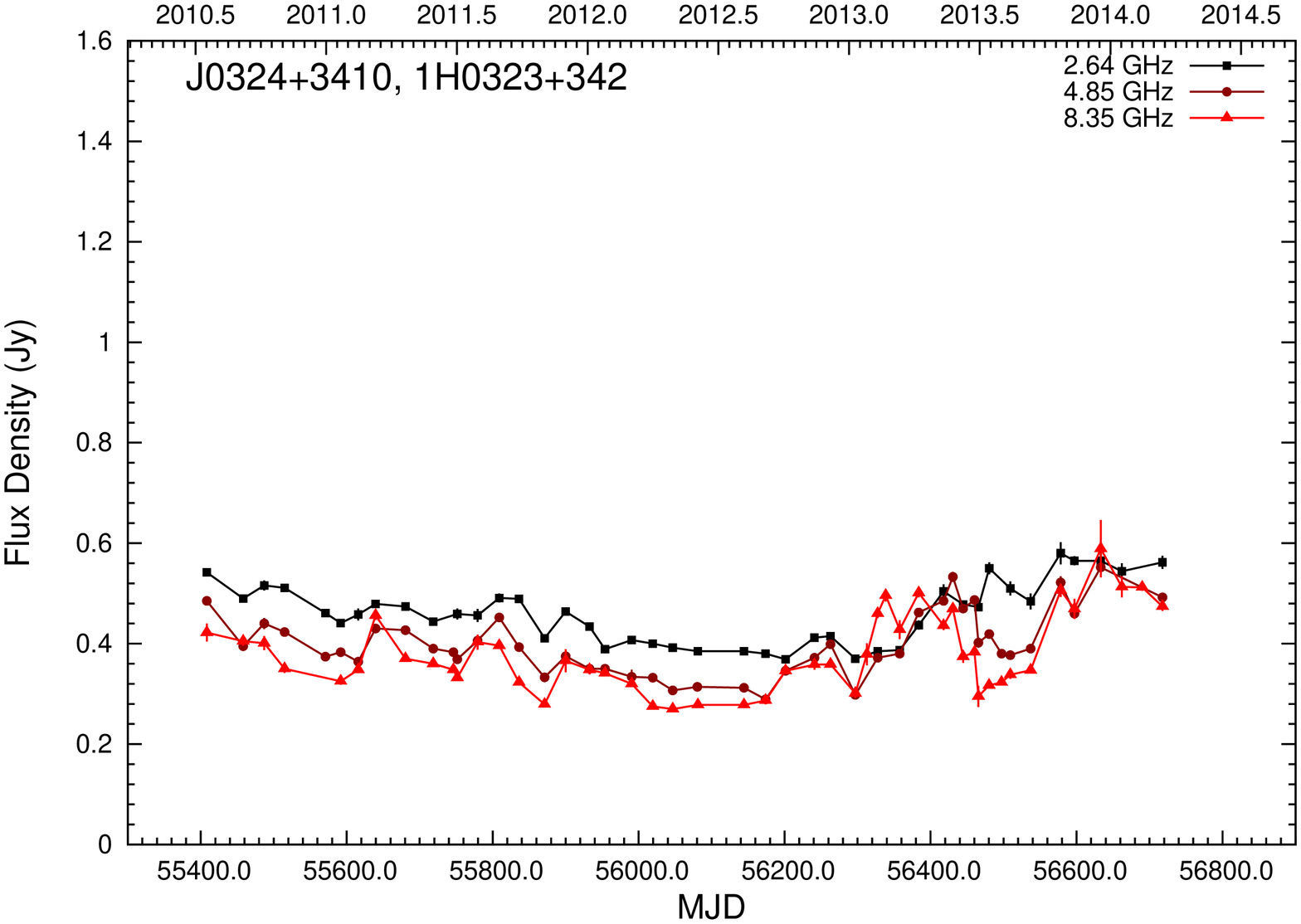} &\includegraphics[width=0.3\textwidth,angle=-90]{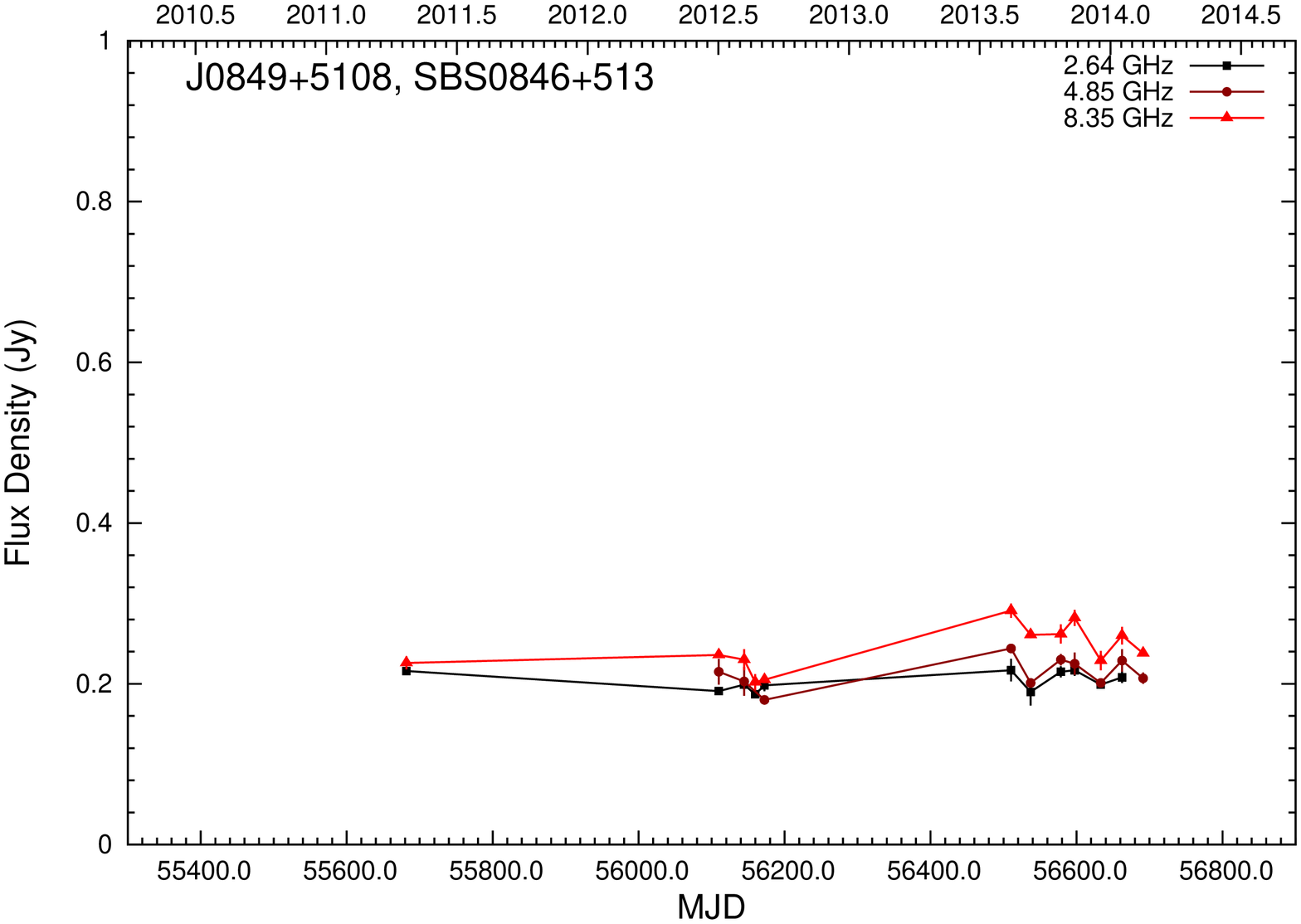}\\ 
\includegraphics[width=0.3\textwidth,angle=-90]{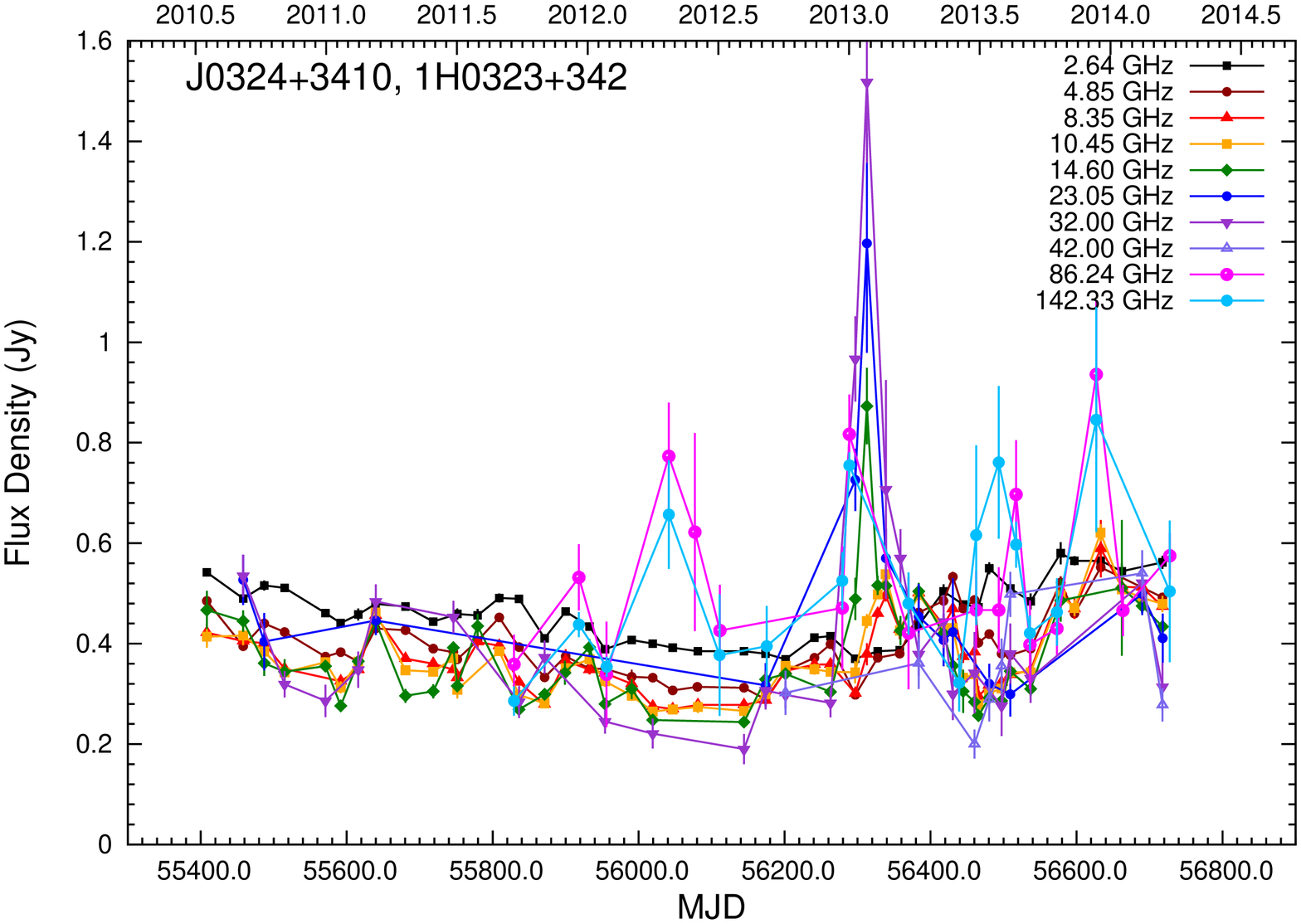} &\includegraphics[width=0.3\textwidth,angle=-90]{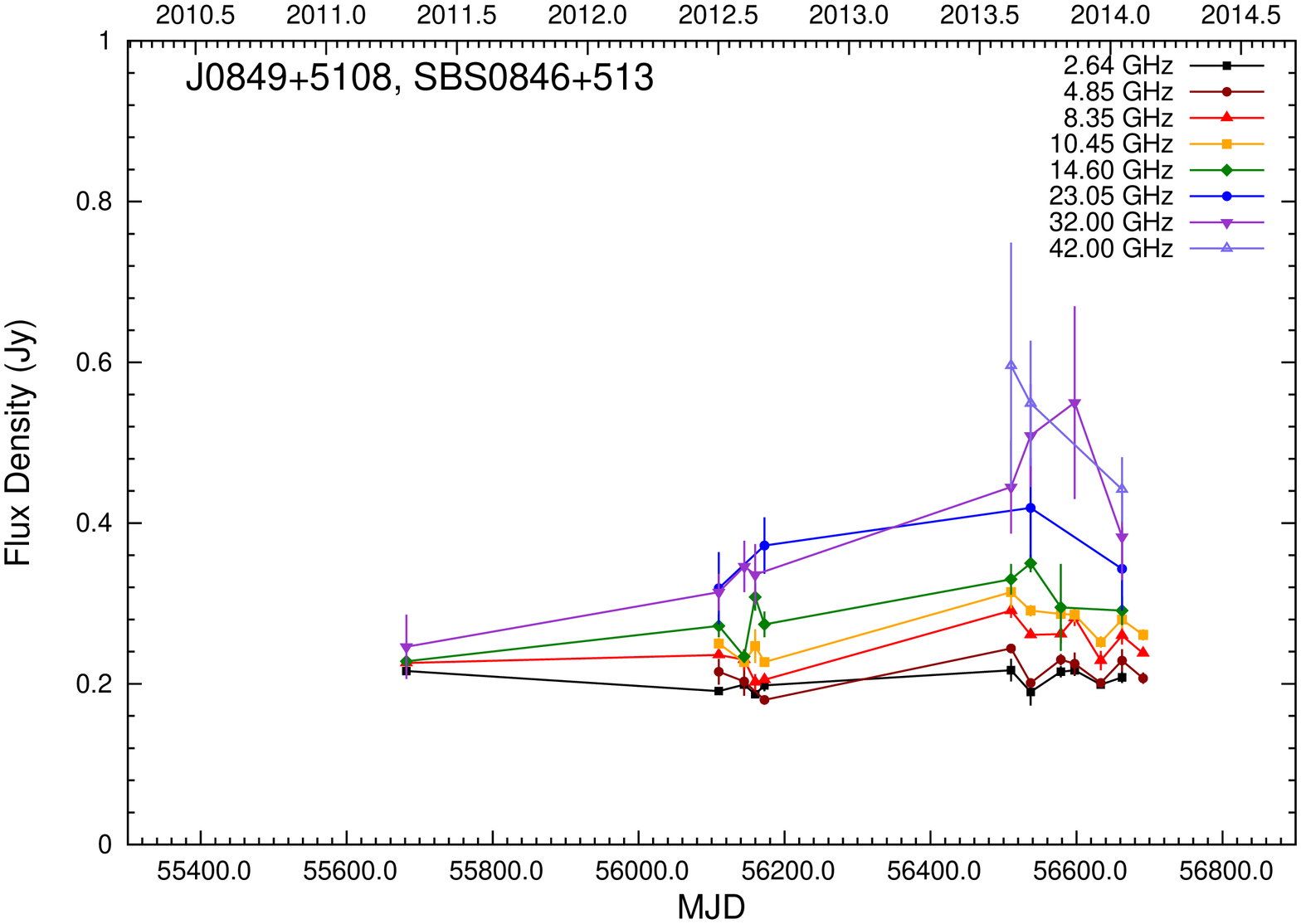} \\
\end{tabular}
\caption{Radio light curves available for J0324$+$3410 (left column) and J0849$+$5108 (right column) at all
  available frequencies. From top to bottom we present the light curves three different frequency bands: low:
  2.64, 4.85 and 8.35~GHz, intermediate: 10.45, 14.60 and 23.05~GHz, and high: 32, 43.05, 86.24 (when
  available) and 142.33~GHz (when available). At the very bottom the datasets are shown over-plotted together
  for comparison. For the same reason, for each source and axes the boundaries are kept identical. Lines
  connecting the data points have been used everywhere to facilitate visual inspection. Each frequency is
  consistently represented by the same colour and symbol. Only data points with a signal-to-noise ratio better
  than 3 have been used. }
\label{fig:lc_lmh_03n08}
\end{figure*}
% -----------------------------------------------------------------------
% -----------------------------------------------------------------------
% the plots here have ONLY 3 sigma data points UPDATED-OK: FINAL January 24
%\setlength{\unitlength}{2cm}
% done with 3 sigma points only on Sep  9 2014
\begin{figure*}[] 
\centering
\begin{tabular}{cc}
\includegraphics[width=0.3\textwidth,angle=-90]{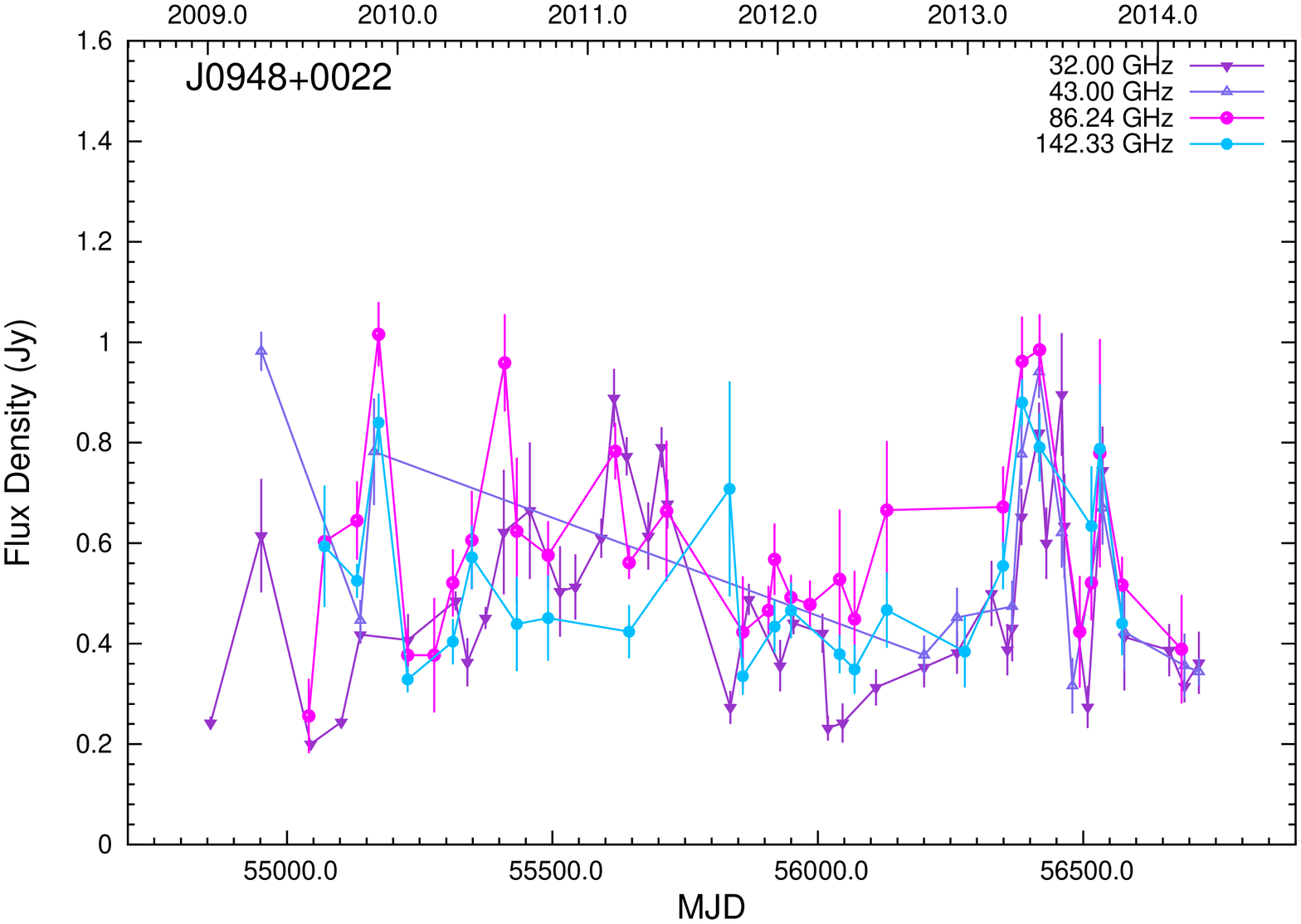} &\includegraphics[width=0.3\textwidth,angle=-90]{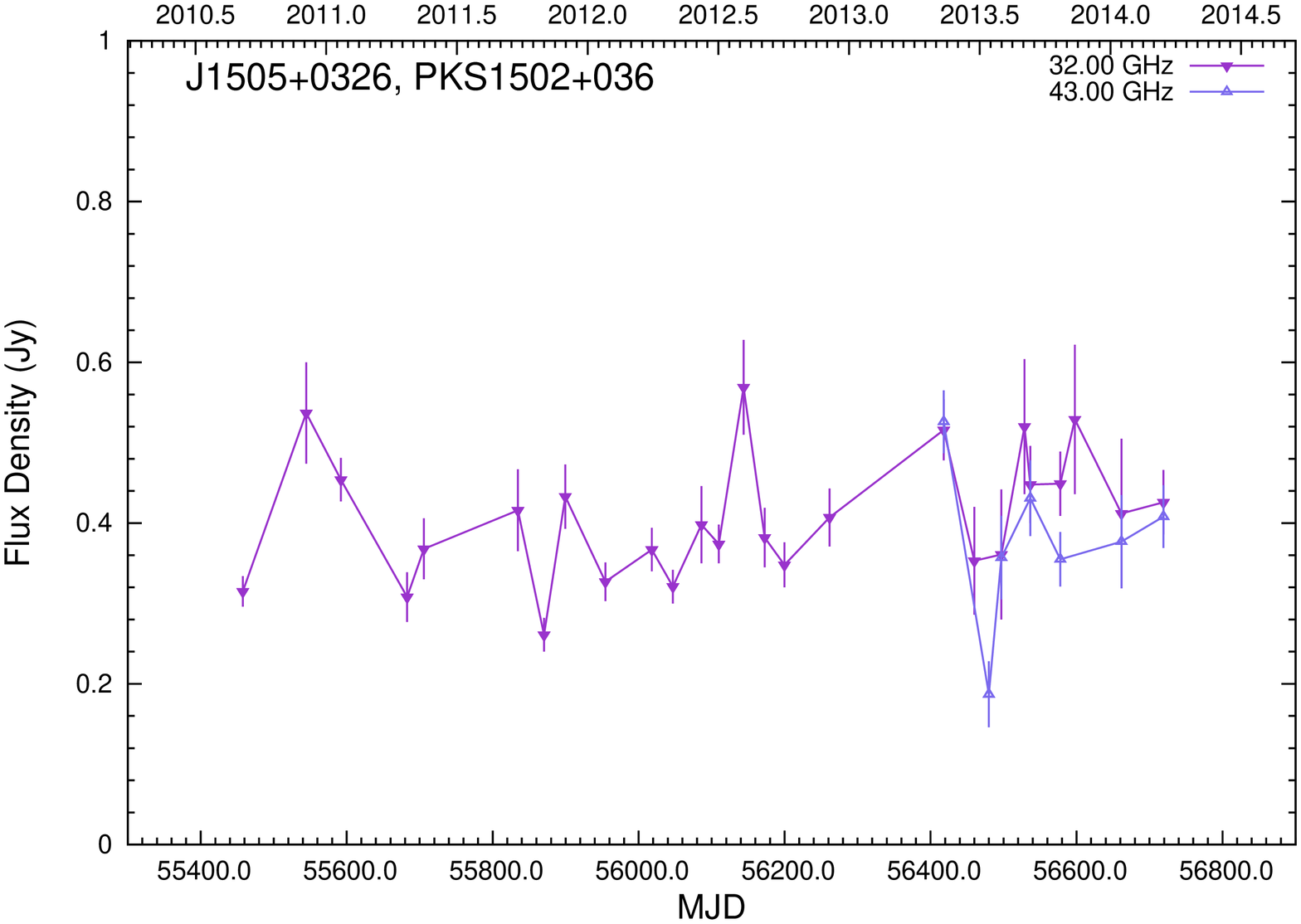}\\ 
\includegraphics[width=0.3\textwidth,angle=-90]{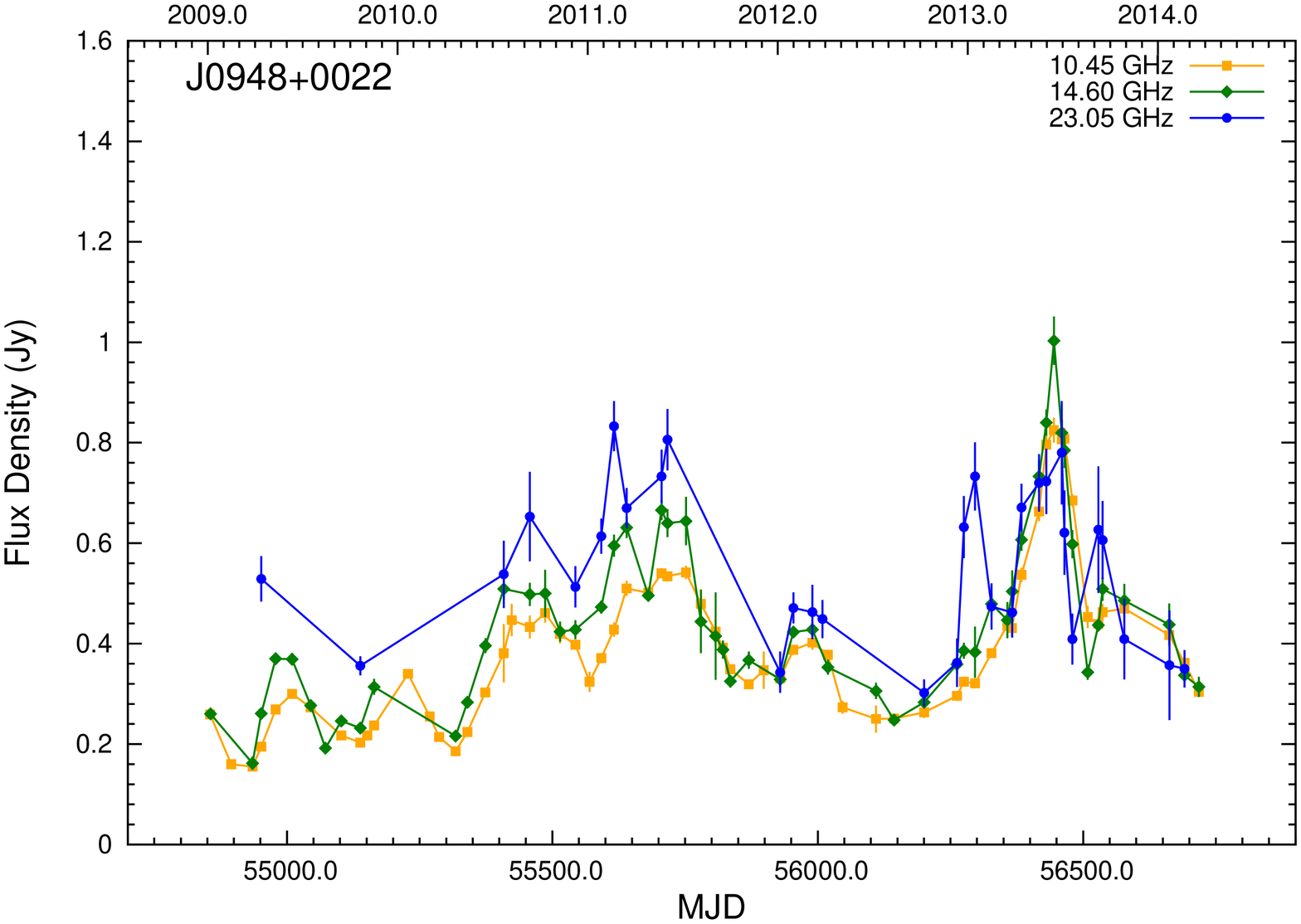} &\includegraphics[width=0.3\textwidth,angle=-90]{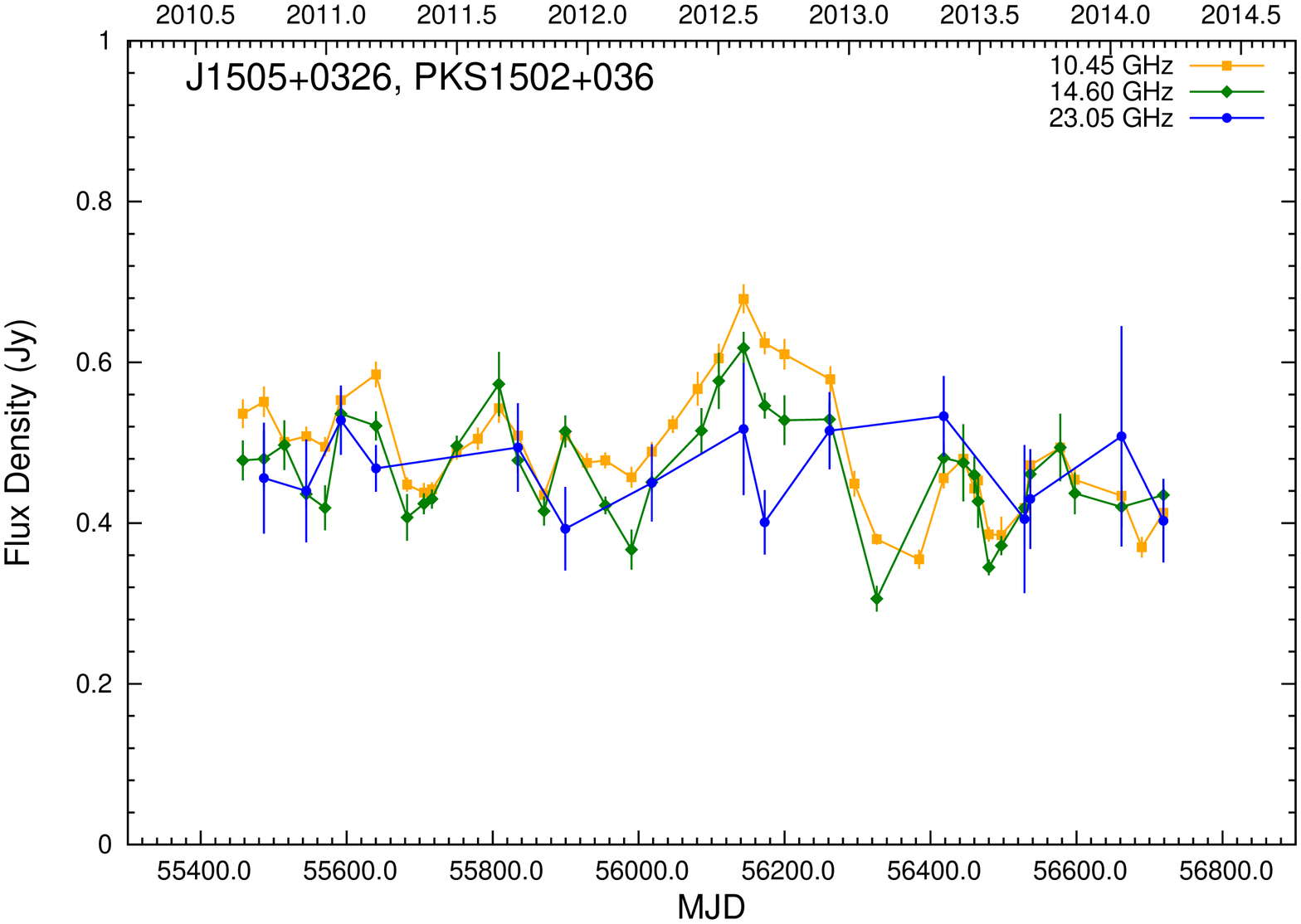} \\
\includegraphics[width=0.3\textwidth,angle=-90]{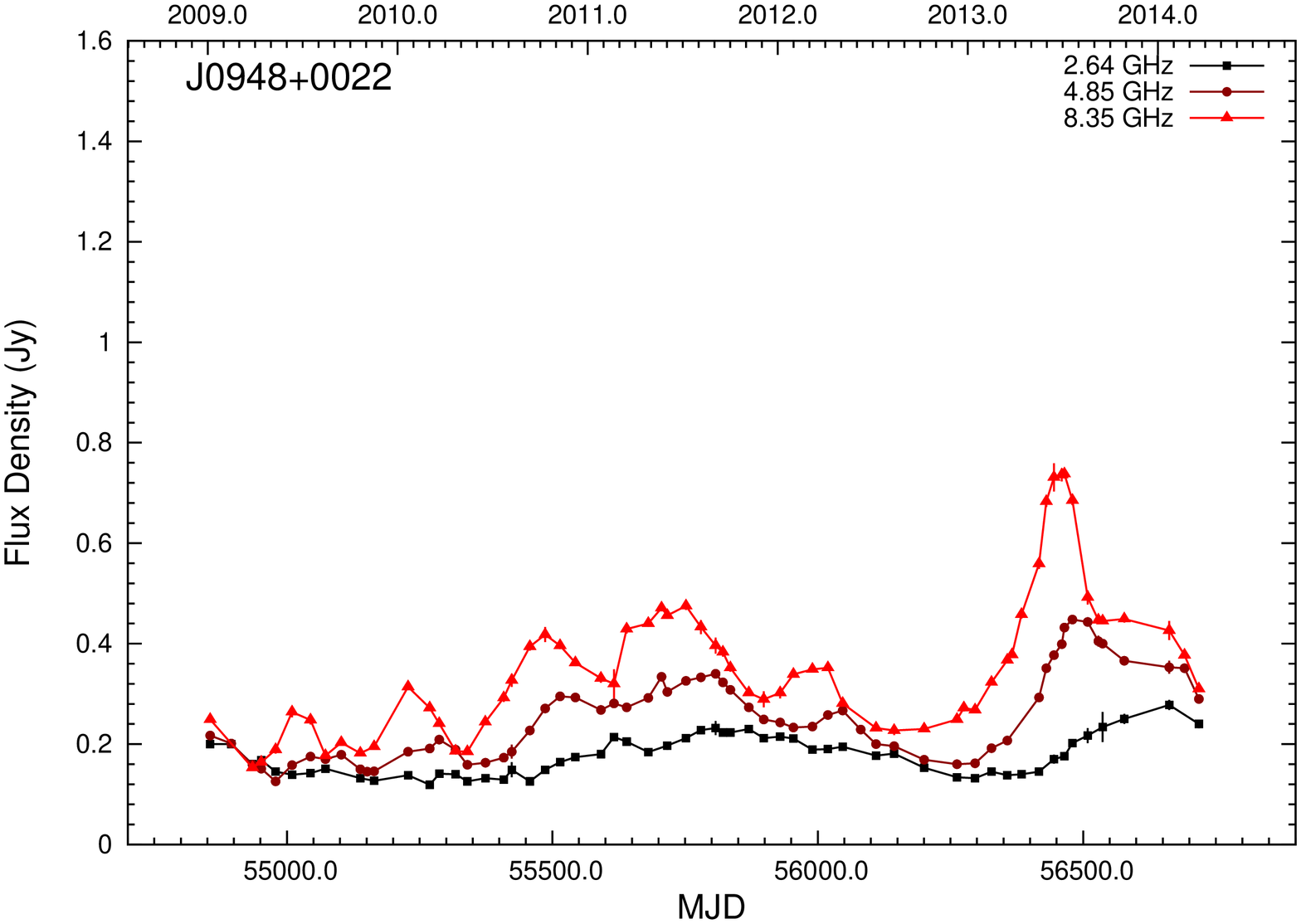} &\includegraphics[width=0.3\textwidth,angle=-90]{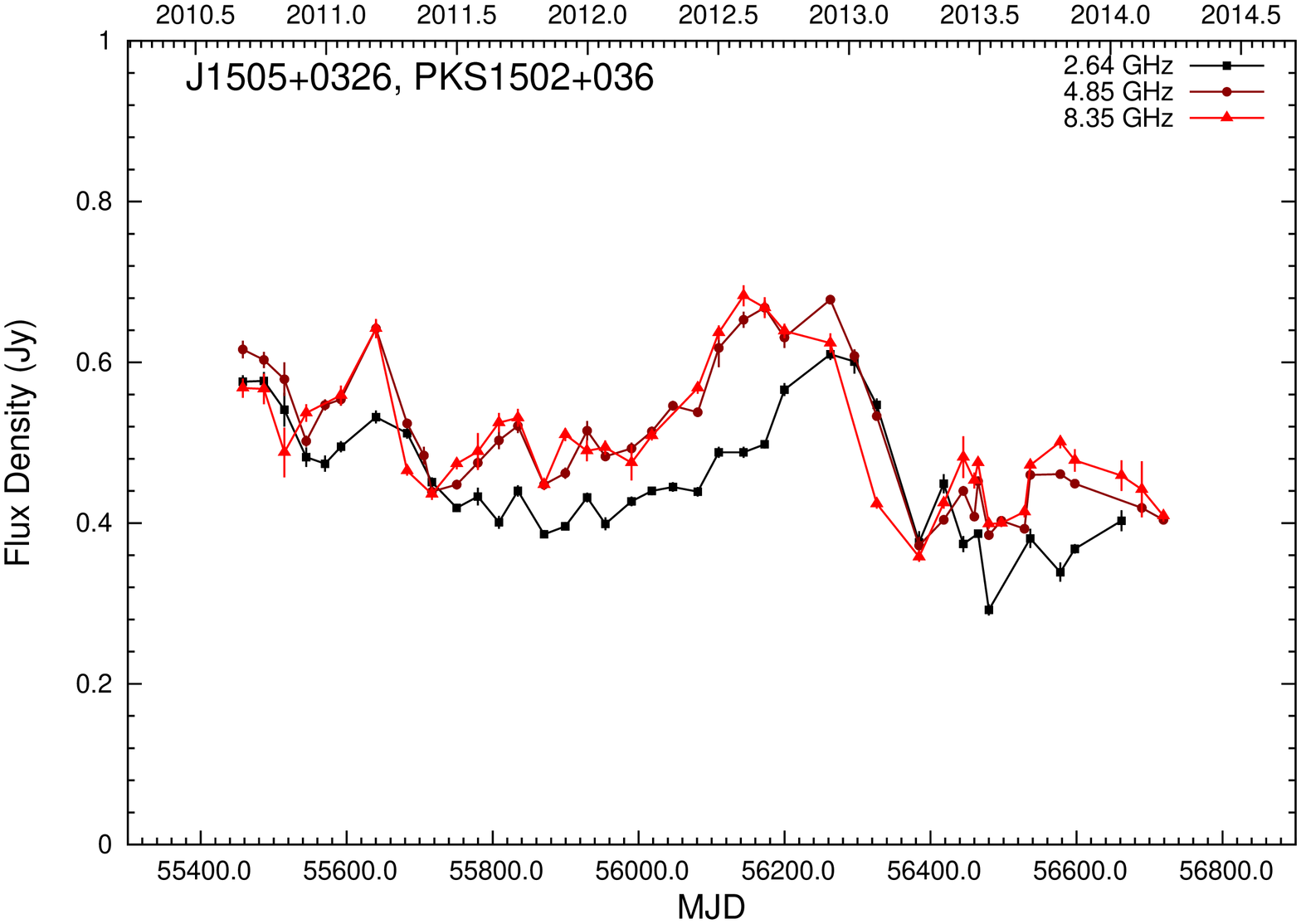}\\ 
\includegraphics[width=0.3\textwidth,angle=-90]{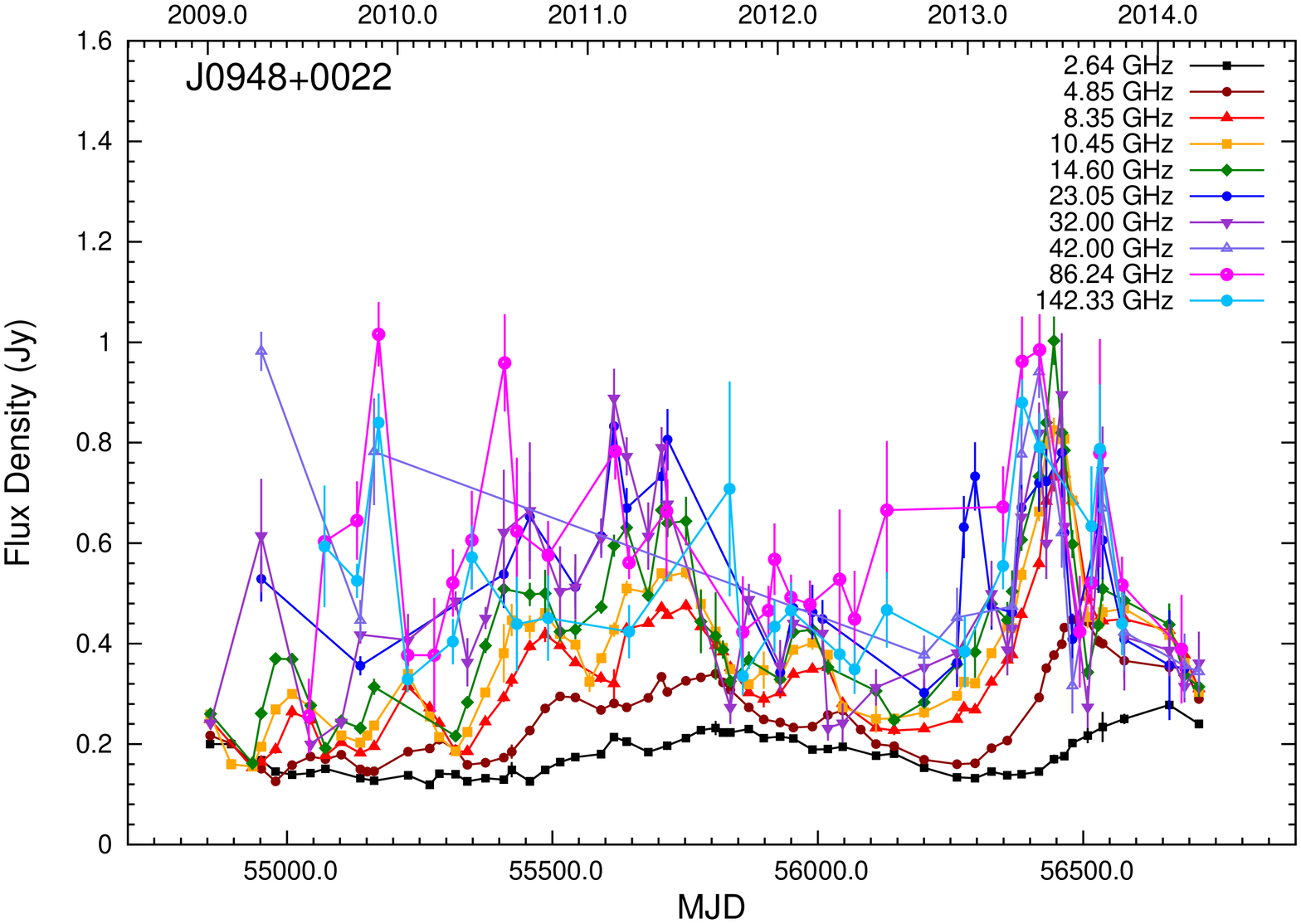} &\includegraphics[width=0.3\textwidth,angle=-90]{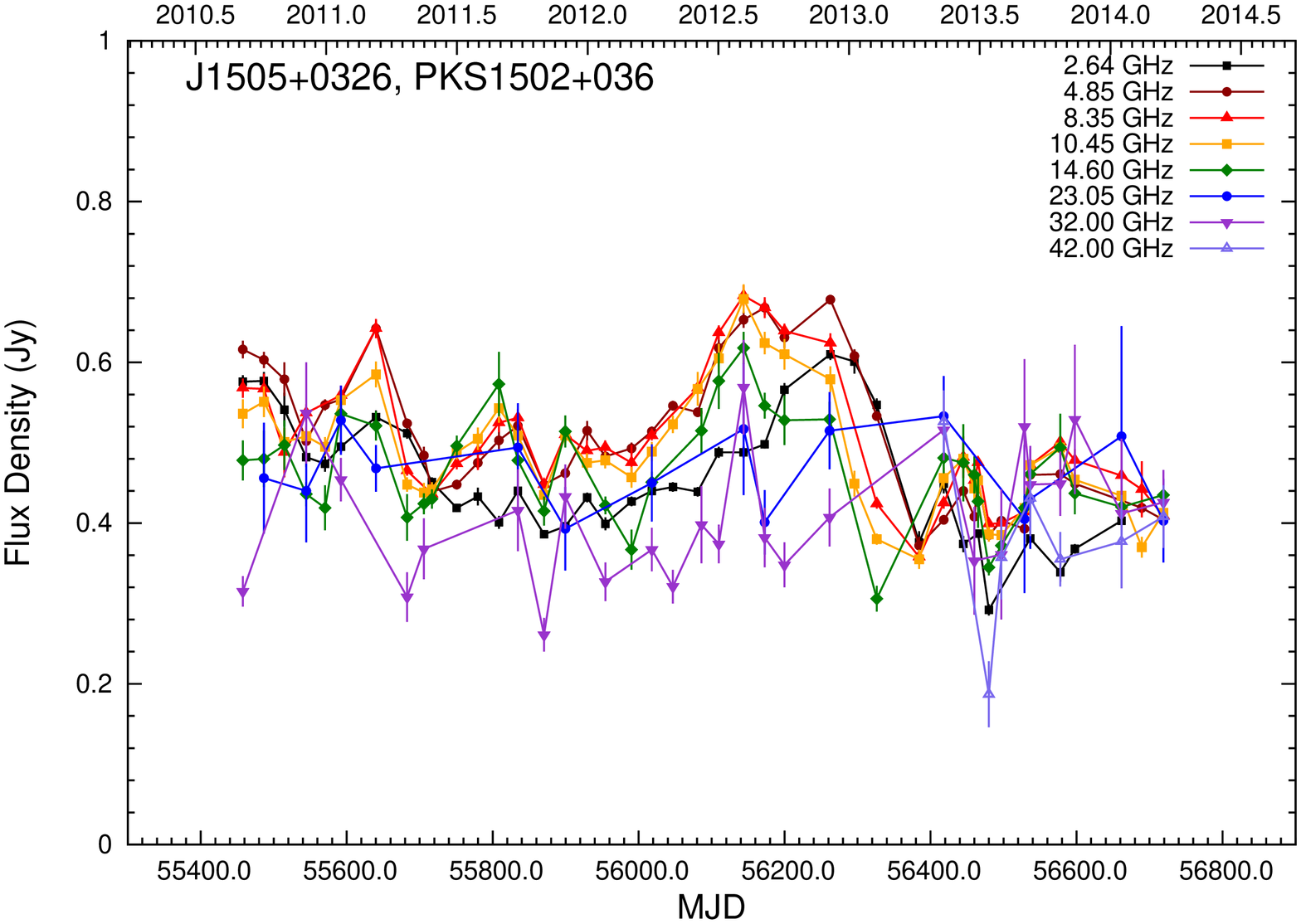} \\
\end{tabular}
\caption{Radio light curves available for J0948$+$0022 (left column) and J1505$+$0326
  (right column) at all available frequencies similar as is Fig.~\ref{fig:lc_lmh_03n08}.}
\label{fig:lc_lmh_09n15}
\end{figure*}
% -----------------------------------------------------------------------
%It was already mentioned earlier, that 
The dataset discussed here is the result of the longest-term multi-frequency radio monitoring for RL and
$\gamma$-ray loud NLSy1s available. It covers the period between early-2009 and mid-2014.  In
Figs.~\ref{fig:lc_lmh_03n08}~and~\ref{fig:lc_lmh_09n15} we present the available light curves at all observing
frequencies. We only show measurements with a signal-to-noise ratio better than 3. All the light curves are
also available online at the CDS. Table~\ref{tab:lc_sample} is an example of an online light-curve file for
J0324$+$3410 at 4.85~GHz. It contains the following information: Column~1 lists the MJD of a measurement,
Col.~2 gives the flux density in units of Jy, and Col.~3 lists the associated error.
%-----------------------------------------------------------------------
\begin{table}[]
  \caption{Example of a light-curve file as published online at CDS. The current
    extract is for J0324$+$3410 at 4.85~GHz.}     
  \label{tab:lc_sample}  
  \footnotesize
  \centering                    
  \begin{tabular}{ccc} 
    \hline\hline                 
    MJD        &$S$ &Error\\
               &(Jy)&(Jy) \\
    \hline    \\     
  55408.300   &0.485  &0.006\\
  55458.148   &0.395  &0.007\\
  55487.059   &0.440  &0.011\\
  55515.026   &0.423  &0.005\\
  55570.904   &0.374  &0.005\\
  \ldots      &\ldots &\ldots\\

   \\ \hline 
  \end{tabular}
\end{table}
% -----------------------------------------------------------------------

The mean sampling -- averaged over all frequencies -- ranges from one measurement every 24 days for
J0324$+$3410 and J0948$+$0022 (approximately the cadence of the most commonly used calibrator, 3C\,286), and
29 days for J1505$+$0326, to more than 100 days for J0849$+$5108. Except for J0849$+$5108, for which the
dataset does not cover a substantial number of activity cycles, the remaining three sources show intense
variability at practically all frequencies, a behaviour commonly seen in blazars
\citep[e.g.][]{2011ApJS..194...29R,2011JApA...32....5A,2012arXiv1205.0539B,2004NewAR..48..399B}. The
phenomenologies seen in these light curves vary significantly and mostly in terms of the observed amplitude of
the different events. In these terms, J1505$+$0326 shows very weak outbursts. J0324$+$3410 shows at least two
major events at the high-end of the bandpass at around MJD 56300 and 56600, which disappear at lower
frequencies. J0948$+$0022 shows frequent recursive events of activity that occur considerably fast; their
characteristic times are of the order of 40--50 days at the highest frequencies (see
Table~\ref{tab:allflares}).  Most of the observed events can be identified at all frequencies (apart from the
very low end of the bandpass) and with a time delay between frequencies (e.g. see event around MJD 56500 for
J0948$+$0022), which is indicative of intense spectral evolution that is also routinely seen in blazar light
curves \citep[e.g.][]{2013A&A...552A..11R}. The variability amplitude is frequency dependent, following the
typical fashion: higher frequencies vary more. For J0324$+$3410 for example, the amplitude of the most
prominent event is more than twice as large at 32.00~GHz as at 14.60 and more than six times lesser at
10.45~GHz; yet another indication that the variability mechanisms seem to be the same as those acting in
typical blazars \citep[e.g.][]{1992A&A...254...80V}. Table~\ref{tab:summary} reports some characteristic
parameters for these datasets. Specifically, for each frequency and source we report the total number of data
points N, the dataset length $\Delta$t, the mean flux density $\left<S\right>$, the standard deviation around
that mean $\sigma$, and the modulation index $m$.

In the following we roughly describe the most noteworthy phenomenological characteristics of the light curves
separately for each source.
% --------------------------------------------------------------------
% 
% The data show here are updated until October 2013 or something included. Th eIRAM I
% dunno. Upadte in january 2014. ONLY datapoint with S/N >=3 are sed! ALL OK
%
% UPADTE after the mepoch of 2 May 2014 and S/N 3 is only thre
% DONE: UPDATED-OK: FINAL May 27
% locked
\begin{table*}[]
  \caption{\label{tab:summary}Summary of the mean flux densities and corresponding
    standard deviations for each observing frequency and source presented in
    Figs.~\ref{fig:lc_lmh_03n08} and \ref{fig:lc_lmh_09n15}. For every entry we also report the
    number of data points N and the modulation index $m=100\cdot\sigma/\left<S\right>$ 
    as a measure of the apparent variability amplitude. Only data points
    meeting the condition of S/N$\ge3$ are included.}
  \centering
  \begin{tabular}{lllrrrrrrrrrrr}
    \hline\hline
    Source   &Observable       &Units&       &2.64  &4.85  &8.35  &10.45 &14.60  &23.05 &32.00    &43.05    &86.24    &142.33  \\
    \hline\\
J0324$+$3410&N                 &     &       &42 &45 &48 &48 &43 &14 &27 &8 &17 &17 \\
            &$\Delta$t         &(days)&      &1310 &1310 &1310 &1310 &1310 &1260 &1260 &517 &899 &899 \\      
            &$\left<S\right>$  &(Jy) &       &0.461 &0.400 &0.380 &0.377 &0.378 &0.500 &0.432 &0.353 &0.541 &0.518 \\
            &$\sigma$          &(Jy) &       &0.061 &0.064 &0.077 &0.083 &0.113 &0.230 &0.272 &0.115 &0.172 &0.164 \\
            &$m$               &(\%) &       &13 &16 &20 &22 &30 &46 &63 &32 &32 &32 \\\\
J0849$+$5108&N                 &     &       &11 &10 &12 &11 &9 &4 &8 &3     &$\ldots$ &$\ldots$ \\
            &$\Delta$t         &(days)&      &981 &581 &1009 &581 &980 &552 &980 &152   &$\ldots$ &$\ldots$ \\      
            &$\left<S\right>$  &(Jy) &       &0.203 &0.214 &0.244 &0.266 &0.287 &0.363 &0.391 &0.529 &$\ldots$ &$\ldots$ \\
            &$\sigma$          &(Jy) &       &0.012 &0.019 &0.028 &0.028 &0.040 &0.043 &0.103 &0.079 &$\ldots$ &$\ldots$ \\
            &$m$               &(\%) &       &6 &9 &12 &11 &14 &12 &26 &15    &$\ldots$ &$\ldots$ \\\\
J0948$+$0022&N                 &     &       &56 &63 &63 &65 &57 &31 &43 &14 &30 &23 \\  
            &$\Delta$t         &(days)&      &1863 &1863 &1863 &1863 &1862 &1739 &1862 &1766 &1644 &1502 \\      
            &$\left<S\right>$  &(Jy) &       &0.177 &0.255 &0.356 &0.390 &0.442 &0.555 &0.488 &0.569 &0.596 &0.530 \\  
            &$\sigma$          &(Jy) &       &0.039 &0.086 &0.142 &0.157 &0.173 &0.155 &0.188 &0.226 &0.193 &0.169 \\  
            &$m$               &(\%) &       &22 &34 &40 &40 &39 &28 &38 &40 &32 &32 \\\\
J1505$+$0326&N                 &     &       &37 &42 &39 &43 &37 &15 &26 &7 &$\ldots$ &$\ldots$ \\  
            &$\Delta$t         &(days)&      &1204 &1261 &1261 &1261 &1261 &1233 &1261 &301 &$\ldots$ &$\ldots$ \\      
            &$\left<S\right>$  &(Jy) &       &0.456 &0.507 &0.503 &0.488 &0.465 &0.463 &0.408 &0.377 &$\ldots$ &$\ldots$ \\  
            &$\sigma$          &(Jy) &       &0.076 &0.085 &0.080 &0.072 &0.066 &0.050 &0.079 &0.103 &$\ldots$ &$\ldots$ \\  
            &$m$               &(\%) &       &17 &17 &16 &15 &14 &11 &19 &27 &$\ldots$ &$\ldots$ \\\\  \hline
  \end{tabular}
%  \tablefoot{The top panel shows likely members of Pismis~11. The second
%    panel contains likely members of Alicante~5. The bottom panel
%    displays stars outside the clusters.\\
%    \tablefoottext{a}{Frequecny in GHz}\\
%    \tablefoottext{b}{Mean flux density in Jy}\\
%    \tablefoottext{c}{Standard Deviation of around the mean flux density in Jy: measure of the variability amplitude.}
%  }
\end{table*}
%--------------------------------------------------------------------

\subsection{J0324$+$3410}
\label{sec:rlcs0324} 
The multi-frequency light curves of J0324$+$3410 are shown in Fig.~\ref{fig:lc_lmh_03n08}.
% Of the four studied RL NLSy1s,
It shows a clearly different behaviour between high- and low-frequency bands despite the rather sparse
sampling. Of the main events present in the 86.24 and 142.33~GHz light curves -- around MJD 56040, 56300 and
56620 -- the first one disappears already at 10.45~GHz, the second is still present at 8.35~GHz, and the last
can be traced down to 4.85~GHz. These events also show different characteristics such as rise and decay
times. Generally, the source activity is very moderate at the lowest frequencies.

The neighbouring frequencies show associated events and strong evidence of intense spectral
evolution. Differences between similar frequencies can be regarded as unusual because the events disappear
fast over frequency. At the highest frequencies, flux density variations by factors of about 5 or more are
present, which are absent at the lowest frequencies. The spectral evolution may be an even better proxy of
this acute behaviour, as we discuss later.

Finally, we note that the baseline at all frequencies is practically flat, an indication that all the
variability incidents evolve fast and dissipate in a dominating relic jet that either does not display any
signs of variability or does so at a very slow pace; the dynamics of the radio SED shape point toward such an
interpretation, as well.
% TODO \red{+++ Nicolas analysis results: time scales etc} \red{+++ PDF, PSD, GMM
% normality test, KDE kernel density estimator}

\subsection{J0849$+$5108}
\label{sec:rlcs0846}
The very few available SEDs for this source indicate that this is another interesting and very active
source. Unfortunately, the lack of a large enough and adequately sampled dataset prevents any systematic
quantitative analysis. Nevertheless, the source shows activity cycles at all frequencies,which evolve
systematically slower at lower frequencies (see the slow rising trend that becomes faster toward higher
frequencies). It remains to be studied whether there is spectral evolution and what its characteristics
are. Clearly, longer time-baselines are needed before any sensible results can be reached.

\subsection{J0948$+$0022}
\label{sec:rlcs0948}
J0948$+$0022 is among the best studied sources because it has been the first RL NLSy1 to be detected by {\it
  Fermi}. The light curves shown in Fig.~\ref{fig:lc_lmh_09n15} clearly indicate intense and repetitive
variability that is prominent at all frequencies except possibly at 2.64~GHz, where most of the outbursting
events have already smeared out. The light curves show variability corresponding to factors of more than 3
even at frequencies as low as 10.45~GHz or below. As an example, the light curve at 8.35~GHz undergoes a
series of clearly discernible outbursts at MJDs, approximately 55020, 55240, 55490, 55725, 55985, and
56450. These events, which are single flares or sub-flares, appear in practically every densely enough sampled
light curve except for that at 2.64~GHz (where the events have already disappeared). Even more interestingly,
the moment of occurrence of an event appears at progressively later times as the frequency decreases, which
indicates opacity effects.  That is, a local maximum in a light curve corresponds to the instant at which the
emitting plasma radiation becomes optically thin at the observing frequency.
% which then can escape. 
If the flare is associated with the emergence of an adiabatically expanding plasmon, this instant should
indeed appear at progressively later times as the frequency drops. This claimed spectral evolution is very
clearly seen in Fig.\ref{fig:specs} as we discuss below. Finally, the pace at which the observed flux density
($dS/dt$) increases during a flare is clearly a function of frequency, with higher frequencies showing much
larger derivatives.
% TODO 140712: 
%\red{(+++ has it been
%  seen by fins for example? quantify, comment, ref)} \red{what I was doing last: was
%  trying to fit simple smooth function to get the peak at frequencies and then plot the
%  pace at which frequencies become op thin and see if different events do that at
%  different pace and also find the DS/dnu - also examine periodicity}

\subsection{J1505$+$0326}
\label{sec:rlcs1505}
The last source in our list of targets is J1505$+$0326. Its mean flux density of 507~mJy at 4.85~GHz, makes it
the brightest member of the sample. Significant variability is also present, but with qualitatively different
characteristics. The most important difference -- at least compared with J0948$+$0022 -- is that the clear
spectral evolution with the delay of the peak as a function of frequency is not as obvious here. Pairs of
adjacent frequencies such as 4.85 and 8.35~GHz seem to show events in phase, although the lowest frequency,
2.64~GHz, does indeed lag by clearly discernible time spans. At intermediate frequencies -- for example at
10.45~GHz -- the amplitude of variability reaches moderate factors of 1.5. Finally, frequencies below
10.45~GHz show a long-term decaying trend modulated by the faster variability discussed above.

\section{Flare decomposition: parametrising variability}
\label{sec:flaredecomp}
The parametrisation of flux density outbursts -- in terms of amplitude and time scale -- is commonly used as a
method to constrain the variability brightness temperature $T_{\rm{var}}$ associated with the event, on the
basis of causality arguments. The variability brightness temperature can only provide a lower limit of the
intrinsic brightness temperature $T_{\rm{B}}$. Values of $T_{\rm{var}}$ in excess of independently calculated
limiting values of the brightness temperature are generally attributed to Doppler boosting, which provides a
handle for the computing limiting Doppler factors, $D$. Assuming an equipartition brightness temperature upper
limit of $5\cdot10^{10}$~K \citep{Readhead1994ApJ}, one can estimate the Doppler factors required to explain
the observed excess.

%With the Doppler factor being a function of the viewing angle and the bulk velocity of the emitting plasma,
%this logistic proxies the mechanical properties of the emitting material emphasising the importance of such
%studies. 
The combination of variability Doppler factors with Very Long Baseline Interferometry (VLBI) measurements of
the apparent speeds allows computing the plasmoid bulk velocity and jet viewing angle
\cite[e.g.][]{1999ApJ...521..493L}. Here, we are equally interested in all these properties for our four RL
NLSy1s, and most importantly, in investigating the possible differences in the characteristics of different
flares in the same light curve, rather than retrieving the characteristics of an average behaviour.

A practice commonly followed in variability studies is the implementation of time-series analysis methods that
are designed to reveal such quantities; for example the structure function analysis \citep{Simonetti1985ApJ}
and the discrete correlation function \citep{1988ApJ...333..646E}. One of the most important caveats of such
methods, however, is that
%despite their seemingly straightforward approach 
they are extremely sensitive to parameters that are difficult to determine in moderately sampled light curves
such as the onset of a flaring episode or the shape of the temporal behaviour of the measured flux density. In
fact, even minor changes in such parameters can result in differences in the estimation of the variability
brightness temperature beyond an order of magnitude. Furthermore, in most cases, these tools are designed to
detect a dominant behaviour that smears out possible significant differences in the characteristics of
individual flares of the same source and even at different observing frequencies.

To overcome such complications, we introduce a novel method for
\begin{enumerate}
\item[a.] first creating ``cumulative'' light curves by conveniently shifting and
  re-normalising the observed light curves. This operation is meant to highlight the
  flares that are detectable at a wide range of frequencies,
\item[b.] and subsequently subjecting those light curves to all necessary operations to extract the desired
  parameters (i.e. flare onset, duration, amplitude, etc.).
\end{enumerate}
The guiding principles while
developing this approach were to
\begin{enumerate}
\item avoid complications introduced by the superposition of simultaneously acting processes. For example,
  time-series analysis methods  often return unrealistically long timescales only as the result of having --
  for example -- a long-term almost-linear trend underlying much faster events.
\item to accommodate a generic approach in the treatment of every flare. That is, to parametrise each event
  independently (for each source and frequency) and investigate the possibility of different behaviours (and
  possibly variability mechanisms) acting in the same source at different times. For J0324$+$3410,
  for example, the most prominent event seems to demand such an approach because its
  phenomenology is very different from the rest (Fig.~\ref{fig:0324_0948decomp})
\end{enumerate}
All the details of the method are discussed in Appendix~\ref{sec:themethod}.

% {\bf The reason why we invet Nics method is (a) because we do not like the SF aba;sys but qe cannot say
% thus ebecasue we use it eslewjhere. F example th eonset of the flare is
% the problem and yje fact tha SF treats everything equally. (b) We treat every flare ib
% the source indvidally. for einsatnce in 0324 the brightest flare behaves comletely
% different from others the most intesnse and makes it obvious that deserves a special
% treamtent ++ add the MJD of the evnt for the redaer. Other reasons: (c) SF gives very long
% timescales because it does a mix of trends, flares etc (d) the new approcah gives us the
% chnace to give an actually analytical description of the flux developement over time for each event so that
% the fits are different function for each flare but also for thesame flare different function for different
% frequexny. }

\subsection{Results of the flare decomposition}
\label{sec:decomp_results}
Since the variability brightness temperature $T_{\rm{var}}$ comprises only a lower limit of the
brightness temperature $T_{\rm{B}}$, the higher the $T_{\rm{var}}$ estimate, the better $T_{\rm{B}}$
is constrained. Henceforth, from all the different estimates of $T_{\rm{var}}$ we compute in the following
analysis, the highest value is regarded as the most meaningful one.
 
The method described above was applied to J0324$+$3410, J0948$+$0022, and J1505$+$0326, for
which sufficiently long datasets were available and which showed a significant number of activity ``cycles''
(5, 6 and 4, respectively).  
%The steps have been iterated until the best shape for each flare had been
%identified 
%\red{NICOLA: lars suggests to show an example that demonstrates the
%  steps}. 
Table~\ref{tab:flaredecomp_results} summarises the results of this analysis. The highest frequencies are
missing clearly because of the lack of data points at these bands.

%{\bf \\RECIPE:
%\begin{itemize}
%\item are the flaresz different from each other at least fro the Tb oint of view?
%\item what values do e get for the Tb imits 
%\item Note that despite the variet y of Tb the D do not change signifficantly! 
%\end{itemize}
%}

\subsubsection{Flare decomposition for J0324$+$3410}
Table~\ref{tab:allflares} summarises the results of the flare decomposition method for all the significant
events seen in the source light-curves shown in Fig.~\ref{fig:lc_lmh_03n08}. As significant events are
regarded flux variations that are above the noise level and can be detected at several wavelengths.  As can
be seen there, the variability of J0324$+$3410 is characterised by five fast variations shorter than 66 days (even at
the lowest frequencies) with a relatively low amplitude; the largest amplitudes of the flares is generally reached at
14.60~GHz.

The flare occurring at around MJD~56313 shows exceptional characteristics. Its maximum is reached at
32.00~GHz, and at that frequency its amplitude is around ten times larger than the average of the other
flares. The extraordinary phenomenology of this event alone could justify the introduction of an alternative
analysis method like the one described here. A classical Structure Function analysis, for example, would have
smeared the event out.  Below 10.45~GHz it moderates its behaviour and displays amplitudes comparable to that
of other events. Its temporal behaviour is similar to that of the immediately subsequent event which peaks
around MJD~56356 at 32.00~GHz. The maximum time delay between 32.00 and 2.64~GHz is around 100 days for both
flares, considerably longer than the estimated delays (between 20 and 70 days) for the other detected
events. This might be an indication that the spectral evolution of the flares is temporarily modified by the
onset of the main outburst.

The largest variability brightness temperatures are measured at 2.64~GHz. They
all exceed $2\times10^{12}$~K, implying a minimum Doppler boosting factor of $\sim4.3$.  In
Fig.~\ref{fig:0324_0948decomp} we plot the time delays, the variability brightness temperature and the
amplitude parameters for each flare separately.
%{\bf \\NOTES:
%\begin{itemize}
%\item Nicola: +++ write this paragraph In the case of J0324$+$3410 flare No 1 -- the major event at MJD 56313
%  -- has characteristics very different from those of the others (see the evolution of the peak flux at
%  different wavelengths, Fig. 0324flareamp). Also the variations of the time delays are remarkable: it would
%  be interesting to check if these variations are random or show some regularity with time. But we don't have
%  enough information to draw any conclusion yet.
%\end{itemize}
%}

\subsubsection{Flare decomposition for J0948$+$0022}
For J0948$+$0022, six prominent outbursts were identified in the light curves
(Fig.~\ref{fig:lc_lmh_09n15}, Table~\ref{tab:allflares}), making the source a prototype for applying the
discussed method.  As can be seen in Fig.~\ref{fig:0324_0948decomp}, for five of the six events the largest
amplitude is seen at 32.00~GHz, while for flare 3 it is seen at 14.60~GHz. This is the result of high-frequency
components that frequently appear at higher frequencies and expand fast towards lower
bands. This behaviour is in accordance with the spectral evolution and the variability. The amplitude of the
flares at different frequencies is shown in the upper panel of Fig.~\ref{fig:0324_0948decomp}. Interestingly,
\begin{enumerate}
\item flares 1 and 2 are remarkably similar to each other, but very different from flares 4, 5 and 6. The
  latter three are relatively isolated events, while the former two occur at the peak of the outburst
  phase. It is likely that this is exactly the reason for the difference in the amplitudes.
\item flare 3 is different from all the others, with a large excess of flux at 14.60 and 10.45~GHz. The
  indications for an unusual spectral evolution of this event may be the interpretation of this phenomenology.
\end{enumerate} 

On the other hand the time delays, between different frequencies shown in the middle panel of
Fig.~\ref{fig:0324_0948decomp} reveal a mildly different behaviour only for flare 5. Unfortunately, a
gap in the 2.64~GHz data approximately where the peak is expected prevents us from excluding that the real
time delays have to be shifted by about 30 days. Similarly, flare 3 at 2.64~GHz and flare 5 at 32.00 and 23.05~GHz
cannot be well constrained because of inadequate sampling.

The largest $T_{\rm{var}}$ is measured at 14.6~GHz. In the lower panel of Fig.~\ref{fig:0324_0948decomp} we
show $T_{\rm{var}}$ versus $\nu$ for the different flares. Most light curves give values of $T_{\rm{var}}$
roughly between 2 and $13\times10^{12}$~K. The differences from flare to flare are small (drops in the flux
density are partially compensated for by decreasing timescales), so that they cannot provide independent
estimates of the variability brightness temperature. The flare models at different wavelengths, however, are
independent of each other and hence the relatively small spread of $T_{\rm{var}}$ at different frequencies can
be regarded as significant.
%an indication for the reliability of the
%results. 
The inferred values of $T_{\rm{var}}$ imply variability Doppler factors of the order of 5 to 7. The most
prominent event occurred at around MJD~56444 (flare 3) and gave a $T_{\rm{var}}$ of more than $10^{13}$~K
implying a $D$ of more than 8.7.

\subsubsection{Flare decomposition for J1505$+$0326}
\label{sub-sec:decom1505}
Finally, for J1505$+$0326 we detected four main events during the monitored period as shown in
Table~\ref{tab:allflares}. They are all relatively fast and of moderate amplitude, more similar to the
characteristics of J0324$+$3410 than to those of J0948$+$0022. The highest variability brightness
temperatures reach values of $\sim2\times10^{13}$~K and are detected at 2.64~GHz, implying Doppler factors of
almost 10.4. The spectral evolution of the flares seems to follow a standard pattern, with a long delay
($\approx$70 days) between 32.00 and 14.60~GHz and a considerably shorter one ($\approx$40 days) between 14.60
and 2.64~GHz.
%-----------------------------------------------------------------------
\begin{table}[]
  \caption{Variability parameters $T_\mathrm{var}$ and $D$ for the three sources whose datasets were long
    enough, and the frequency at which they are located. Here we report the highest values of $T_\mathrm{var}$.}     
  \label{tab:flaredecomp_results}  
  \footnotesize
  \centering                    
  \begin{tabular}{lrrr} 
    \hline\hline                 
    Source     &\multicolumn{1}{c}{$\nu$} &\multicolumn{1}{c}{$T_\mathrm{var}$} &\multicolumn{1}{c}{$D$} \\
                      &\multicolumn{1}{c}{(GHz)}   &\multicolumn{1}{c}{(K)} & \\
    \hline    \\     
    J0324$+$3410 &2.64 &$25\times10^{11}$ &4.3 \\ 
    J0948$+$0022 &14.60 &$13\times10^{12}$ &8.7 \\
    J1505$+$0326 &2.64 &$26\times10^{12}$ &10.4 \\
   \\ \hline 
  \end{tabular}
\end{table}
% -----------------------------------------------------------------------

\section{Radio spectral energy distributions}
\label{sec:spec_evolution}
% -----------------------------------------------------------------------
% the plots here have ONLY 3 sigma data points UPDATED-OK: FINAL January 24
% ea-140924: ONLY DATA points with 3 sigma are used in theplots. How: I ran the ipython oer all files in the:
% "spectra" folder of each source folder (e.g. J0948+0022 etc). I nfound less than 3sigma and commented the
% lines out in the Spectra-date.dat fiels and ten I ran plot_spectra.py programme there in. 
\begin{figure}[] 
\centering
\begin{tabular}{c}
\includegraphics[width=0.3\textwidth,angle=-90]{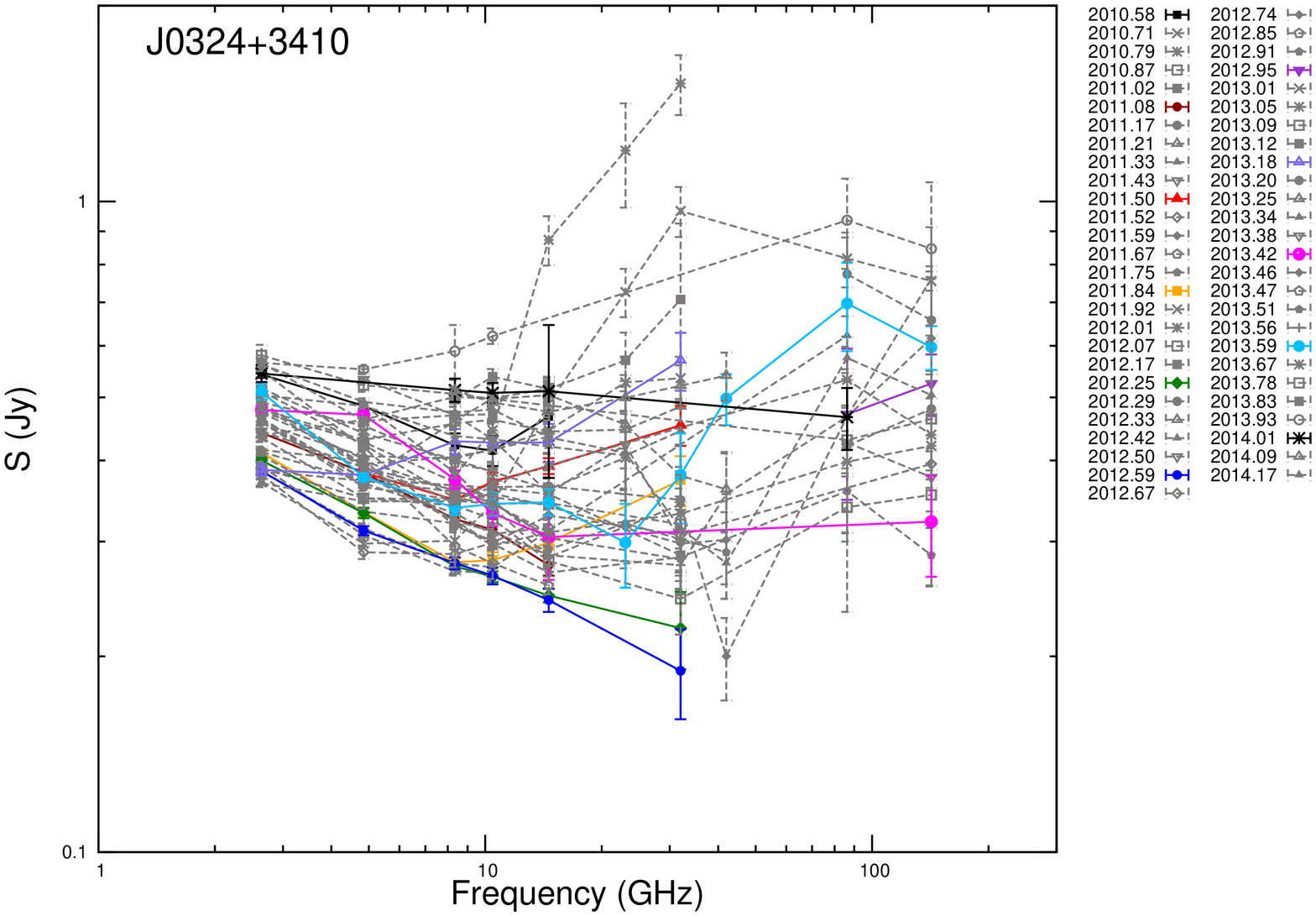}\\
\includegraphics[width=0.3\textwidth,angle=-90]{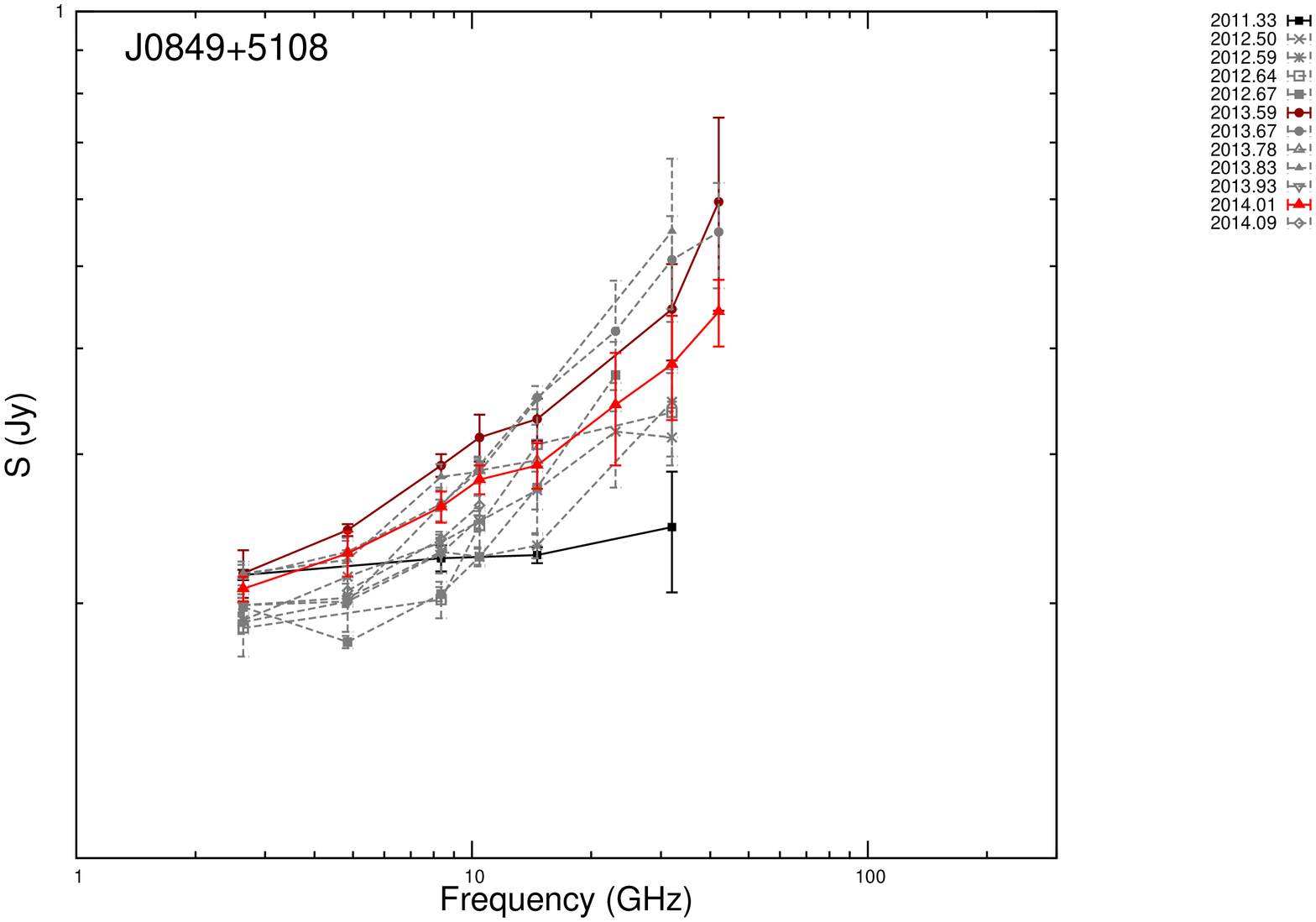} \\
\includegraphics[width=0.3\textwidth,angle=-90]{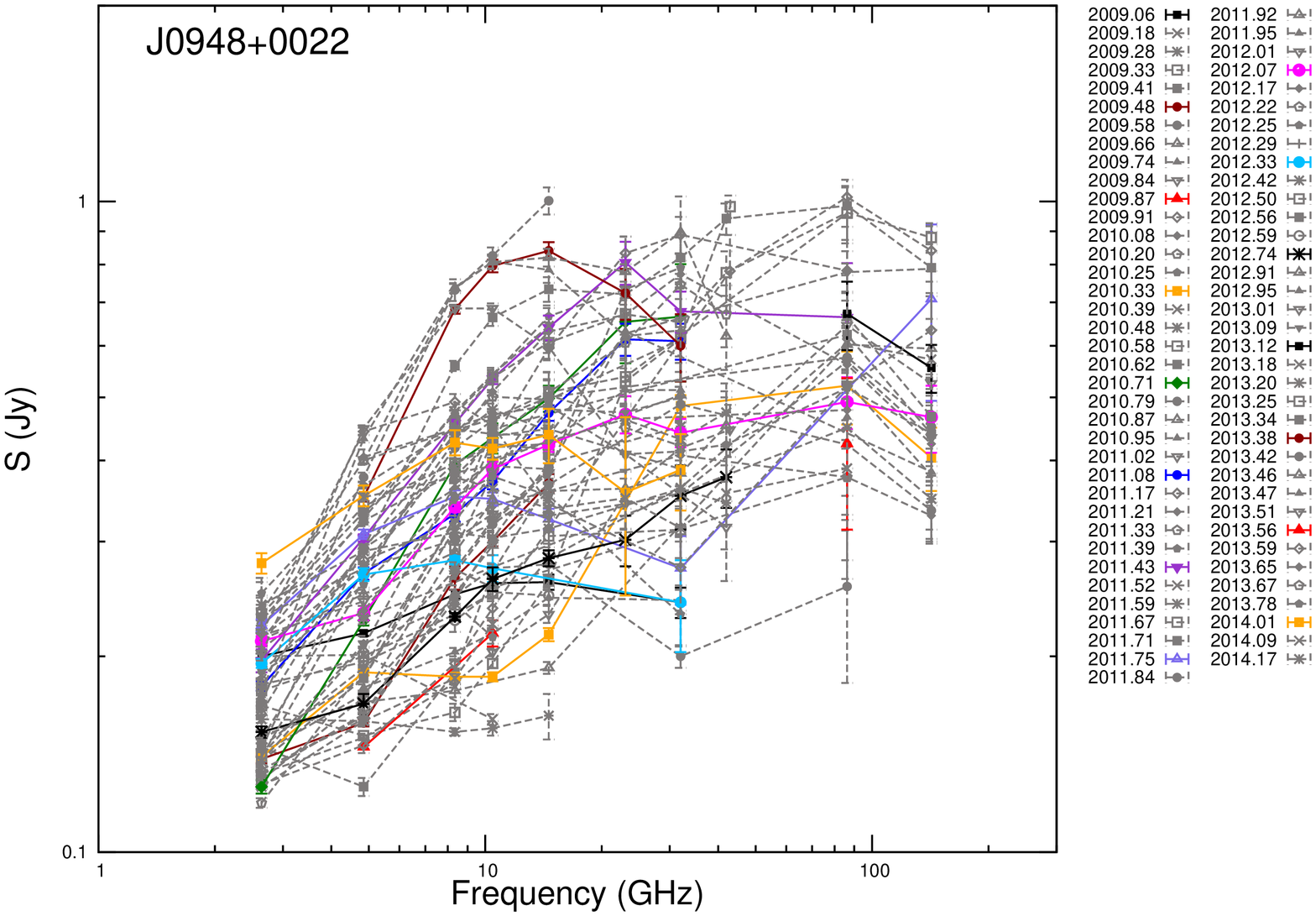} \\
\includegraphics[width=0.3\textwidth,angle=-90]{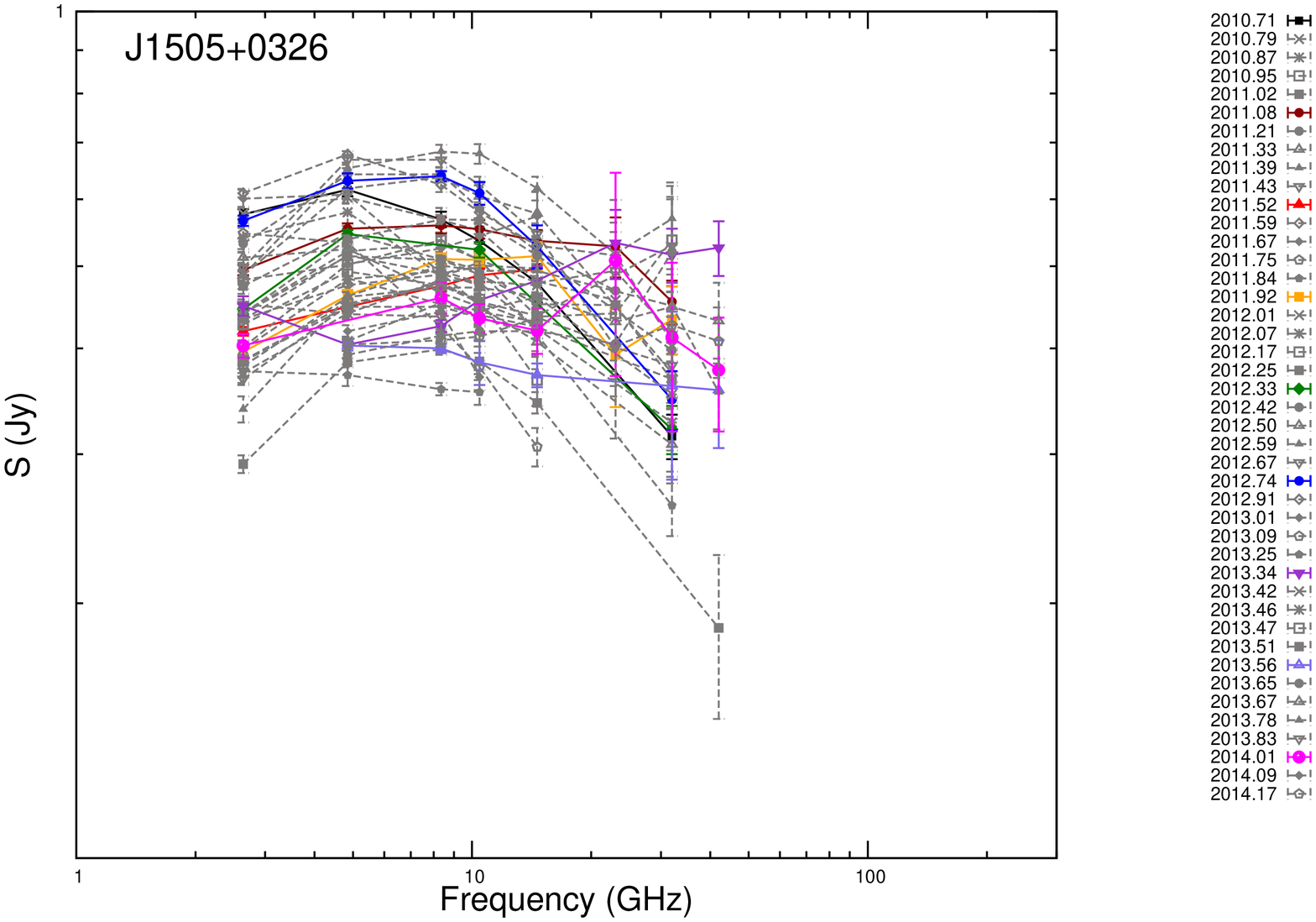} \\
\end{tabular}
\caption{Radio SEDs of the four monitored NLSy1 galaxies. The data points are connected with
  straight segments to guide the eye. The legend denotes the fractional year for each SED.}
\label{fig:specs}
\end{figure}
% -----------------------------------------------------------------------
Figure~\ref{fig:specs} shows the densely sampled broadband radio SEDs. There only data points with an
S/N$\ge 3$ are shown. For guidance to the eye, the data points are connected with segments, and to avoid
crowding the plots, only every fifth SED is coloured\footnote{Animated SEDs can be found at
  \url{http://www3.mpifr-bonn.mpg.de/div/vlbi/fgamma/}}. In Table~\ref{tab:specs} a summary of their most
characteristic parameters is given. These are the total number of available SEDs of at least two different
frequencies, N; the average cadence quantified as the mean time-separation between two consecutive SEDs,
$\frac{max(\mathrm{MJD})-min(\mathrm{MJD})}{N}$; the coherence,that is, the mean time-separation between the
earliest and latest measurement in a SED, $max(\mathrm{MJD})- min(\mathrm{MJD})$; and the spectral index in
three sub-bands computed from the mean flux densities in those bands. The spectral index is defined as
$\alpha=d\ln S/d\ln\nu$, where $S$ is the measured flux density at a frequency $\nu$. As can be seen there,
the mean cadence (excluding J0849$+$5108) is shorter than a month, and the combined Effelsberg-IRAM spectra
are made of measurements synchronous within less than two days. It is reasonable to assume that the shapes of
the individual SEDs are free of source variability effects.

From the plots in Fig. ~\ref{fig:specs} it is evident that the mere sample of these four sources displays the
main classes of spectral variability patterns described in \cite{2012arXiv1202.4242A}. There, the variability
patterns were interpreted as the result of the same underlying system consisting of a steep spectrum component
attributed to a large-scale jet populated with high-frequency synchrotron components that evolve according to
\cite{Marscher1985ApJ}. For \cite{2012arXiv1202.4242A} the different observed classes were governed by the
amount of ``activity'' that the observing bandpass allows us to see, and was also caused by different
intrinsic properties. Except for J1505$+$0326, where the spectral evolution is not discernible, all other
three NLSy1 would be classified as type 1 or 2 in that classification scheme because they show an intense
spectral evolution with or without a steep -- presumably quiescent state -- spectrum.
% --------------------------------------------------------------------
%
% The data show here are updated until October 2013 or something included. Th eIRAM I
% dunno. Update in January 2014. ONLY datapoint with S/N >=3 are used! ALL OK
%
% low: 1<= nu <=10 GHz, med: 10<= nu <=25 GHz hig: 25<= nu <=250 GHz
% Explanations:
% the colum for the numbers of spetra N and Dtau I get from teh files J0324+3410.log
% WRONG: I think this is all coming from teh statistics.txt
% he spectral indices I ran AverageListNewNEw on the SpInd-J0324+3410.dat by running mi.py
%For the cadence I take the file Spectra_avrg-J0324+3410.dat, I read the JD and I find min
%and max, do the difference and divde with the nimber N of spetcra that I have.
%
\begin{table*}[]
  \caption{\label{tab:specs} Mean characteristics of the observed radio SEDs shown in
    Fig.~\ref{fig:specs}. The spectral indices are computed in three sub-bands, and 
    along with the mean values, the extremes are reported as well.}
\centering
  \begin{tabular}{lcccccccccccccc}
    \hline\hline
    Source       &N   &Cadence &Coherence   &\multicolumn{3}{c}{low: 2.64 - 8.35~GHz} &
    &\multicolumn{3}{c}{mid: 10.45 - 23~GHz} & &\multicolumn{3}{c}{high: $\ge$43.05~GHz}  \\
\cline{5-7}
\cline{9-11}
\cline{13-15}
                 &    &(days)  &(days)      &min &mean &max                      & &min &mean &max                     & &min &mean &max                      \\
    \hline\\
% ea-140919: updated all values
    J0324$+$3410 &55  &25   &2.0  &$-$0.49 &$-$0.21 &$+$0.12 & &$-$0.53 &$-$0.04 &$+$1.57 & &$-$0.67 &$+$0.07 &$+$0.77 \\ % OK updated on Spet 19, 2014
    J0849$+$5108 &12  &84   &0.02 &$+$0.04 &$+$0.18 &$+$0.44 & &$+$0.08 &$+$0.32 &$+$0.66 & &$+$0.28 &$+$0.63 &$+$1.08 \\
    J0948$+$0022 &75  &26   &2.2  &$-$0.13 &$+$0.51 &$+$1.26 & &$-$0.83 &$+$0.27 &$+$1.02 & &$-$1.73 &$-$0.03 &$+$1.58 \\
    J1505$+$0326 &43  &29   &0.2  &$-$0.22 &$+$0.13 &$+$0.34 & &$-$0.66 &$-$0.18 &$+$0.18 & &$-$0.86 &$-$0.24 &$+$0.07 \\\\
    \hline
  \end{tabular}
  \tablefoot{Column description: 2: the total number of available SEDs; 3: the mean cadence i.e. the separation
    between two consecutive spectra measured in days; 4: is the $max(\mathrm{MJD})- min(\mathrm{MJD})$ within one single SEDs; 5, 6, 7: are the minimum, the mean and the maximum spectral indices for the low band and so forth.}
%  \tablefoot{The top panel shows likely members of Pismis~11. The second
%    panel contains likely members of Alicante~5. The bottom panel
%    displays stars outside the clusters.\\
%    \tablefoottext{a}{Frequecny in GHz}\\
%    \tablefoottext{b}{Mean flux density in Jy}\\
%    \tablefoottext{c}{Standard Deviation of around the mean flux density in Jy: measure of the variability amplitude.}
%  }
\end{table*}
%--------------------------------------------------------------------

\subsection{J0324$+$3410}
\begin{figure}[] 
\centering
\includegraphics[trim=80pt 30pt 30pt 20pt  ,clip,width=0.35\textwidth,angle=-90]{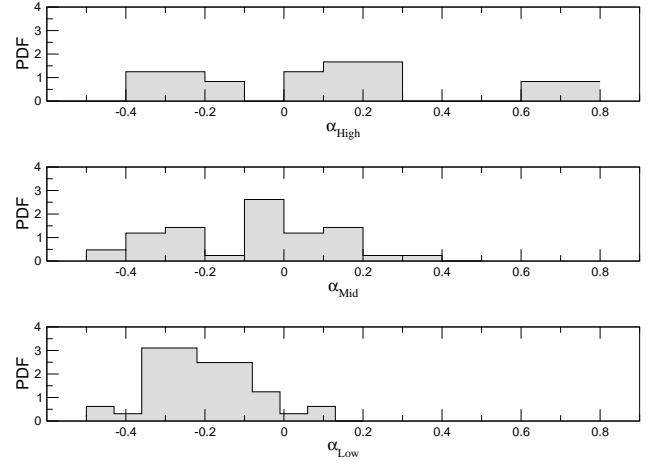} 
\caption{Probability density functions of the spectral indices observed for J0324$+$3410 in three different
  sub-bands: low, middle, and high. The bin size is selected to be approximately one standard
  deviation of the plotted indices. }
\label{fig:spind0324}
\end{figure}

For J0324$+$3410 Fig.~\ref{fig:specs} clearly shows the emergence of two distinct
components: 
\begin{itemize}
\item A steep spectrum component that appears at frequencies below 10.45~GHz and mostly retains
  a constant negative spectral index that reaches values as low as almost $\approx
  -0.5$ (see Table~\ref{tab:specs}) and could be the optically thin part of a synchrotron
  self-absorbed (SSA) component. 
\item A sequence of high-frequency components (HFC) that
  introduce intense variability as a result of their spectral evolution.
\end{itemize}
A convenient way to understand the reason for the observed behaviour is to examine the activity in terms of
spectral index distribution in three different frequency sub-bands:
\\\\
\begin{tabular}{rl}
{Low}:      &for $\nu \in {[{~~2.64}~,~~~~8.35]}$~GHz\\
{Middle}:   &for $\nu \in {[{10.45}~,~~23.05]}$~GHz\\
{High}:     &for $\nu \in {[{32.00}~,142.33]}$~GHz.\\
\end{tabular}
\\\\\\
In Fig.~\ref{fig:spind0324} we show the spectral index distributions in these sub-bands. The histograms are
normalised so that they comprise the probability density functions (PDF) of the three indices. The bin
size was selected to be similar (if not equal) to the mean standard deviation computed over the corresponding
dataset. This representation compresses the behaviour and the phenomenology discussed earlier and immediately
indicates the sub-band where the activity is mostly concentrated.
%% -----------------------------------------------------------------------
% DONE: added on May 6, 2014
% -----------------------------------------------------------------------
 
As can be seen in Fig.~\ref{fig:specs}, the region above roughly 10.45~GHz hosts all the source activity in
terms of spectral evolution. It appears to be intensely variable, with the outbursting events emerging at the
upper end of the band and transversing downwards following a characteristic evolutionary path; first a
build-up at high frequencies, progressively harden as the flux increases, and a descent towards the left end
of the bandpass while shifting the peak to lower flux densities to eventually diffuse in the persistent steep
spectrum below 10.45~GHz.

This scheme agrees at least qualitatively with the evolutionary scenario proposed by
\cite{Marscher1985ApJ}. The succession of these components and the superposition of their evolving spectra
results in a flat average spectrum -- with a spectral index practically around 0.0 -- which at times can be
as hard as $+1.6$ (see Table~\ref{tab:specs}).

The steep spectrum component that dominates the SEDs at lower frequencies is not static. It shows variability
in a self-similar fashion and mostly retains a steep character. The spectral index is positive for less than
7\% of the time. The lowest indices ($\approx -0.5$, see also Fig.~\ref{fig:spind0324}) indicate the
canonical values seen in large-scale relaxed jets \citep[e.g.][]{angelakis2009AnA}. A natural way to explain
this behaviour is to assume that below about 10~GHz, the high-frequency components are optically thin. They
transverse that part of the bandpass and slowly fade away and gradually contribute less and less power. This
lets the overall spectrum appear to move downwards in flux density while preserving a rather constant index. This
interpretation implies that high-frequency components with similar intrinsic parameters are generated
(i.e. magnetic field and particle density). At the very low frequencies (2.64 and 4.85~GHz), the moderate
spectral variability indicates an underlying large-scale jet. Essential in favour of this
scenario would be the detection of significant polarisation in that part of the emission, as we discuss in
Sect.~\ref{sec:polarization}.

The upper end sub-band ($\nu\ge 43.05$~GHz) can become as steep as of $\alpha \approx -0.7$
(Fig.~\ref{fig:spind0324}), denoting that the transition from optically thick to thin regime
of the HFC occurs in that spectral region. This occurs rather rarely; only 14\% of the
cases show such a steep index, most likely because as an event reveals its optically thin
part, a new one emerges, that fast hardens the otherwise softening spectrum. 

In the middle sub-band ($10.45 \le \nu\le 23$~GHz), however the source spends half the time
in soft and half in hard state since the ``mixing'' of thin and thick events is prominent
there. In any case, the transition from thick to thin has very serious consequences
because if proven true, it should be accompanied with exactly 90\degr rotations of the
projected electric vector position angle (EVPA) as is described by
\cite{2014arXiv1401.2072M}.

\subsection{J0849$+$5108}
The limited dataset available for J0849$+$5108 (see Fig.~\ref{fig:specs}) already reveals a very interesting
behaviour. First to be noted is the complete absence of any steep spectrum signature from the SEDs that have
been gathered so far. The spectrum oscillates between flat and hard within all three sub-bands. As an example,
the low sub-band spectral index varies between $+0.02$ and $+0.44$. In the high sub-band it reaches values
higher than $+1$. Furthermore, the events always show their optically thick part, which -- in the framework of
incoherent synchrotron processes -- is expected to be associated with an insignificant degree of linear
polarisation. Clearly, more duty cycles are needed to allow better quantification of its spectral behaviour.

\subsection{J0948$+$0022}
\begin{figure}[] 
\centering
%                       t     l    b     r
\includegraphics[trim=80pt 30pt 30pt 20pt  ,clip,width=0.35\textwidth,angle=-90]{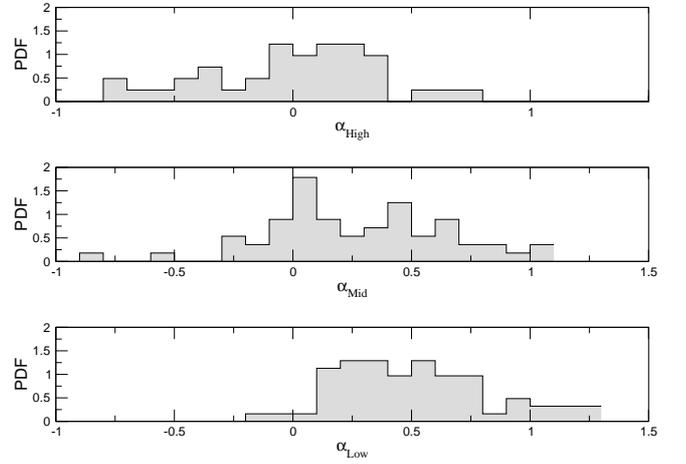} 
\caption{Probability density functions of the spectral indices observed for
  J0948$+$0022 in three different sub-bands: low, middle, and high. The bin size
  is selected to be approximately one standard deviation of the plotted indices.  }
\label{fig:spind0948}
\end{figure}

The complete absence of a quiescence spectrum also holds for J0948$+$002. The spectra appear to be highly
inverted and intensely variable. An impressive spectral evolution with HFCs emerging in the high sub-band is
observed. The HFCs often transit to optically thin phases within the middle sub-band. Interestingly, the
lowest band spectral index ($2.64\le \nu\le 8.35$~GHz) oscillates between a minimum of $-0.13$ and a maximum
of $+1.26$, while the middle- and high-band indices oscillate between $-0.83$ and $+1.02$ and $-1.73$ and
$ +1.58$, respectively (c.f Fig.~\ref{fig:spind0948}). This implies that we witnes the spectral evolution of
the HFCs, but fail to detect the quiescent steep spectrum, if any. VLBI studies (e.g. Karamanavis et al. in
prep.) also reveal only a compact core structure in agreement with the previous finding.

Furthermore, the relatively fast pace at which the spectral evolution occurs is interesting and rather
exceptional. In fact, significant variability has previously  been detected with a mean cadence of 25
days. This is also seen in the light curves, as discussed in
Sect.~\ref{sec:rlcs0948}. Figure~\ref{fig:spind0948} again gives the probability density functions of the
three spectral indices as for J0324$+$3410.
%% -----------------------------------------------------------------------
% DONE: added on May 6 2014 
% -----------------------------------------------------------------------

\subsection{J1505$+$0326 }
\label{sec:seds1505}
The behaviour of J1505$+$0326 is somewhat different from the behaviour of the sources discussed
previously. Most of the time, it shows a convex spectrum that has a global maximum in the vicinity of 5 -
10~GHz and occasionally seems to be modulated by the evolution of HFCs that become optically thin. For that
reason, it could be classified as type 5b in the scheme proposed by \cite{2012JPhCS.372a2007A}. The low
sub-band spectral index reaches moderately hard states ($\max \left(\alpha\right)=+0.34$), while the high
sub-band can become very soft ($\min \left(\alpha\right)=-0.86$).
%{\bf \\NOTES:
%\begin{itemize}
%\item HYPER SpInd vs Flux 
%\item HYPER comment: komossa et al 2006 say: DSSJ094857.3+002225 and RX J16290+4007 have
%  in- verted radio spectra but we know show that they are really variable it is not that
%  inverted simply! 
%\item HYPER studies done so far are focusing on low freqiuecies dominated by jet relic
%  emission opt. thin that never chanegs. We sample the heratgh of the activity! 
%\item HYPER find archival data from low frequencies to show the steep spectrum
%\item Komossa 2006 says: On the basis of the radio morphology and spectra, Komossa et
%  al. [52] also suggested that some RLNLS1s of their sample could be similar to Compact
%  Steep Spectrum (CSS) radio sources.
%\item HYPER histos of low spind etc for each source and get the extremes
%\item HYPER : the Damando and oriety papers have kinematic studies for 0846. Use it! 
%\item an item
%\item spectral evolution do a fits that rebeca did 
%\item TODO: take all spectra and find the min and max spectral idex at low and high band
%  after subtracting the steep spec. In the 0324 wher yu see it use you data. for the rest
%  iethret archival or -0.5. Then Table the spindex and the gamma of the electrons
%\end{itemize}
%}

\section{Jet powers}
\label{sec:jetpow}
% -----------------------------------------------------------------------
% 
%\begin{figure}[ht!] 
%\centering
%\begin{tabular}{l}
%\includegraphics[width=0.295\textwidth,angle=-90]{figures/JP_J0324p3410.ps}\\ 
%\includegraphics[width=0.3\textwidth,angle=-90]{figures/JP_J0849p5108.ps}\\
%\includegraphics[width=0.3\textwidth,angle=-90]{figures/JP_J0948p0022.ps}\\
%\includegraphics[width=0.3\textwidth,angle=-90]{figures/JP_J1505p0326.ps}\\
%\end{tabular}
%\caption{Temporal evolution of the radiative jet power. From top to bottom: J0324$+$3410,
%  J0849$+$5108, J0948$+$0022, J1505$+$0326. Time starts on 2009 January 1 (i.e. MJD 54832.0).}
%\label{fig:jetpow}
%\end{figure}
% -----------------------------------------------------------------------
Here we wish to quantify the power delivered by the four NLSy1s in our sample in the
radio bands and examine whether this could be delivered by a jet similar to those in typical blazars.
 
Within the jet theoretical framework proposed by \cite{1979ApJ...232...34B} and empirical relations that have
been proposed for relativistic jets in AGNs and galactic binaries by
\cite{2011RAA....11.1266F,2012AIPC.1505..574F,2014IJMPS..2860188F}, the radio measurements can accommodate the
computation of the jet power \citep[see also][]{2014arXiv1409.3716F}. In fact, it is possible to calculate
both the radiative and kinetic (electron, proton, magnetic) jet power from the radio core emission.  In a
conical jet, the optically thick radio emission is linked to the jet power as
\begin{equation}
L_{\rm radio} \propto P_{\rm j}^{17/12},
\end{equation}
By studying a sample of 88 $\gamma$-ray AGN (FSRQs, BL Lac Objects, RLNLS1s), \cite{2014IJMPS..2860188F} found
that the best normalisation coefficients values are
\begin{equation}
\log P_{\rm j,rad} = (12 \pm 2) + (0.75 \pm 0.04)\log L_{\rm radio, core}
\end{equation}
\begin{equation}
\log P_{\rm j,kin} = (6 \pm 2) + (0.90 \pm 0.04)\log L_{\rm radio, core},
\end{equation}
where $P_{\rm j,rad}$ and $P_{\rm j,kin}$ are the radiative and kinetic (proton, electron, and magnetic field)
powers of the jet, while $L_{\rm radio, core}$ is the $K$-corrected luminosity of the radio core
at 14.60~GHz. The $K$-correction was made by multiplying the monochromatic flux by the factor of
$(1+z)^{\alpha-1}$, assuming a flat radio spectral index ($\alpha\approx 0$, see also
Sect.~\ref{sec:spec_evolution}).

Although  it would be possible to use different frequencies as well, we persist in using
14.60~GHz for a number of reasons, such as to be consistent with the measurements
used to extract the used normalisation coefficients; the good sampling of the corresponding light curve; the
fact that this band is less contaminated by extended emission, which gives more reliable estimates of the radio
core emission; it allows comparison with other projects such as MOJAVE \citep{Kellermann2004ApJ,2009AJ....137.3718L,2013AJ....146..120L}.
%\begin{itemize}
%\item Simplicity of calculations (one estimate of the jet power suffices).
%\item It is an intermediate frequency between the source activity region and the low sub-band
%  where the events start disappearing. It is therefore a measure of both the power
%  attributed to the jet and that to variability events.  
%\item 14.6~GHz is the best-sampled frequency in the intermediate band.
%\item It allows comparison with other projects such as MOJAVE \citep{Kellermann2004ApJ}
%  which includes also structural information particularly useful in the case of kinematic
%  studies (Furhmann et al. in prep.)
%\end{itemize}

% Figure~\ref{fig:jetpow} shows the radiative jet power curves for the sources discussed
% here and for the sub-set of data that has been used which covers \red{+++ add exact
% dates}. 
In Table~\ref{tab:jet_pow} we provide the naturally weighted averages of the radiative $P_{\rm{j, rad}}$,
kinetic $P_{\rm{j, kin}}$, and total powers $P_{\rm{j, tot}}$. The averaging was made in the power rather
than the flux density domain. In the same table we present the most recent estimates for the black-hole masses
\citep{2014arXiv1409.3716F}.
%archival black-hole masses to demonstrate the fact that they appear
%to be systematically smaller than typically seen in AGNs. In the cases of J0324$+$3410,
%J0948$+$0022 and J1505$+$0326, the reported black-hole mass logarithms are taken from
%\cite{2009ApJ...707L.142A}. For J0849$+$5108 we estimated the black-hole mass with the
%method proposed by \cite{2006AJ....132..531K}, where the 5100~$\AA$ luminosity was
%approximated with the luminosity from the SDSS ``g'' filter magnitude. The FWHM was taken
%from \cite{2008ApJ...685..801Y}.
%-----------------------------------------------------------------------
\begin{table}[]
  \caption{Radiative and kinetic jet powers calculated from the 14.60~GHz observations
    averaged over the period discussed here. Literature values of their black-hole masses are as well.
    % TODO 140713:
    % \red{may be add averaged values at other frequencies to see contributions}
  }     
  \label{tab:jet_pow}  
  \footnotesize
  \centering                    
  \begin{tabular}{lcccc} 
    \hline\hline      
    % ea-140921: ATTENTION: teh values all were updated with luigis new numbesr send to me
    % in May After I send him the final data set. The eerros were
    % NOT uoadetd !
    Source     &$P_{\rm{j, rad}}$ &$P_{\rm{j, kin}}$ &$P_{\rm{j, tot}}$ &$\log \rm{M}^{1}$ \\
               &(erg~s$^{-1}$)  &(erg~s$^{-1}$) &(erg~s$^{-1}$) &($\log \rm{M}_{\sun}$)\\
    \hline    \\     
    J0324$+$3410 &$1.7\times10^{43}$ &$3.0\times10^{43}$ &$4.7\times10^{43}$ &7.6 \\ 
    J0849$+$5108 &$6.6\times10^{44}$ &$2.5\times10^{45}$ &$3.1\times10^{45}$ &7.5 \\
    J0948$+$0022 &$7.9\times10^{44}$ &$3.0\times10^{45}$ &$3.7\times10^{45}$ &7.9 \\
    J1505$+$0326 &$5.0\times10^{44}$ &$1.7\times10^{45}$ &$2.2\times10^{45}$ &7.3 \\
   \\ \hline 
  \end{tabular}
  \tablebib{(1) \cite{2014arXiv1409.3716F}}
\end{table}
% -----------------------------------------------------------------------
%for four of the five sources studied in the present work. One -- J$1246+0238$ -- was
%observed only once on MJD $55681.882$ (2011 April 30), but it was too weak for a
%meaningful campaign. Anyway, the calculated powers are: $P_{\rm jet,rad}=(2.7\pm2.1)\times
%10^{43}$~erg~s$^{-1}$, $P_{\rm jet,kin}=(5.1\pm4.8)\times 10^{43}$~erg~s$^{-1}$, which are
%just upper limits.

\subsection{J0324$+$3410}
% updated after the second run that Luigi did in May 2014
The prominent event seen in the source light-curves at around MJD~56300 (January 2013)
gave a corresponding increase in the radio power at that time. \cite{2013arXiv1306.4017T}
reported a change in the X-ray spectrum from January to February 2013 associated with that
event, with the emergence of a hard tail during the second epoch. The average values of
the radiative $P_{\rm{j, rad}}$ and kinetic jet power $P_{\rm{j, kin}}$ -- as measured in
the present work -- are $1.7\times 10^{43}$ and $3.0\times
10^{43}$~erg~s$^{-1}$. The same quantities calculated from the SED modelling
by \cite{2009ApJ...707L.142A} after the first year of the satellite operations were  
$6.3\times 10^{42}$ and $2.2\times 10^{44}$~erg~s$^{-1}$.

\subsection{J0849$+$5108}
% updated after the second run that Luigi did in May 2014
Unfortunately, the limited dataset available for J0849$+$5108 is not adequate to reveal the
source activity characteristics. Furthermore, available measurements are not simultaneous
to the $\gamma$-ray active phase it went through in 2010 \citep{2011nlsg.confE..24F}.
% some multi wavelength campaigns were organised (D'Ammando et al. 2012, 2013a).
The radiative jet power reported by \cite{2012MNRAS.426..317D,Dammando:2013hu} is around
$(5-50)\times 10^{44}$~erg~s$^{-1}$ (the maximum refers to the outbursts), which is consistent with our
  calculations, which gave an average of $6.6\times 10^{44}$~erg~s$^{-1}$.
%\cite{2012MNRAS.426..317D,Dammando:2013hu} do
%not give any estimate of the kinetic power, prohibiting a relevant comparison.

\subsection{J0948$+$0022}
\label{subsec:power0948}
% updated after the second run that Luigi did in May 2014
J0948$+$0022 was extensively studied by \cite{2012A&A...548A.106F}, who discussed the results of more than
three years of multi-wavelength data over the period between 2008 and 2011. They presented the
calculations of jet powers in terms of SED modelling. With the current work -- which comprises the
  natural extension of that monitoring to mid-2014 -- we can compare these findings with those of the
currently applied method for the overlapping period. A caveat of the SED modelling has been the delay
of the radio emission with respect to the optical-to-$\gamma$-ray emission, as is generally
expected. \cite{2014MNRAS.441.1899F} found that the average time delays at the source rest frame can span
from less than 10 to more than 70 days, systematically decreasing towards higher frequencies,  which
provides yet another argument in support of using the 14.60~GHz for the current
calculations. 

The radiative powers computed by \cite{2012A&A...548A.106F} were based on SEDs that were selected on the basis
of the $\gamma$-ray activity. During the lowest activity SED, the jet radiative power was estimated to be
$P_{\rm{j,rad}}=0.4\times 10^{45}\mathrm{~erg~s}^{-1}$, while during the highest activity state it was
$P_{\rm j,rad}=7.9\times 10^{45}\mathrm{~erg~s}^{-1}$. From our radio dataset, the radiative power ranges
between $0.4$ and $1.8\times 10^{45}\mathrm{~erg~s}^{-1}$. For the jet kinetic power, the SED modelling
applied by \cite{2009MNRAS.397..985G} showed that it ranges from $0.2$ to $4.1\times10^{47}$~erg~s$^{-1}$,
while based on the 14.60~GHz measurements of the present work, it varies in the range between $1.5$ to
$7.8\times 10^{45}$~erg~s$^{-1}$. This result is very interesting when the fundamental assumptions are
considered. The kinetic jet power -- as calculated from the SED modelling -- relies on the assumption that
there is one proton associated with each electron \citep{2000ApJ...534..109S}. If that were not the case and
instead there were one proton for more electrons, the kinetic power would be much lower, which is what our
finding indicates.

\subsection{J1505$+$0326}
% updated after the second run that Luigi did in May 2014
The multi-frequency light curves of J1505$+$0326 show at least one major event with mild spectral
evolution. Despite its mild variability, %during the whole campaign.
we can compare, our values with those from SED modelling reported in \cite{2009ApJ...707L.142A}.
$1.0\times 10^{44}$ and $1.6\times 10^{46}$~erg~s$^{-1}$ are the values of the radiative and kinetic jet
powers. The averages we calculated are $5.0\times 10^{44}$ and $1.7\times 10^{45}$~erg~s$^{-1}$,
respectively. The almost constant activity of the source is consistent with what was reported by
\cite{2013MNRAS.433..952D}, who analysed the monitoring at 15~GHz performed by the Owens Valley Radio
Observatory (OVRO).

% TODO Notes July 2, 2014
%
%===================================================================
%\section{Radio Luminosities (or powers) \red{[---Stefanie---]}}
%\red{zhou et a2003 gives alot of backgorund compare guves table and everuthhing!!!!}
%
%{\bf \\NOTES:
%\begin{itemize}
%\item HYPER Damando in paper 2 of the 0846 says how he computes the jet powers in seciotn
%  5 sed modeling and gives deremer 2009 and finke 2008
%\end{itemize}
%}

%===================================================================
\section{Polarisation}
\label{sec:polarization}
Assuming that the dominant radiation mechanism is incoherent synchrotron processes and that the
  variability is attributed to evolving synchrotron self-absorbed components, it is expected that significant
linear polarisation is detected. In this context, the degree of polarisation and the EVPA would depend
  on the emitting plasma opacity and hence the spectral index
\citep{1970ranp.book.....P,1980rgrt.book.....P}. Assuming highly ordered and intense magnetic fields
\citep{1967Natur.216..147S}, as those expected during flares, even circular polarisation could occur. Furthermore,
the orientation of the electric vector should also depend both on the optical thickness of the material and
on the assumed magnetic field configuration. Multi-band polarisation studies are then a unique probe of the
underlying physical conditions at the emission region.

For the sample discussed here, \cite{1994A&AS..106..303N} found optical polarisation from
J0324$+$3410, while \cite{1981ApJ...243...60M} and \cite{1984PASP...96..402S} detected
``high and variable optical polarisation'' from J0849$+$5108. This has been seen as an
indication that a relativistic jet is operating at the source. Similarly, \cite{2012ApJ...760...41D} have
detected significant polarisation in the radio.

% TODO 140713: \red{Ryosuke Itoh (Hiroshima University): "Study of Relativistic Jets with
% Optical Polarimetry in Various Timescales"He is talking about SED vs polarisation, He
% discusses 0948 very high polarisation and in intra night very variable and gives also
% 30\% in Radio by Doi}

\subsection{Radio polarisation} 
\label{subsect:radiopol}
\cite{2013hell.confQ..34A} explored the possibility of detecting radio linear polarisation at the
intermediate frequency of 8.35~GHz. The data in Table 3 there were corrected for instrumental polarisation and
indicated that all four sources display detectable polarisation of a remarkably low
magnitude. Exceptionally, J0324$+$3410 showed values higher than those typically observed in AGN at these
frequency bands ($\sim4$~\% at 5~GHz, e.g. \citealt{Klein2003AnA}). Its fractional linear polarisation at this
frequency is about 6~\% with an EVPA of about 36\degr, while at 4.85~GHz it reaches values of 7~\%
with an EVPA of about 44\degr. This phenomenology would agree with assuming that the optically thin
emission dominates towards lower frequencies. The 4.85~GHz emission is less contaminated by the optically
thick part of an HFC and displays a higher fractional polarisation. At both frequencies the EVPA
appears to be almost perpendicular to the jet orientation (the jet axis is about 120\degr, Karamanavis et al. in prep.),
so that the projected magnetic field appears to lie almost parallel to the jet, possibly implying a helical jet with a
very long helix step.

Similar arguments hold for J0849$+$5108, although the effect is much more moderate. Unlike J0324$+$3410, the
other NLSy1s show a very low polarisation degree. As is indicated by their spectral variability pattern, the
sources undergo flaring events that show their optically thick part at 8.35~GHz. 
%Polarisation
%measurements above 43.05~GHz and above 10.45~GHz for J0948$+$0022 and J1505$+$0326, respectively, should
%disclose a significant degree of polarisation. 

\subsection{$R$-band polarisation} 
\label{subsect:opticalpol}
First attempts to monitor the $R$-band linear polarisation in 2013 with the RoboPol instrument
\citep{2014MNRAS.442.1706K, 2014MNRAS.442.1693P} showed
%that in $R$-band all four sources
%behave differently urging for a careful investigation. The first indications hint that all
%J0324$+$3410 , J0849$+$5108 and J0948$+$0022 show polarisation at significant statistical
%levels. Specifically, 
that J0324$+$3410 and J0948$+$0022 present a rather low fractional polarisation that does not exceed a few percent
($\approx 3$~\%). \cite{2014arXiv1405.3731I} found similar values for the former case, while the latter
showed past events of polarisation values that reached almost 40~\% \citep{2013ApJ...775L..26I}. For
J0849$+$5108 we detected polarisation levels of about 10~\%.
%J1505$+$0326 on the other hand was attempted
%only two times with no detection of polarisation whatsoever.

%A thorough linear and circular multi-band radio as well as, linear $R$-band
%polarisation analysis is underway and will be discussed elsewhere (Angelakis et al. in prep.). 

%{\bf \\NOTES:
%\begin{itemize}
%\item HYPER : SpInd - polarization
%\item HYOE see 2013arXiv1304.1786M 2013arXiv1301.0657E because the y measure Optical variability of the source
%  polarization un optical
%\item include OVRO polarization and whatever else exists. IRAM!
%\item give polarization LC
%\item do RM analyss
%\item in 0324 we see even ibxrease of polarization towards low ferqiunies this may be du
%  to lees contamination from the High freq copoent
%\end{itemize}
%}
%
%

%===================================================================
\section{Discussion}
\label{sec:discussion}
The main goal for the analysis has been to study the relativistic jets that emit  the observed
radio power and to quantify its properties. 
%Specifically, The motivation for the current work has been the
%analysis and understanding of the physics driving the radio emission in the four sources we are examining, on
%the basis of the multi-frequency cm and mm monitoring. 
The discussed dataset comprises the longest term multi-wavelength radio monitoring datasets of the known $\gamma$-ray-detected NLSy1s to date.
%Still, it must be noted that the number of activity cycles included is not yet
%sufficiently large to extract solid conclusion. 

The focus was placed  mostly on the variability properties of the radio emission. In this context, we follow
different paths. 
% We examining the archival radio loudness of our sources.
We introduced a flare decomposition algorithm to quantify the variability amplitude, the involved timescales,
and its frequency dependence for each flare separately.  We then examined whether variability retains the same
characteristics over source, frequency or event.  Subsequently, the variability parameters (amplitude and time
scale) served as the base for computing the limiting values for the variability brightness temperatures and the
corresponding Doppler factors. In the frequency domain, much attention was given to  the spectral evolution seen
in each source, chiefly because of the signatures that the variability mechanisms leave in the evolutionary
path followed by the flaring events. Furthermore, the power delivered by source in the radio regime was computed at
one selected frequency and was compared with the powers seen in the jets of typical blazars. Finally, the radio
and the optical polarisation was investigated as a signature of relativistic jet emission.

% The archival radio loudness classifies 3 of our sources as very RL and one as only moderately RL.
The mean flux densities are between 200 and 600~mJy. All sources appear significantly variable at cm and mm
bands, except maybe J0849$+$5108, for which it is rather premature for such a statement to be made. The flare
decomposition algorithm revealed several events in each light curve. The characteristics of different
events can vary even in the same light curve, implying that either the underlying mechanism is
not the same, or that each time different evolutionary stages are detected. The delays between the peaks
at different frequencies of the same event range between a few days, tenths of days, and 180 days, showing a
power-law frequency dependence. The flare timescales range between $\approx 30$ and 200--300~days. The
computed highest variability brightness temperature $T_{\rm{var}}$ is between $10^{12}$ and
$10^{13}$~K. Consequently, values ranging from 4.3 to 10.4 were found for the corresponding Doppler
factor, $D$.

From examining the behaviour in the frequency domain, it is immediately clear  that intense
variability occurs in the same ways as routinely seen in blazars. Specifically, the variability events seem
to appear at high frequencies and evolve downwards in frequencies, following an evolutionary path seen in
blazar flares. Interestingly, the power of the events is a significant part of the overall power.
Apart form the case of J0324$+$3410 that shows signs of a steep underlying spectrum the rest give no hints for
such in our bandpass.  

The 14.60~GHz radio power 
%-- assuming that it is produced by a jet -- 
ranges roughly between
$10^{43}$ and $10^{45}$~erg~s$^{-1}$ in agreement with what should be expected from a relativistic jet.  The
8.36~GHz polarisation data indicated low degree of fractional polarisation for three of our
sources. For J0324$+$3410 however it reached values up to almost 6~\%. Strong indication that we are dealing
with the non-thermal emission originating at a relativistic jet which is even dominating over a core emission
despite the source compactness at VLBI scales.  Despite the interesting findings of first
attempts to conduct optical linear and radio linear and circular polarisation study (that we have already
discussed in Sect.~\ref{sec:polarization} and will discuss below), a methodical and complete polarisation
analysis will be presented in subsequent publication after longer optical polarisation time baselines are
completed. 
%For the reader's convenience, the points worth discussing are gathered in a separate paragraph for
%each source before a complete picture is drawn.

\subsection{J0324$+$3410} 
As we extensively discussed in Sect.~\ref{sec:rlcs}, the source light-curve has similar characteristics
as blazars with the only exception of the moderate average flux densities. In combination with the picture
drawn from the radio -- almost monthly sampled -- SEDs and their measured spectral indices, this points to the
scenario of a jet within which high-frequency components are generated. The HFCs
subsequently evolve transversing the possible opacity phases, before disappearing
with an expansion towards low frequencies \citep{Marscher1985ApJ,turler2000AnA...361..850T}. The emergence of
later events of similar phenomenology indicates that the physical conditions are stable (particle densities,
magnetic field strength) at the location of the emission and could be in accordance with a reoccurring
activity that would result from a stable or, a slowly evolving acceleration region. The variability of
J0324$+$3410 is characterised by fast variations -- shorter than 65 days, even at the lowest frequencies -- with
relatively low amplitude. From the flare decomposition exercise we estimate that the highest brightness
temperature limits are all higher than $2 \times 10^{12}$~K, implying a Doppler boosting factor of at least 4.3,
rather low compared with the variability values seen in blazars \citep[e.g.][]{2009A&A...494..527H}.
%It is
%noteworthy that Fuhrmann et al. (in prep.) are combining the variability Doppler factors with kinematic VLBI
%studies to extract the jet angle and velocity.

An essential element of the exact processes operating at the source would be studying of polarisation
behaviour at radio and optical wavelengths.
% Particularly important for the radio emission -- that can recursively undergo a transition from optically
% thick to thin (contrary to the optical), would be the investigation of the polarisation light curves.
% significantly larger than those typically observed with single-dish measurements in AGN around these
% frequency bands ($\sim4$~\% at 5~GHz) \red{(JIANNIS: a good reference)}).
As was discussed earlier, J0324$+$3410 displayed fractional linear polarisation of about 6~\% and 7~\% at
8.35~GHz and 4.85~GHz, respectively \citep{2012nsgq.confE..58A}. This increase towards lower frequencies might
be an indication that optically thin emission -- arguably from a relativistic jet -- becomes
progressively dominant towards lower frequencies. At higher frequencies the emission is dominated by optically
thick (and hence less polarised) emission of an evolving HFC. In the optical regime the detected 
significant polarisation has been seen as evidence of an, at least mildly, relativistic jet
\citep{2008ApJ...685..801Y}. Our $R$-band RoboPol monitoring \citep{2014MNRAS.442.1706K,2014MNRAS.442.1693P}
showed that the optical polarisation of the source is very low and shows fluctuations between 1 and 3~\%. Its
$R$-band magnitude is almost stable at roughly 15.3~mag. A possible explanation for this unusually
low value of the optical polarisation may be the contamination of the emission with an unpolarised stellar
component that would result a lower net polarisation.

Finally, the total jet powers computed from our 14.6~GHz dataset agree well with those computed with SED
modelling that accounts for all different emission components. In conclusion, there is compelling evidence
that the source radio emission is dominated by a jet emission that shows characteristics often seen in
blazars, but not necessarily as powerful.

\subsection{J0849$+$5108} 
Unfortunately, the limited number of observations that have been performed on the source do not allow a proper
light curve analysis. The radio SED, however, already reveals clear signs of -- at least mild -- spectral
evolution indicative of recursive activity, a typical characteristic of radio blazars. Its spectral index
shows variability at all sub-bands (see Sect.~\ref{sec:spec_evolution}) and reaches values of $+1$, indicating
that our bandpass samples a part of the spectrum where most of the radio activity occurs. The dominance of the
assumed optically thick emission could also explain the very low radio polarisation.
%Longer datasets will confirm this assumption if the spectrum becomes soft with and increase in
%the polarisation. 
%\cite{2008ApJ...685..801Y} reported the detection of significant optical polarisation by
%cite{1994A&AS..106..303N}. As it will be discussed in a separate publication, the first robopol attempts also
%indicate significant $R$-band polarisation.

\subsection{J0948$+$0022} 
The detection of J0948$+$0022 as a $\gamma$-ray emitter \citep{2009ApJ...699..976A} triggered the monitoring
programme we discussed here. Consequently, its light curves are the best sampled. It is very radio-loud
\citep{2003ApJ...584..147Z}, in contrast to the typical NLSy1s, and displays light curves with clear intense
variability. Its radio SED shown in Fig.~\ref{fig:specs} reveals a very interesting case with sequential HFC
appearing at high frequencies, evolving through the bandpass before fading with the emergence of a new
component. The lowest band spectral index shows values as inverted as $+1.26$, while the high band spectral index can be
as steep as $-1.73$, implying that the source activity occurs within our bandpass. A careful
cross-correlation of spectral index trains with the associated radio polarisation will be key to really
proving the presence of the HFC that cause the variability; this would favour a typical blazar-like
jet.

From the flare decomposition it appears that the best constraint in the variability brightness temperature is
set by the 14.60~GHz measurements and gives values $13\times 10^{12}$, implying a variability Doppler factor of
$\sim 8.7$. Karamanavis et al. (priv. comm.) have been trying to detect moving components in 15~GHz
MOJAVE data for the purpose of measuring apparent speeds. However, the source appears to be unresolved. In
conclusion, it could be said that this source resembles most the typical blazar behaviour of the four
sources in our sample.

In Sect.~\ref{subsec:power0948} we discussed that the jet powers computed from SED modelling
\citep{2012A&A...548A.106F} are more than an order of magnitude larger that those computed here.
Interestingly, this could be an indication that the jet composition is different from the composition assumed
  for SED modelling.

Recently, \cite{2014arXiv1410.7144D} reported the detection by {\it Fermi} of a strong $\gamma$-ray flare
  observed between December 2012 and January 2013 in the 0.1--100~GeV energy range. Its peak occurred at MJD
  56293. A quasi-simultaneous flare was observed from optical to X-rays, suggesting a tight connection of
  the corresponding emitting regions.  The examination of OVRO 15~GHz light curves showed no flaring activity
  at the time of the $\gamma$-ray flare. A strong radio peak was instead observed in June 2013. The authors
  hypothesised that there might be a delay between the $\gamma$-ray and the radio activity of about five months, due to opacity
  effects and propagation of the shock along the jet.  The case becomes very interesting when examining our
  light curves, which show the detection of a radio-flare quasi-simultaneous to the $\gamma$-ray flare. As can
  be seen in Fig.~\ref{sec:rlcs}, the 23.05~GHz data show a very prominent flare. It starts at the beginning of
  December 2012, peaks between December 15 and January 4, 2013, and ends at the beginning of February
  2013. The flare is weaker in the 32.00~GHz light curve, while at 14.60~GHz it is practically
  invisible. Unfortunately, the insufficient sampling of highest frequencies in our programme prevent us from
  quantifying the flare dependence in frequency. Our data suggest that the $\gamma-$ray flare
  propagates to radio frequencies down to $\sim$23.05~GHz with almost no time delay, showing that the
  propagation of the shock along the jet is fast. At lower frequencies, the flare drops drastically, probably
  because of opacity effects. The examination of the radio SED implies a subtle HFC that barely reaches the
  centre of our bandpass before it disappears. In any case, the sudden disappearance of the event as a
  function of frequency demands more investigations of the mechanisms behind the very broadband emission
  (from $\gamma$ rays to radio) and its evolution.

\subsection{J1505$+$0326} 
The radio SED of J1505$+$0326 differs from the usual behaviour seen in the other NLSy1s in our
sample. Its spectrum is most often convex with its peak in the range of 5--10~GHz. Occasionally,
there appear signs of evolution caused by an HFC entering the bandpass.  It is essential, however, that the
total number of available cycles does not allow deciding whether the spectrum is intrinsically
convex or whether it possibly changes at a slow pace.

The flare decomposition detected four rather short events and have moderate
amplitude despite the source's radio-loudness. The highest variability brightness
temperature limit was computed at 2.64~GHz and yielded values of $\sim 2.6\times 10^{13}$~K, 
implying a boosted relativistic jet with Doppler factors of  more than 10.

\section{Conclusions}
\label{sec:conclusions}
Before we conclude, we emphasise that our findings cannot necessarily
be assumed to be  representative of all RL, $\gamma$-ray NLSy1 galaxies. The most outstanding findings
of this work are as follows:
\begin{enumerate}
\item We presented the most systematic radio monitoring of RL and $\gamma$-ray-detected
  NLSy1s to date. The obtained light curves spread over periods longer than five years. The acquired
  dataset includes eight frequency bands between 2.64 and 142.33~GHZ and is available online for
  the community for further studies. 
\item The light curves show the typical phenomenology seen in blazars, although
  they generally have  lower flux densities. All of them
  show variability events that  in most cases have a smaller amplitude, however. Sometimes very prominent and
  energetic events appear and dominate the light-curve appearance.
\item We introduced an alternative method for detecting flares and quantifying their
  characteristics, which -- contrary to the traditional methods -- is approaching the problem on the basis of
  individual flares. In that context,  the properties of each flare can change over frequency; similarly, the
  different events in the same light curve can appear with different properties, which indicates either different
  variability mechanisms or a different flare evolution stage. 
\item The flare decomposition method  showed that the computed variability brightness temperatures
  are  in general moderate -- in contrast to the majority of blazars --, implying Doppler factors that do not exceed about 10. The operating
  jet then must be moderately relativistic. 
\item The same method showed that the duration of the events is somewhat shorter than what is seen in blazars, 
  resulting in slightly more frequent events. 
\item For the three cases where longer periods have been covered, the radio SEDs clearly show intense
  variability that is mostly characterised by clear spectral evolution of the events. The evolution seems to
  occur slightly faster than in blazars -- as was also indicated by the results of the flare decomposition --
  but follows patterns possibly caused by evolving internal shocks. Furthermore, in one case the shape of the
  radio SED towards the lower frequencies gives a direct indication of a quiescent jet (steep spectrum,
  steady).
\item The jet power estimations gave values  in accordance with jet outputs comparable to the least energetic
  blazars (BL\,Lac objects). For J0948$+$0022 the comparison with SED modelling results indicates
  lower values that may be  attributed to different p-e ratios in the jet. 
\item The first multi-frequency radio polarisation results showed that our sources have shown very low or negligible
  polarisation. In contrast, J0324$+$3410 a displayed polarisation degree of 6 and 7~\% at 8.35 and
  4.85~GHz. This is rather high, especially given the compactness of the source, and could be explained by
 the  jet dominating the core in the total power. The increase towards lower
  bands is in accordance with that assumption as more and more optically thin emission from an underlying jet
  becomes detectable. 
From the above it becomes rather evident that 
\item a mildly relativistic boosted jet must be operating at these sources and probably is responsible for the
  observed radio emission. This  jet lie, probably towards the low end of the energy distribution, however. 
\item The characteristics of its radio emission are qualitatively similar to those in blazars, but are generally
  less intense. The spectral evolution seems to occur  at a faster pace,  involving shorter timescales
  probably related to the systematically smaller black-hole masses or higher accretion rates. 
\item In the particular case of J0948$+$0022, a flare detected by {\it Fermi} towards the end of 2012
    seems to be present in our 23.05~GHz dataset, but is practically absent at frequencies below that. This unusual
    behaviour deserves further investigation of the possible mechanism that could cause such an abrupt flare
    disappearance.
\end{enumerate}

\begin{acknowledgements}
  Our study is based on observations carried out with the 100~m telescope of the MPIfR
  (Max-Planck-Institut f\"ur Radioastronomie) and the IRAM 30~m telescope. IRAM is
  supported by INSU/CNRS (France), MPG (Germany) and IGN (Spain). This research has made
  use of the NASA/IPAC Extragalactic Database (NED) which is operated by the Jet
  Propulsion Laboratory, California Institute of Technology, under contract with the
  National Aeronautics and Space Administration. IM and VK are supported for this
  research through a stipend from the International Max Planck Research School (IMPRS) for
  Astronomy and Astrophysics at the Universities of Bonn and Cologne. NM is funded by an ASI fellowship under contract number I/005/11/0
\end{acknowledgements}

%\bibliographystyle{aa} % style aa.bst
%\bibliography{/Users/mangel/work/Literature/MyBIB/References.bib} % your references Yourfile.bib

%\Online

\begin{appendix} %First online appendix

\section{flare decomposition method}
\label{sec:themethod}
The detection and parametrisation of different flares (which are then used to compute the associated
variability brightness temperature) requires solving several problems, most of which have to do with the
paradox that before an independent parameterisation of all flares is reached, we need to identify their average
shape and spectral evolution, minimise the arbitrariness of the results. The problems to be
addressed can be summarised as follows:
\begin{enumerate}
\item We ned to identify of the strongest flare, accounting for the fact that its amplitude may strongly depend on
  frequency so that the strongest flare at one frequency may be a minor event at others.
\item We also need to localise of the same flare at different frequencies, given its spectral
  evolution.
\item The flares need to be modelled efficiently using the minimum number of free parameters.
\item Finally, we have to test and possibly correct the model to optimise the results.
\end{enumerate}
The implementation of these steps was made as follows:
\begin{enumerate}
\item We estimated the average time delays between the flux density variations detected in different light
    curves. The main complication in this procedure is the fact that using a standard cross-correlation
  function \cite[see][]{1988ApJ...333..646E,1992ApJ...384..453L} on pairs of light curves does not ensure
  consistent results. For example, assuming the estimated delays between three light curves $a$, $b$, and $c$
  to be $\tau_{a,b}$, $\tau_{a,c}$ and $\tau_{b,c}$, it might well be that $\tau_{a,c} \neq \tau_{a,b} +
  \tau_{b,c}$.  A solution of  this problem is to include all the possible pairs of light curves when
  calculating the time delays. By visual inspection, a maximum time delay between light curves,
  $\tau_{max}$, is identified. Then, for every pair of light curves $i$ and $j$ with $i<j$, a cross-correlation
  function $r_{i,j}(\tau_{i,j})$ is calculated for any time delay $\tau_{i,j}$ in the range ($-\tau_{max},
  \tau_{max}$). For $n$ monitored frequencies
  % , and given a set of time delays $(\tau_{1,2}, \tau_{2,3}, ..., \tau_{n-1, n})$,
an overall correlation function can be calculated as
\begin{equation}
  R(\tau_{1,2}, \tau_{2,3}, ..., \tau_{n-1, n})=\prod_{i,j} r_{i,j}(\tau_{i,j}),
\end{equation}
where
\[
\begin{array}{lp{0.8\linewidth}}
  \tau_{i,j}=\tau_{i,i+1}+\tau_{i+1,i+2}+...+\tau_{j-1,j},  & \\
\end{array}
\]
We computed $R$ for any possible set of time delays, excluding all the combinations involving
$r_{i,j}(\tau_{i,j})<0$. The most significant set of time delays was selected as the one providing the highest
overall correlation value.
\item We composed  a ``cumulative'' light curve, in which the strongest flare was identified. The
  light curves were shifted in time by the best fit delays found in the previous step, so that their variations
  were aligned. Their fluxes were normalised by their standard deviation, to have similar variations
  at all wavelengths. Assuming for example that all the variations at 86.24~GHz have an amplitude of 1~Jy and at
  2.64~GHz have an amplitude of 0.~Jy, they all count the same for the identification of the flares. This way, we
  ensured that our detection algorithm was not biased towards the highest frequencies. The resulting light
  curves were merged to create one well-sampled total ``cumulative'' light curve. The merging of the light
  curves does not require averaging; the points from all the light curves are combined regardless of
  the frequency at which they were observed. The previous steps are meant to make all points equivalent, giving us
  the opportunity create a well-sampled light curve in which every data point from every frequency
  participates with the same weight. To construct the cumulative light curve, one could
  identify the following steps:
\item We identified the strongest flare. The cumulative light curve was
  smoothed (with an adaptive smoothing factor) and localised the highest flux density.
\item We localised the same flare at different frequencies. Knowing the average time delays between
  variations occurring at different frequencies, the strongest flare can be identified in each of the actual
  light curves. Gaussian fit models, different for each wavelength, were applied to the actual light curves to
  determine the amplitude and timescale of the strongest flare; they were then subtracted from the original
  flux densities, creating a new set of light curves (one per wavelength) for which all the steps from step 3 on
  are repeated. The procedure was iterated until no other significant flares are detected.
\item  Model test. The residual light curves were
  checked to verify that no peaks or dips are systematically present in the position of
  the subtracted flares. Their existence would indicate that the flares are not properly
  modelled; new models would be created and the whole procedure repeated until a
  satisfactory result is reached.
\end{enumerate}

% "synthetic light curve" > better "cumulative light curve"? because that it what it
% really is: a cumulative light curve, produced by conveniently shifting and
% re-normalizing the real light curves...
This approach has the advantage of individually localising and modelling
each event. Once a flare is detected in a cumulative light curve, the average time delays
returned by step 1 allow us to infer its approximate position in the real light
curves. The accurate position is then identified from the flux peak closest to
the approximate position. In this way, we are able to estimate the actual time delays as a
function of wavelength for each flare separately, and therefore detect possible variations
in their spectral evolution. Naturally, the procedure also provides information about how
the amplitude and the rising and decaying times of the flares vary with wavelength, which 
leads to a full parametrisation of the flux density variations observed in the sources;
this is necessary to compute brightness temperatures.

\section{Results of the flare decomposition}
Here we append the results of the flare decomposition method described in
Sect.~\ref{sec:decomp_results}.
\longtab{2}{
\begin{longtable}{rrcccrc}
\caption{\label{tab:allflares} Results of the flare decomposition for all three sources.}\\
\hline\hline
    \multicolumn{1}{c}{$\nu$} &\multicolumn{1}{c}{Time} &Amplitude  &Peak     &Time  &\multicolumn{1}{c}{$T_\mathrm{var}$} &$D$\\
                              &\multicolumn{1}{c}{delay}         & &position &scale &        &\\
    \multicolumn{1}{c}{(GHz)} &\multicolumn{1}{c}{(d)}      &(Jy)   &(MJD)    &(d)
    &\multicolumn{1}{c}{($10^{11}$~K)} & \\    
\hline
\endfirsthead
\caption{continued.}\\
\hline\hline
    \multicolumn{1}{c}{$\nu$} &\multicolumn{1}{c}{Time} &Amplitude  &Peak     &Time  &\multicolumn{1}{c}{$T_\mathrm{var}$} &$D$\\
                              &\multicolumn{1}{c}{delay}         & &position &scale &        &\\
    \multicolumn{1}{c}{(GHz)} &\multicolumn{1}{c}{(d)}      &(Jy)   &(MJD)    &(d)
    &\multicolumn{1}{c}{($10^{11}$~K)} & \\    
\hline\\
\endhead
\\
\hline
\endfoot
\\\multicolumn{7}{c}{J0324$+$3410}\\\\
\multicolumn{7}{c}{Flare 1}\\                            
32.00 &$-$100  &1.284  &56313  &46  &4.6 &2.2\\            
23.05 &$-$101  &0.744  &56312  &38  &8.0 & 2.7\\            
14.60 &$-$105  &0.514  &56308  &33  &18  & 3.6\\            
10.45 & $-$82  &0.205  &56331  &47  &6.9 & 2.6\\            
8.35  & $-$81  &0.133  &56333  &65  &3.9 & 2.2\\            
4.85  & $-$52  &0.105  &56362  &64  &8.6 & 2.9\\            
2.64  &   0  &0.081  &56413  &61  &25  & 4.3\\\\          
\multicolumn{7}{c}{Flare 2}\\                            
32.00 &$-$108   &0.139  &56356  &32  &1.0 & 1.3\\           
23.05 &$-$109   &0.157  &56355  &30  &2.7 & 1.9\\           
14.60 & $-$78   &0.180  &56385  &29  &8.1 & 2.7\\           
10.45 & $-$79   &0.157  &56384  &45  &5.8 & 2.4\\           
8.35  & $-$82   &0.130  &56381  &65  &3.8 & 2.2\\           
4.85  & $-$30   &0.132  &56433  &66  &10.2& 3.1\\          
2.64  &   0   &0.086  &56463  &63  &24 & 4.3\\            
\\                                                       
\multicolumn{7}{c}{Flare 3}\\                            
32.00 &$-$73   &0.121   &55747  &30  &1.0 & 1.3\\           
14.60 &$-$45   &0.146   &55775  &29  &6.5 & 2.5\\           
10.45 &$-$8    &0.114   &55812  &45  &4.2 & 2.2\\            
8.35  &$-$29   &0.108   &55791  &64  &3.1 & 2.0\\           
4.85  &$-$14   &0.108   &55806  &64  &8.9 & 3.0\\           
2.64  &  0   &0.084   &55820  &63  &24 & 4.2\\            
\\                                                       
\multicolumn{7}{c}{Flare 4}\\                            
32.00 &$-$24   &0.116   &55638  &30  &1.0 & 1.3\\           
14.60 &$-$26   &0.158   &55636  &29  &7.1 & 2.6\\           
10.45 &$-$23   &0.145   &55639  &45  &5.4 & 2.4\\           
8.35  &$-$25   &0.133   &55638  &65  &3.9 & 2.2\\           
4.85  & $-$2   &0.110   &55661  &64  &9.0 & 3.0\\           
2.64  &  0   &0.081   &55662  &61  &25 & 4.3\\            
\\                                                       
\multicolumn{7}{c}{Flare 5}\\                            
32.00 &$-$73   &0.072   &56183  &26  &0.8 & 1.2\\           
14.60 &$-$57   &0.137   &56199  &27  &7.0 & 2.6\\           
10.45 &$-$43   &0.128   &56213  &45  &4.7 & 2.3\\           
8.35  &$-$15   &0.108   &56240  &65  &3.1 & 2.0\\           
4.85  &  5   &0.093   &56260  &64  &7.7 & 2.8\\           
2.64  &  0   &0.070   &56256  &61  &21 & 4.1\\            
\\\multicolumn{7}{c}{J0948$+$0022}\\\\                   
\multicolumn{7}{c}{Flare 1}\\                            
32.00 &$-$175   &0.400   &55617   & 44   & 53   & 7.6 \\     
23.05 &$-$179   &0.320   &55614   & 50   & 69   & 7.1 \\     
14.60 &$-$153   &0.305   &55640   & 61   & 102  & 8.1\\    
10.45 &$-$148   &0.249   &55644   &103   & 57   & 6.7\\      
8.35  & $-$92   &0.198   &55700   &103   & 76   & 6.6\\      
4.85  & $-$87   &0.098   &55705   &146   & 51   & 5.9\\      
2.64  &   0   &0.064   &55792   &268   & 34   & 5.3\\      
\\                                                       
\multicolumn{7}{c}{Flare 2}\\                            
32.00 &$-$168   &0.337   &55705  &  42   & 49   & 7.4\\      
23.05 &$-$159   &0.332   &55715  &  50   & 71   & 7.2\\      
14.60 &$-$153   &0.309   &55721  &  61   & 103  & 8.1\\    
10.45 &$-$128   &0.267   &55745  & 103   & 61   & 6.9\\      
8.35  & $-$92   &0.185   &55782  & 101   & 73   & 6.6\\      
4.85  & $-$65   &0.101   &55808  & 146   & 53   & 6.0\\      
2.64  &   0   &0.058   &55874  & 260   & 32   & 5.2\\      
\\                                                       
\multicolumn{7}{c}{Flare 3}\\                            
32.00 &$-$171   &0.331   &56406  &  42   & 48   & 7.4\\      
23.05 &$-$168   &0.277   &56409  &  50   & 60   & 6.8\\      
14.60 &$-$133   &0.408   &56444  &  62   & 133  & 8.7\\    
10.45 &$-$133   &0.400   &56444  & 110   & 81   & 7.5\\      
8.35  &$-$125   &0.306   &56452  & 108   & 106  & 7.3\\    
4.85  & $-$96   &0.138   &56481  & 151   & 68   & 6.4\\      
\\                                                       
\multicolumn{7}{c}{Flare 4}\\                            
32.00 &$-$173   &0.320   &55446  &  42   & 46 & 7.3\\        
23.05 &$-$161   &0.240   &55458  &  50   & 52 & 6.5\\        
14.60 &$-$146   &0.119   &55473  &  55   & 48 & 6.4\\        
10.45 &$-$139   &0.105   &55479  &  93   & 30 & 5.5\\        
8.35  &$-$131   &0.094   &55487  &  94   & 43 & 5.6\\        
4.85  & $-$96   &0.062   &55523  & 137   & 37 & 5.4\\        
2.64  &   0   &0.037   &55619  & 230   & 26 & 4.9\\        
\\                                                       
\multicolumn{7}{c}{Flare 5}\\                            
14.60 & $-$71   &0.092   &54998  &  52   & 42 & 6.1\\        
10.45 & $-$56   &0.071   &55013  &  91   & 21 & 5.0\\        
8.35  & $-$41   &0.053   &55028  &  87   & 29 & 5.0\\        
4.85  &   7   &0.038   &55076  & 128   & 26 & 4.9\\        
2.64  &   0   &0.026   &55069  & 206   & 23 & 4.7\\        
\\                                                      
\multicolumn{7}{c}{Flare 6}\\                            
32.00 &$-$168   &0.180   &55978  &  37   & 33 & 6.5\\        
23.05 &$-$170   &0.163   &55976  &  45   & 43 & 6.2\\        
14.60 &$-$175   &0.095   &55972  &  52   & 43 & 6.2\\        
10.45 &$-$157   &0.098   &55989  &  93   & 28 & 5.4\\        
8.35  &$-$147   &0.084   &55999  &  92   & 40 & 5.5\\        
4.85  &$-$105   &0.059   &56041  & 137   & 35 & 5.3\\        
2.64  &   0   &0.032   &56146  & 221   & 24 & 4.8\\        
\\\multicolumn{7}{c}{1505$+$0326}\\\\                    
\multicolumn{7}{c}{Flare 1}\\                            
32.00 &$-$104   &0.235   &55545  &  29  & 48 & 7.4\\     
14.60 & $-$37   &0.174   &55611  &  50  & 58 & 7.6\\      
10.45 & $-$10   &0.177   &55639  &  80  & 44 & 6.9\\      
8.35  & $-$13   &0.165   &55636  &  78  & 72 & 6.9\\      
4.85  &  $-$5   &0.148   &55643  &  99  & 111& 7.9\\     
2.64  &   0   &0.112   &55649  & 109  & 233& 10.0\\    
\\                                                       
\multicolumn{7}{c}{Flare 2}\\                            
32.00 &$-$120   &0.210   &56146  &  26  & 52  & 7.6\\     
14.60 & $-$33   &0.188   &56233  &  50  & 62  & 7.8\\     
10.45 & $-$32   &0.175   &56234  &  80  & 44  & 6.9\\     
8.35  & $-$30   &0.160   &56235  &  78  & 70  & 6.8\\     
4.85  &  $-$8   &0.156   &56257  & 103  & 108 & 7.8\\    
2.64  &   0   &0.128   &56266  & 111  & 258 & 10.4\\   
\\                                                       
\multicolumn{7}{c}{Flare 3}\\                            
32.00 &$-$107   &0.143   &56088  &  25  & 39  & 6.8\\      
14.60 & $-$55   &0.211   &56140  &  50  & 70  & 8.1\\      
10.45 & $-$53   &0.197   &56141  &  84  & 45  & 6.9\\      
8.35  & $-$49   &0.172   &56145  &  78  & 75  & 7.0\\      
4.85  & $-$21   &0.138   &56173  &  99  & 103 & 7.7\\     
2.64  &   0   &0.101   &56194  & 107  & 219 & 9.8\\     
\\                                                       
\multicolumn{7}{c}{Flare 4}\\                            
14.60 & $-$27   &0.162   &55809  &  48  & 58  & 7.6\\      
10.45 & $-$29   &0.097   &55808  &  77  & 27  & 5.8\\      
8.35  & $-$16   &0.087   &55821  &  73  & 44  & 5.9\\      
4.85  &  $-$5   &0.092   &55831  &  95  & 75  & 7.0\\      
2.64  &   0   &0.078   &55836  & 103  & 181 & 9.2\\     
\end{longtable}                                                                                                                 
}% End \longtab
%-----------------------------------------------------------------------

% -----------------------------------------------------------------------
\begin{figure*}[] 
\centering
\begin{tabular}{lll}
\includegraphics[trim=0pt 0pt 0pt 80pt  ,clip ,width=0.27\textwidth,angle=-90]{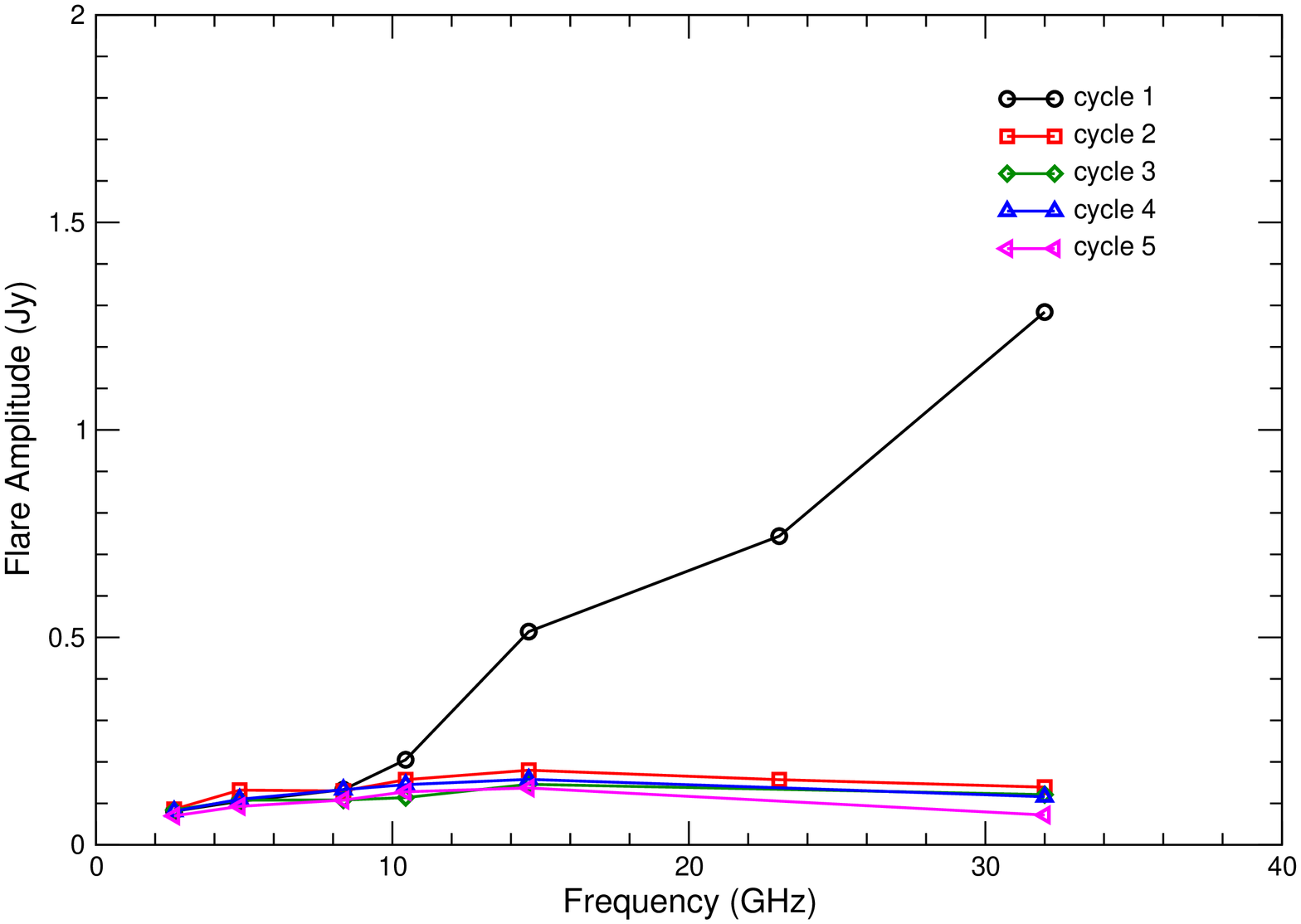} &\includegraphics[trim=0pt 0pt 0pt 30pt  ,clip ,width=0.27\textwidth,angle=-90]{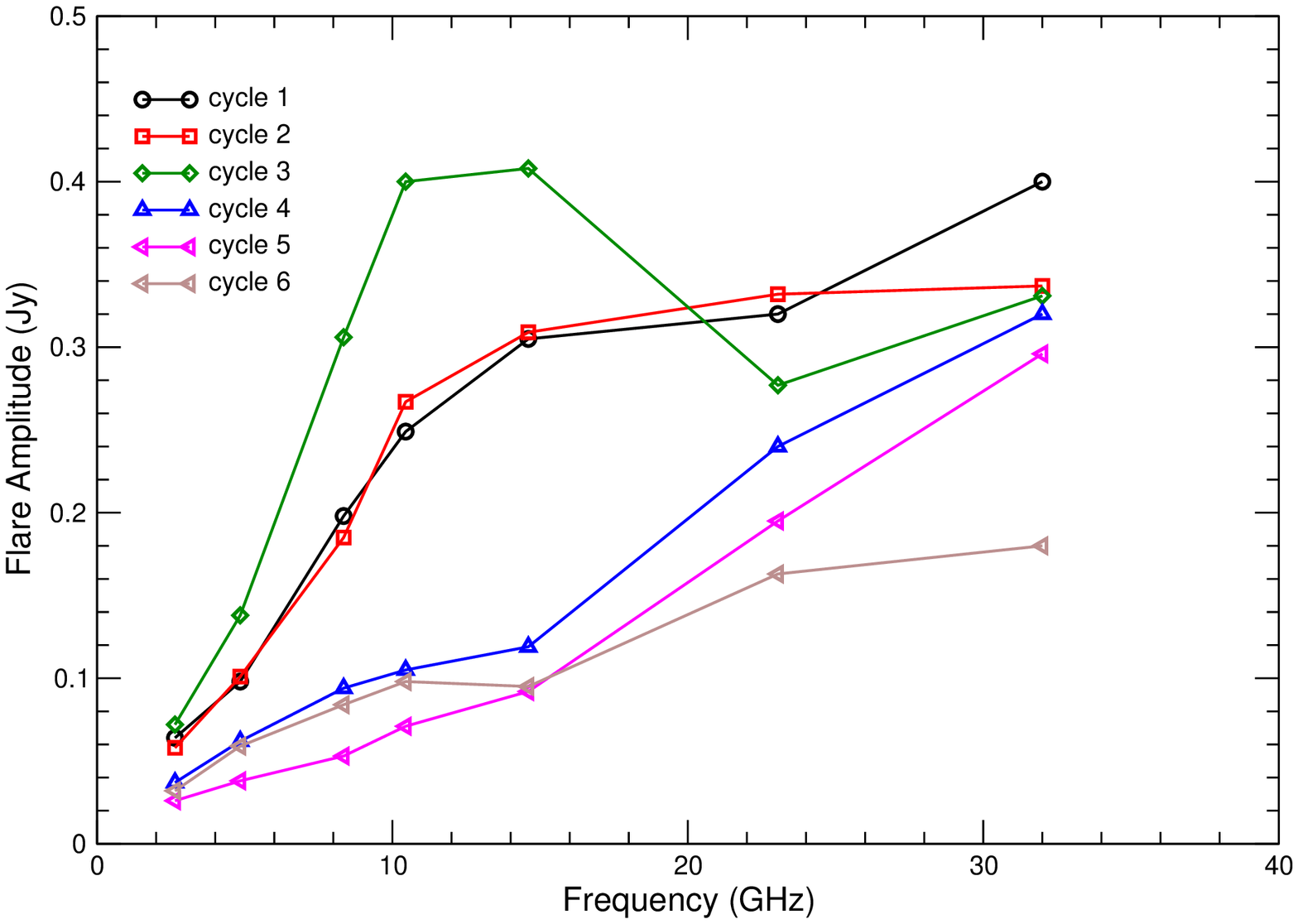}&\includegraphics[trim=0pt 0pt 0pt 30pt  ,clip ,width=0.27\textwidth,angle=-90]{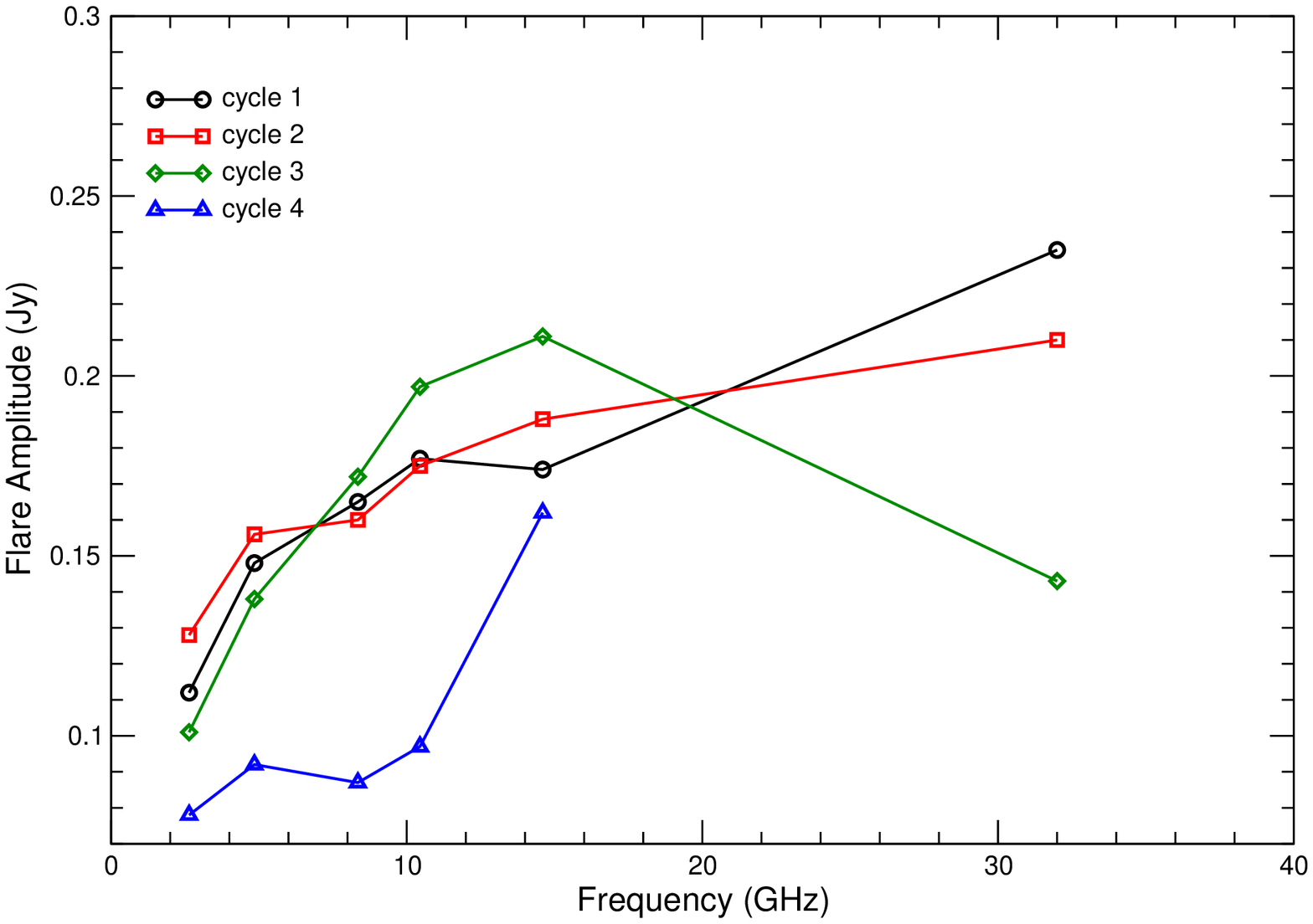}\\ 
\includegraphics[trim=0pt 0pt 0pt 30pt  ,clip ,width=0.27\textwidth,angle=-90]{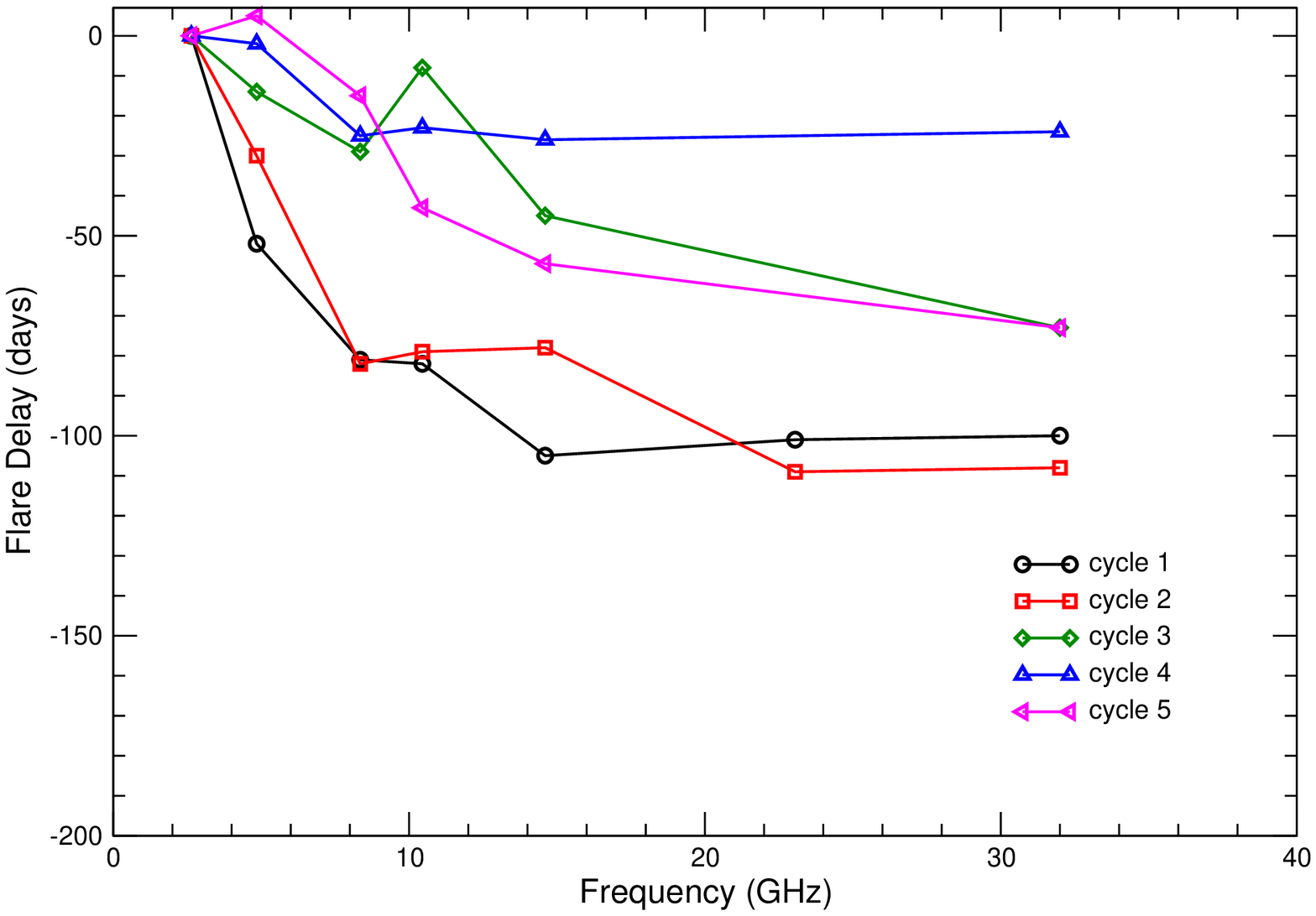} &\includegraphics[trim=0pt 0pt 0pt 30pt  ,clip ,width=0.27\textwidth,angle=-90]{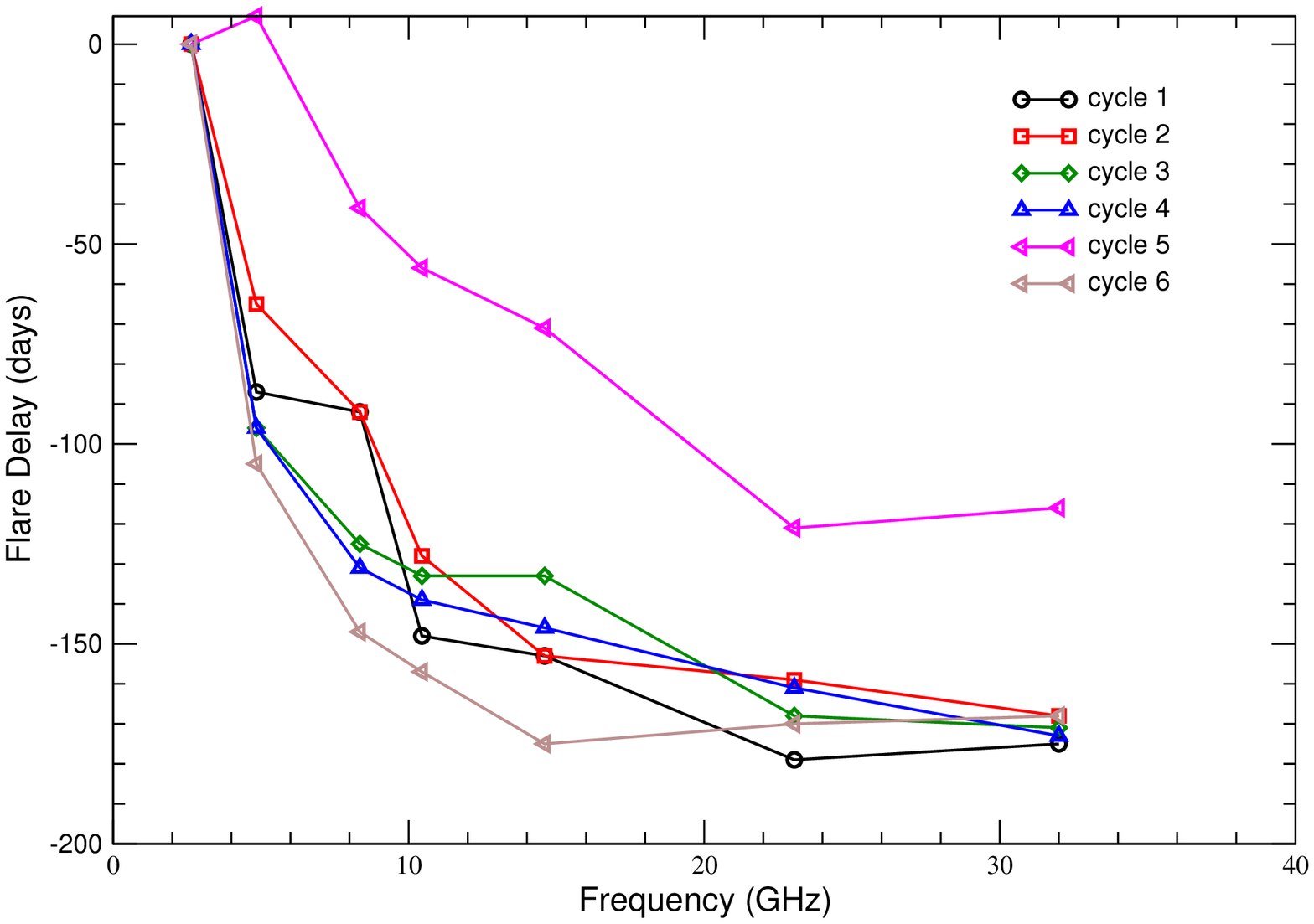}&\includegraphics[trim=0pt 0pt 0pt 30pt  ,clip ,width=0.27\textwidth,angle=-90]{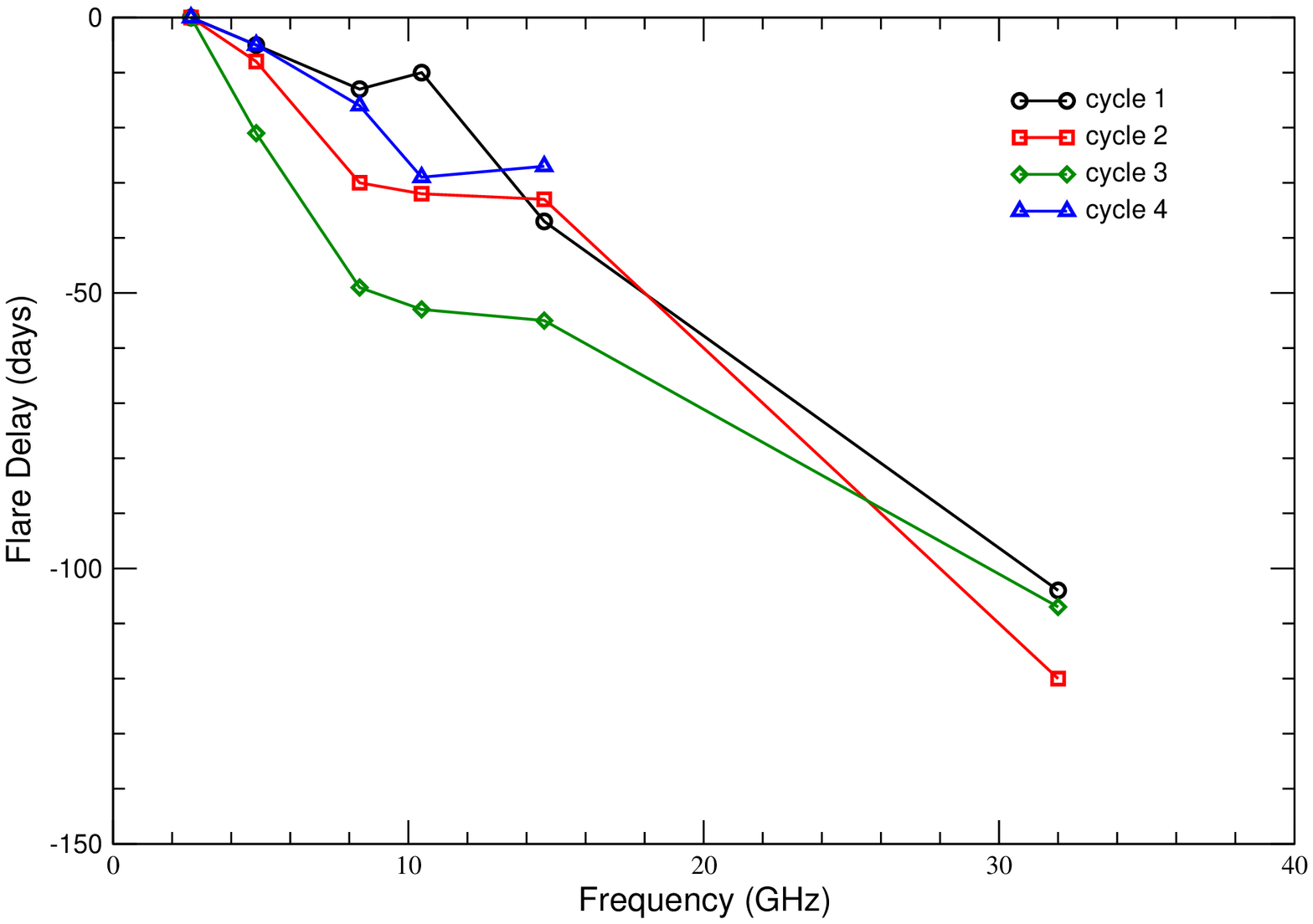}\\
\includegraphics[trim=0pt 0pt 0pt 80pt  ,clip ,width=0.27\textwidth,angle=-90]{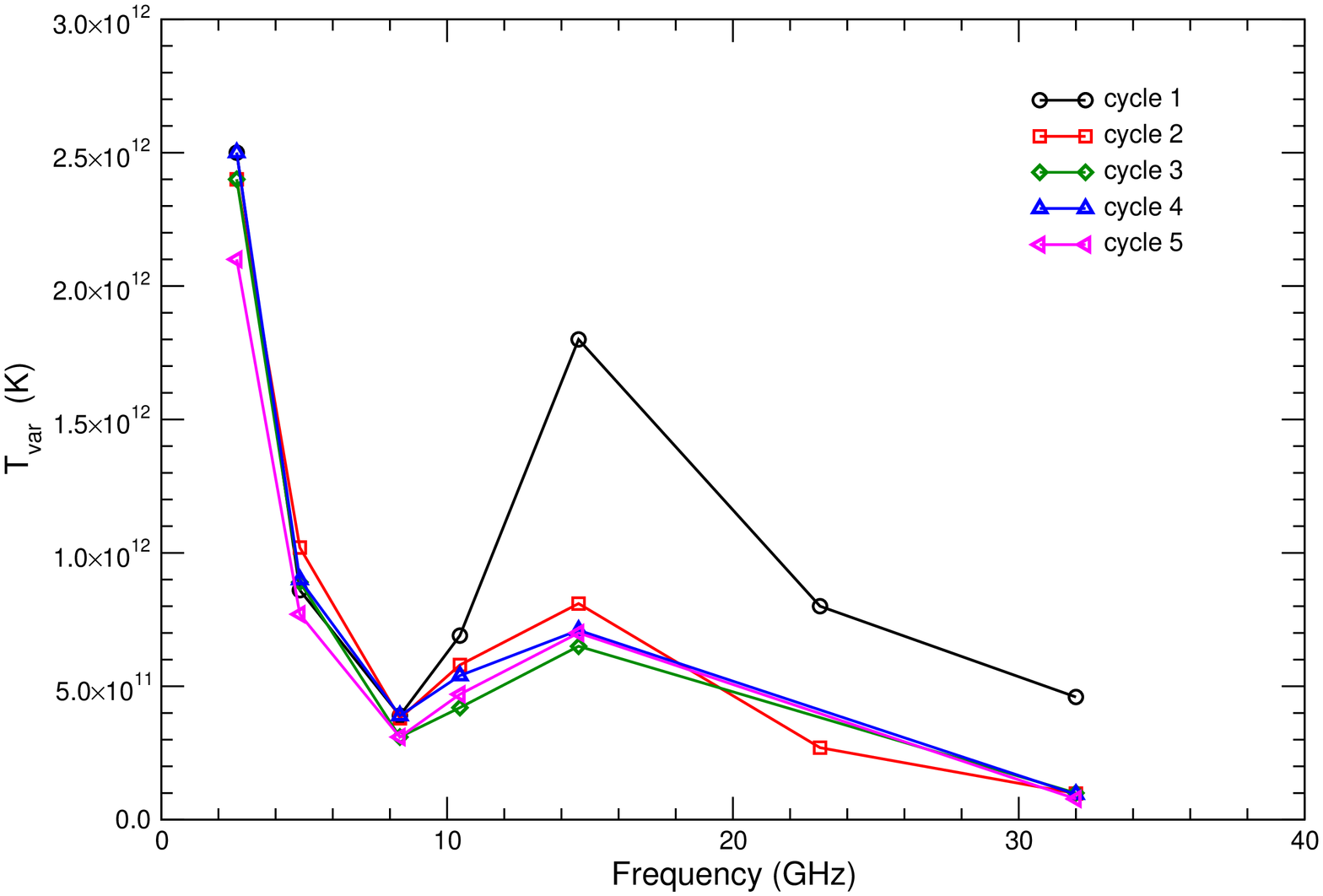}  &\includegraphics[trim=0pt 0pt 0pt 30pt  ,clip ,width=0.27\textwidth,angle=-90]{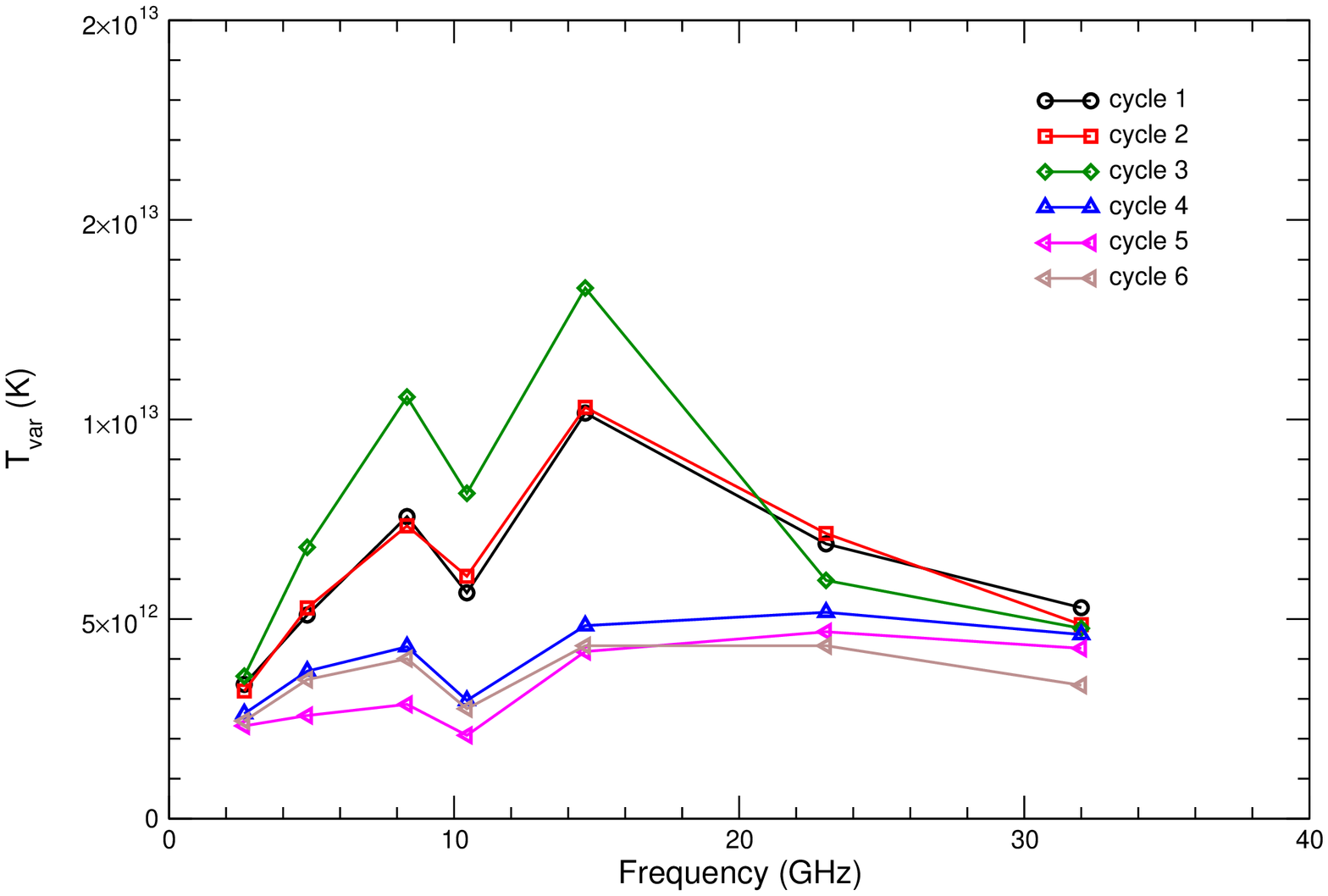}&\includegraphics[trim=0pt 0pt 0pt 80pt  ,clip ,width=0.27\textwidth,angle=-90]{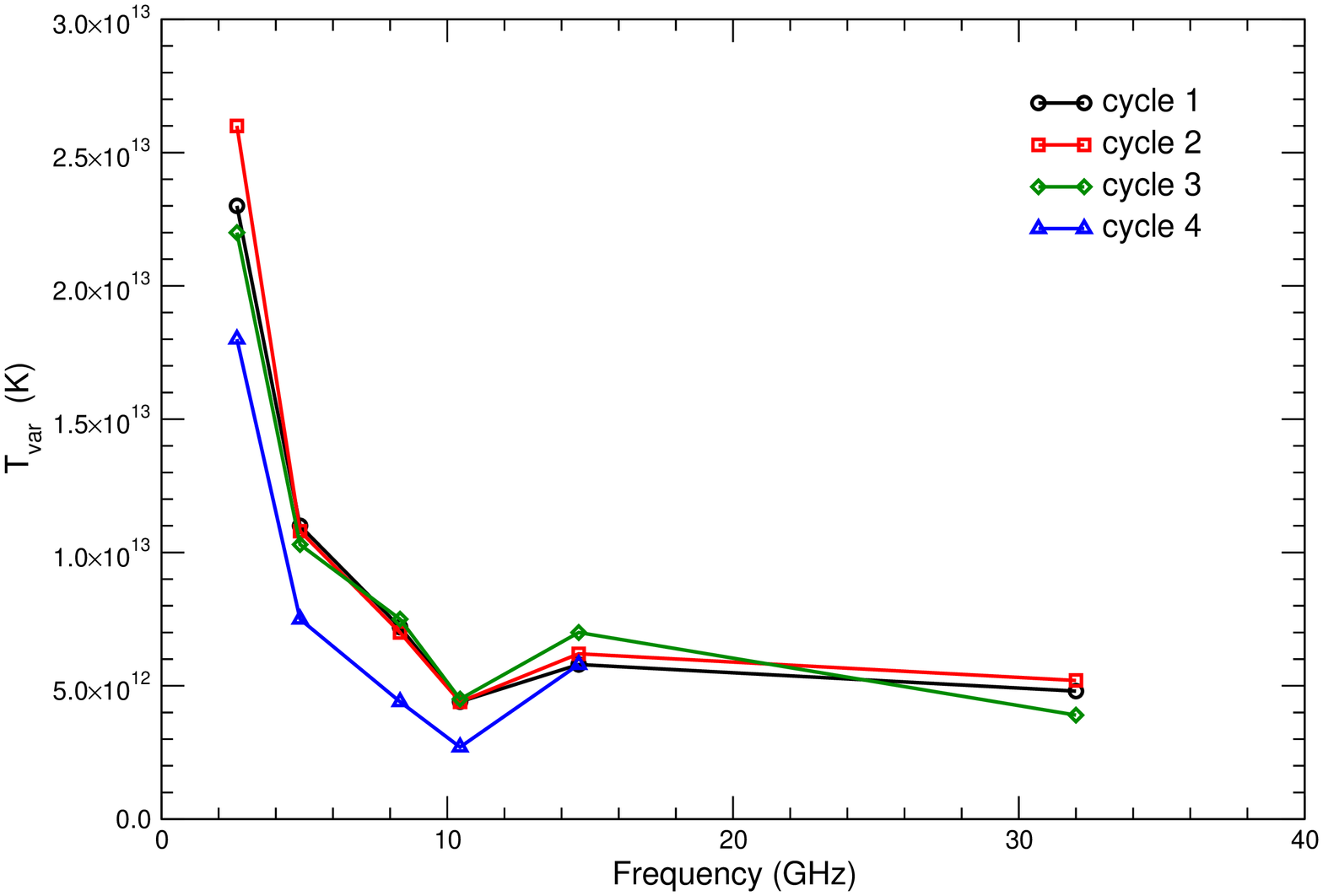}\\ 
\end{tabular}
\caption{Left column: J0324$+$3410:  top panel: the amplitude of the detected
  flares as a function of observing frequency with the cycle labels corresponding to the
  events seen in Fig.~\ref{fig:lc_lmh_03n08}; middle panel: the flare delays; bottom panel: the
  $T_{\rm{var}}$ for the different activity cycles (flares) as a function of observing
  frequency. Middle column: J0948$+$0022. Right column: J1505$+$0326.}
\label{fig:0324_0948decomp}
\end{figure*}
% -----------------------------------------------------------------------

\end{appendix}

\end{document}